\pdfoutput=1
\PassOptionsToPackage{x11names,svgnames}{xcolor}
\documentclass[submission, Phys]{SciPost}
\usepackage{amssymb}
\usepackage{xspace}
\usepackage{graphicx}
\usepackage{afterpage}

\usepackage{subfig}
\usepackage{microtype}
\makeatletter
\g@addto@macro\bfseries{\boldmath}
\let\@to\to
\renewcommand{\to}{\ensuremath{\@to}}

\makeatother

\usepackage{lineno}

\usepackage{todonotes}
\usepackage{xpatch}
\makeatletter
\xpretocmd{\todo}{\@bsphack}{}{}
\xapptocmd{\todo}{\@esphack}{}{}
\makeatother

\usepackage{siunitx}
\DeclareSIUnit{\electronvolt}{\text{e\kern-0.15ex V}}
\DeclareSIUnit{\eV}{\electronvolt}
\DeclareSIUnit{\MeV}{\mega\eV}
\DeclareSIUnit{\GeV}{\giga\eV}
\DeclareSIUnit{\TeV}{\tera\kern-0.1ex\eV}
\DeclareSIUnit{\ifb}{\femto\barn\tothe{-1}}

\newcommand{\rivet}{R\protect\scalebox{0.8}{IVET}\xspace}

\newcommand{\hepdata}{HEPData\xspace}
\newcommand{\contur}{\textsc{Contur}\xspace}

\newcommand{\pT}{\ensuremath{p_\mathrm{T}}\xspace}
\newcommand{\MET}{\ensuremath{E_\mathrm{T}^\mathrm{miss}}\xspace}
\newcommand{\CLs}{\ensuremath{\mathrm{CL}_\mathrm{s}}\xspace}

\newcommand{\B}{\ensuremath{B}\xspace}
\newcommand{\T}{\ensuremath{T}\xspace}

\newcommand{\BT}{\ensuremath{BT}\xspace}
\newcommand{\XT}{\ensuremath{X\mspace{-1mu}T}\xspace}
\newcommand{\BY}{\ensuremath{BY}\xspace}
\newcommand{\BTX}{\ensuremath{BT\mspace{-2mu}X}\xspace}
\newcommand{\BTY}{\ensuremath{BTY}\xspace}
\newcommand{\BTXY}{\ensuremath{BT\mspace{-2mu}XY}\xspace}
\newcommand{\triple}[3]{\text{#1:#2:#3}\xspace}
\newcommand{\WZH}{\triple{$W\!$}{$Z$}{$H$}}
\newcommand{\WZHozz}{\triple{1}{0}{0}}
\newcommand{\WZHzoz}{\triple{0}{1}{0}}
\newcommand{\WZHzzo}{\triple{0}{0}{1}}

\newcommand{\WZHtoo}{\triple{$\tfrac{1}{2}$}{$\tfrac{1}{4}$}{$\tfrac{1}{4}$}}
\newcommand{\WZHzoo}{\triple{0}{$\tfrac{1}{2}$}{$\tfrac{1}{2}$}}

\begin{tabular}{llll}
\swatch{cornflowerblue}~ATLAS $WW$ &
\swatch{navy}~ATLAS $\mu$+MET+jet &
\swatch{cadetblue}~ATLAS $e$+MET+jet &
\swatch{dimgrey}~CMS jets \\
\swatch{silver}~ATLAS jets &
\swatch{gold}~ATLAS $\gamma$ &
\swatch{orange}~ATLAS $\ell\ell$+jet &
\swatch{darkorange}~ATLAS $\mu\mu$+jet \\
\swatch{orangered}~ATLAS $ee$+jet &
\swatch{orangered!90}~ATLAS $ee$+jet
\end{tabular}

\vspace*{2em}

\begin{tabular}{llll}
\swatch{cornflowerblue}~ATLAS $WW$ &
\swatch{blue}~ATLAS $\ell$+MET+jet &
\swatch{cadetblue}~ATLAS $e$+MET+jet &
\swatch{snow}~ATLAS $t\bar{t}$ hadr \\
\swatch{darkorange}~ATLAS $\mu\mu$+jet &
\swatch{orange}~ATLAS $\ell\ell$+jet &
\swatch{gold}~ATLAS $\gamma$
\end{tabular}

\vspace*{2em}

\begin{tabular}{lll}
\swatch{cornflowerblue}~ATLAS $WW$ &
\swatch{navy}~ATLAS $\mu$+MET+jet &
\swatch{cadetblue}~ATLAS $e$+MET+jet \\
\swatch{darkorange}~ATLAS $\mu\mu$+jet &
\swatch{dimgrey}~CMS jets &
\swatch{snow}~ATLAS $t\bar{t}$ hadr
\end{tabular}

\swatch{darkolivegreen} ATLAS\_13\_HMDY\\
\swatch{crimson} ATLAS\_3L\\
\swatch{magenta} ATLAS\_4L\\
\swatch{tomato} ATLAS\_7\_DY\\
\swatch{mediumseagreen} ATLAS\_7\_LL\_GAMMA\\
\swatch{lightgreen} ATLAS\_7\_LMET\_GAMMA\\
\swatch{crimson} ATLAS\_8\_HMDY\_LL\\
\swatch{greenyellow} ATLAS\_8\_MM\_GAMMA\\
\swatch{darkgoldenrod} ATLAS\_GAMMA\_MET\\
\swatch{blue} ATLAS\_LMETJET\\
\swatch{green} ATLAS\_METJET\\
\swatch{snow} ATLAS\_TTHAD\\
\swatch{firebrick} ATLAS\_ZZ\\
\swatch{seagreen} CMS\_13\_HMDY\\
\swatch{saddlebrown} CMS\_3L\\
\swatch{hotpink} CMS\_4L\\
\swatch{lightsalmon} CMS\_EEJET\\
\swatch{deepskyblue} CMS\_EMETJET\\
\swatch{yellow} ATLAS\_GAMMA\\
\swatch{goldenrod} CMS\_GAMMA\_MET\\
\swatch{salmon} CMS\_LLJET\\
\swatch{powderblue} CMS\_LMETJET\\
\swatch{darkgreen} CMS\_METJET\\
\swatch{steelblue} CMS\_MMETJET\\
\swatch{darksalmon} CMS\_MMJET\\
\swatch{wheat} CMS\_TTHAD\\
\swatch{royalblue} CMS\_WW\\
\swatch{indianred} CMS\_ZZ\\
\swatch{brown} LHCB\_7\_LLJET\\
\swatch{white} other\\

\begin{document}

\begin{center}
  \Large
  \textbf{New sensitivity of current LHC measurements\\ to vector-like quarks}
\end{center}

\begin{center}
A.~Buckley$^1$
J.~M.~Butterworth$^2$,
L.~Corpe$^2$,
D.~Huang$^2$,
P.~Sun$^1$
\end{center}

\begin{center}
$^1$ School of Physics \& Astronomy, University of Glasgow,\\ University~Place, G12~8QQ, Glasgow, UK\\
$^2$ Department of Physics \& Astronomy, University College London,\\ Gower~St., WC1E~6BT, London, UK
\\
\end{center}

\begin{center}
\today
\end{center}

\vspace{20pt}

\section*{Abstract}
\textbf{
Quark partners with non-chiral couplings appear in several extensions of the Standard Model.
They may have non-trivial generational structure to their couplings, and may be produced
either in pairs via the strong and EM interactions, or singly via the new couplings of the model.
Their decays often produce heavy quarks and gauge bosons, which will contribute to a variety
of already-measured ``Standard Model'' cross-sections at the LHC.
We present a study of the sensitivity of such published LHC measurements to vector-like quarks, first comparing
to limits already obtained from dedicated searches, and then broadening to some so-far unstudied parameter regions.
}


\noindent\rule{\textwidth}{1pt}
\tableofcontents\thispagestyle{fancy}
\noindent\rule{\textwidth}{1pt}


\begin{center}
\emph{This paper is dedicated to the memory of Puwen Sun, 1990-2020.}
\end{center}

\section{Introduction}
\label{sec:intro}
Quark partners with vector-like couplings to the weak sector (VLQs) may exist with masses close enough to the electroweak scale that
they can help solve the hierarchy problem,
and yet they evade many experimental constraints since their mass is not necessarily acquired via the Brout-Englert-Higgs
mechanism~\cite{Aguilar-Saavedra:2013qpa}.
VLQs can be motivated by a variety of higher-energy extensions to the Standard Model (SM), and
strategies for discovering them have been developed at both the Tevatron~\cite{Atre:2008iu}
and the LHC~\cite{Atre:2011ae,Buchkremer:2013bha}. Here we use the
model-independent framework presented in Ref.~\cite{Buchkremer:2013bha}.
The allowed parameter space is nevertheless
quite tightly
constrained by theory, and searches at the LHC have set lower limits of
\SIrange{1}{1.3}{\TeV}
on their masses, for various
scenarios within this framework~\cite{Aaboud:2018xpj,Aaboud:2018saj,Sirunyan:2018qau,Sirunyan:2018fjh,Sirunyan:2018omb}.

Typically, searches at the LHC have assumed the VLQs couple only
to the third generation of SM quarks, since this is the scenario least
constrained by previous measurements~\cite{Buchkremer:2013bha}. For example, as discussed by Buchkremer \textit{et al},
the absence of flavour-changing neutral currents (FCNC) implies strong constraints on the VLQ coupling to the SM when
more than one generation couples, but these constraints are eased when the
coupling is only to the third generation.
Such studies also tend to focus on the production of one flavour of VLQ at a time.

In this paper we examine the sensitivity of available particle-level measurements (as opposed to dedicated
searches) to VLQs, using the \contur framework~\cite{Butterworth:2016sqg}
\footnote{Making use, where possible, of the updated treatment of correlated experimental uncertainties described
in Ref.~\cite{Brooijmans:2020yij}.}
to inject signal into results of LHC analyses
present in the \rivet library~\cite{Buckley:2010ar,Bierlich:2019rhm}, and derive constraints using the \CLs technique~\cite{Read:2002hq}. We use Herwig~\cite{Bahr:2008pv,Bellm:2019zci}
to inclusively generate all leading-order $2 \rightarrow 2$
processes involving one or more VLQs. Next-to-leading-order predictions are available~\cite{Cacciapaglia:2018qep,Fuks:2016ftf} but are beyond the scope of this study.
This inclusive signal generation implies that a wide array of signatures is covered, allowing
us to move away from some of the above simplifying assumptions to provide more general limits\footnote{Since the data are known to
agree with the SM, we do expect limits rather than a discovery at this stage!}.
For each parameter point, $30,000$ BSM events are generated for each beam condition, corresponding to a luminosity significantly greater
than that of the data over the relevant parameter space.

We first give an overview of the phenomenology of VLQ models, and how they might be expected to be produced and decay at the LHC in Section~\ref{sec:pheno}.
We then benchmark \contur against the LHC searches in Section~\ref{sec:BT}, using $B^{-\frac{1}{3}}$ and $T^{\frac{2}{3}}$ production only and assuming the $X^{\frac{5}{3}}$ and $Y^{-\frac{4}{3}}$ VLQs to be decoupled.
In Section~\ref{sec:all}, we extend beyond this simplest model, and study the
sensitivity for all four VLQs active, with degenerate masses, again for the case
of coupling to only the third generation of SM quarks.  In
Section~\ref{sec:kappa} we look at VLQ production as a function
of the overall coupling $\kappa$ and generic VLQ mass $M_Q$, where single production is of particular
interest when allowing non-zero coupling to the first and second generations of
SM quarks.

A selection of non-standard VLQ and leptoquark signatures in different models
was also studied in contribution~5 of Ref.~\cite{Brooijmans:2020yij}, also
showing some interesting sensitivity to these models.

\section{Overview of VLQ phenomenology}
\label{sec:pheno}

Four types of vector-like quark $Q$ are allowed --- $B$, $T$, $X$, and $Y$ ---
which may be arranged in various weak $SU(2)_L$ multiplets.  The $B$ and $T$ are
present in all theoretically-allowed scenarios~\cite{delAguila:2000aa,Aguilar-Saavedra:2013qpa}
and have
EM charges $-\frac{1}{3}$ and $\frac{2}{3}$, like the SM $b$ and $t$, while the
$X$ and $Y$ (with charges $\frac{5}{3}$ and $-\frac{4}{3}$ respectively) may be
elided in some models. All VLQs have quark-like triplet colour charges. In
addition to their quark-like QCD and EM couplings via the usual covariant
derivative recipe, the VLQs additionally couple 
to SM quarks $q$ and weak
bosons $V \in \{W, Z, H\}$, via new $QqV$ vertices. These interactions are
parametrised by overall couplings
$\kappa$, 
$\xi^V$ parameters controlling the relative strengths of the $V$
couplings to each VLQ, and $\zeta_i$ parameters governing the mix of SM quark
generations $i$ in each coupling.  The relevant parts of the Lagrangian are,
following the notation of Ref.~\cite{Buchkremer:2013bha},
%
\begin{equation}
  \label{eq:vlq}
  \begin{split}
    \mathcal{L} =
    &  \kappa_B \left[
      \sqrt{\frac{\zeta_i\xi^{B}_{W}}{\Gamma^0_{W}}} \frac{g}{\sqrt{2}} [\bar{B}_{L/R} W^-_\mu \gamma^\mu u^i_{L/R}]
      +  \sqrt{\frac{\zeta_i\xi^{B}_{Z}}{\Gamma^0_{Z}}} \frac{g}{2c_W} [\bar{B}_{L/R} Z_\mu \gamma^\mu d^{\,i}_{L/R}]
      -  \sqrt{\frac{\zeta_i\xi^{B}_{H}}{\Gamma^0_{H}}} \frac{M_B}{v} [\bar{B}_{R/L} H d^{\,i}_{L/R}]
    \right] \\
    + &   \kappa_T \left[
      \sqrt{\frac{\zeta_i\xi^{T}_{W}}{\Gamma^0_{W}}} \frac{g}{\sqrt{2}} [\bar{T}_{L/R} W^+_\mu \gamma^\mu d^{\,i}_{L/R}]
      +  \sqrt{\frac{\zeta_i\xi^{T}_{Z}}{\Gamma^0_{Z}}} \frac{g}{2c_W} [\bar{T}_{L/R} Z_\mu \gamma^\mu u^i_{L/R}]
      -  \sqrt{\frac{\zeta_i\xi^{T}_{H}}{\Gamma^0_{H}}} \frac{M_T}{v} [\bar{T}_{R/L} H u^i_{L/R}]
    \right] \\
    +  &  \kappa_X \left[
      \sqrt{\frac{\zeta_i}{\Gamma^0_{W}}} \frac{g}{\sqrt{2}} [\bar{X}_{L/R} W^+_\mu \gamma^\mu u^i_{L/R}]
    \right]
    + \kappa_Y \left[
      \sqrt{\frac{\zeta_i}{\Gamma^0_{W}}} \frac{g}{\sqrt{2}} [\bar{Y}_{L/R} W^-_\mu \gamma^\mu d^{\,i}_{L/R}]
    \right] + \text{h.c.} \, ,
  \end{split}
\end{equation}
where $M_Q$ is the mass of $Q$, $c_W$ is the usual cosine of the weak mixing
angle, $v$ is the Higgs-field vacuum expectation value, and $\Gamma^0_V$ are
functions of $m_V/M_Q$ only. The $\kappa$, $\xi$ and $\zeta$ are defined for
each $Q$ such that $\sum_{V} \xi^{V} = 1$ and $\sum_{i} \zeta_i = 1$, meaning
$\zeta_i\xi^{V} = \textrm{BR}(Q \rightarrow Vq_i)$.  Significantly, EM charge conservation
implies that the $T$ and
$B$ VLQs couple to all three SM weak bosons, while the $X$ and $Y$ couple only
to $W^\pm$.

From equation~\eqref{eq:vlq} it can be seen that the couplings of $B$ and $X$ to $W$ bosons
are the same up to the $\xi^V$ factor: any \WZH mixture other than $1$:$0$:$0$ will
result in a smaller $BW\!q$ coupling than the $XW\!q$ one. The same argument
applies to the $T$ and $Y$. If we ignore the $\Gamma^0$ components, the $B$ and $T$ couplings to $Z$ and
$H$ respectively contain additional factors of $1/\sqrt{2} c_W$ and
$M_Q / gv = M_Q/\sqrt{2}m_W$.
However, the $\Gamma^0$ factors act to ensure that
the $\sum_{V} \xi^{V} = 1$ relationships are preserved regardless of $m_V$ and $M_Q$.

\begin{figure}[t]
  \centering
  \subfloat[]{\includegraphics[]{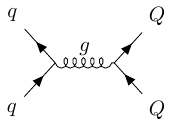}\label{fig:feyndiags:QQ1}}
  \subfloat[]{\includegraphics[]{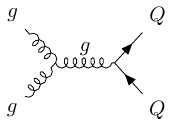}\label{fig:feyndiags:QQ2}}
  \subfloat[]{\includegraphics[]{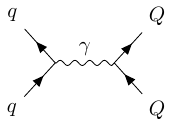}\label{fig:feyndiags:QQ3}}\\
  \subfloat[]{\includegraphics[]{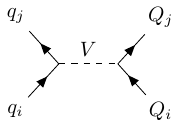}\label{fig:feyndiags:QQ'1}}
  \subfloat[]{\includegraphics[]{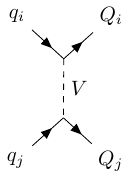}\label{fig:feyndiags:QQ'2}}\\
  \subfloat[]{\includegraphics[]{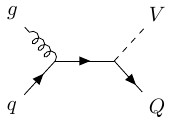}\label{fig:feyndiags:QV1}}
  \subfloat[]{\includegraphics[]{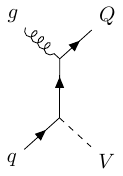}\label{fig:feyndiags:QV2}}
  \subfloat[]{\includegraphics[]{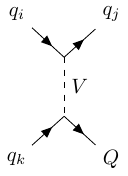}\label{fig:feyndiags:Qq1}}
  \subfloat[]{\includegraphics[]{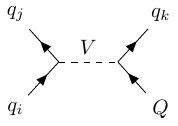}\label{fig:feyndiags:Qq2}}
  \caption{Leading-order Feynman diagrams for production of VLQs $Q$.  The top row
  (\protect\subref*{fig:feyndiags:QQ1}--\protect\subref*{fig:feyndiags:QQ3}) shows VLQ
  pair-production diagrams via strong and EM interactions, which do not depend
  on $\kappa$.  The second row
  (\protect\subref*{fig:feyndiags:QQ'1}--\protect\subref*{fig:feyndiags:QQ'2}) shows
  pair-production of VLQs via a weak boson $V \in \{W,Z,H\}$, which may lead to
  different-flavoured VLQs in the final state.  The third row
  (\protect\subref*{fig:feyndiags:QV1}--\protect\subref*{fig:feyndiags:Qq2}) shows
  single-production of $Q$ in association with a weak boson or
  SM quark $q$.}
\label{fig:feyndiags}
\end{figure}

This combination of couplings means that the VLQs may be pair-produced from SM
initial states via the strong and EM interactions,
singly produced via the weak interaction, and weakly pair-produced by
$t$-channel exchange of an SM weak boson. They only decay via their weak
interaction, into the mixtures of SM weak bosons and quarks governed by the
$\xi$ and $\zeta$ parameters. Examples of leading-order Feynman diagrams for VLQ production are shown in
Figure~\ref{fig:feyndiags}.

In $pp$ collisions at the LHC, the phenomenology of VLQ production depends
strongly on which generations of SM quarks they couple to.  In the rest of this
section, all four VLQs are assumed to have the same mass and $\kappa$ values.
The $\xi$ parameters are chosen such that the $B$ and $T$ couple to the bosons according to
the ratio $\WZH = \WZHtoo$, as motivated by Ref.~\cite{Buchkremer:2013bha}.

\begin{figure}[tbp]
\vspace{-0.4cm}
\subfloat[$T$ coupling to $u,d$ only]{\includegraphics[width=0.33\textwidth]{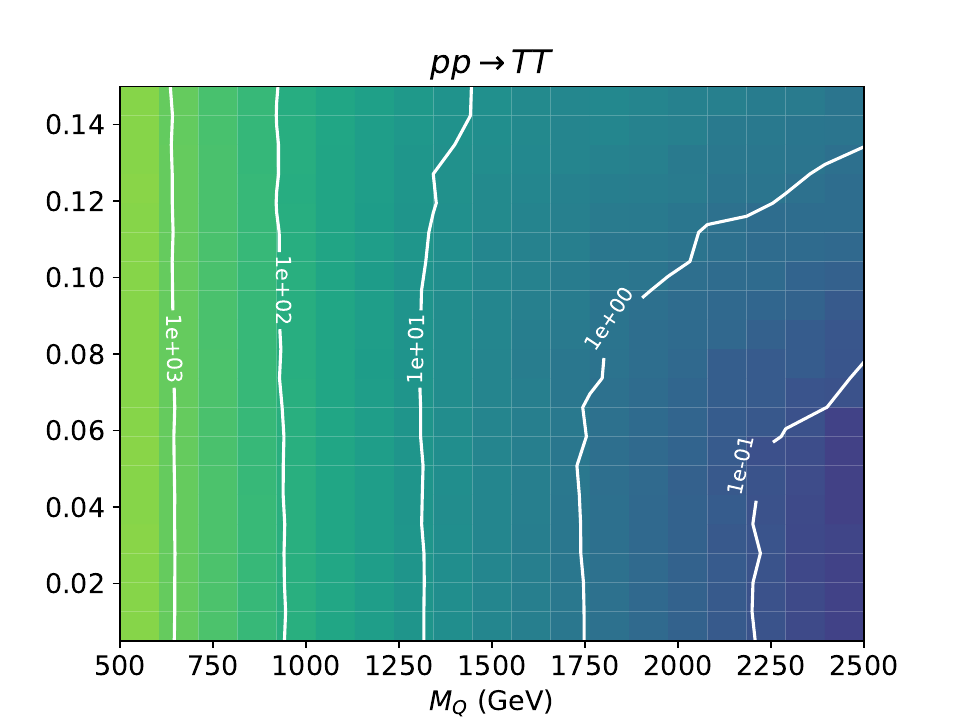}} 
\subfloat[$T$ coupling to $c,s$ only]{\includegraphics[width=0.33\textwidth]{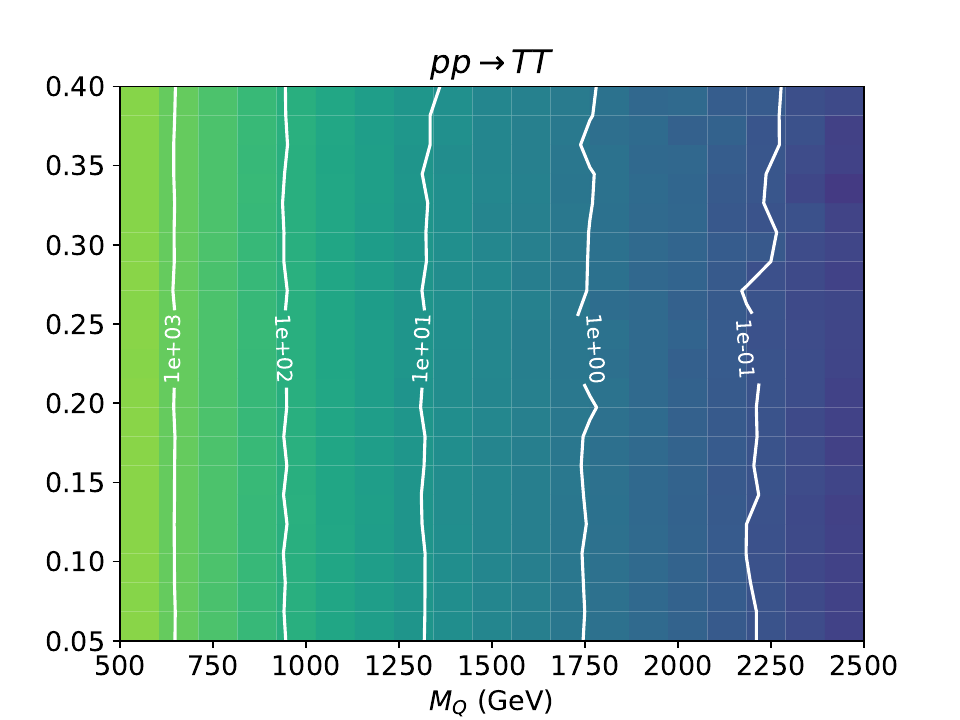}} 
\subfloat[$T$ coupling to $t,b$ only]{\includegraphics[width=0.33\textwidth]{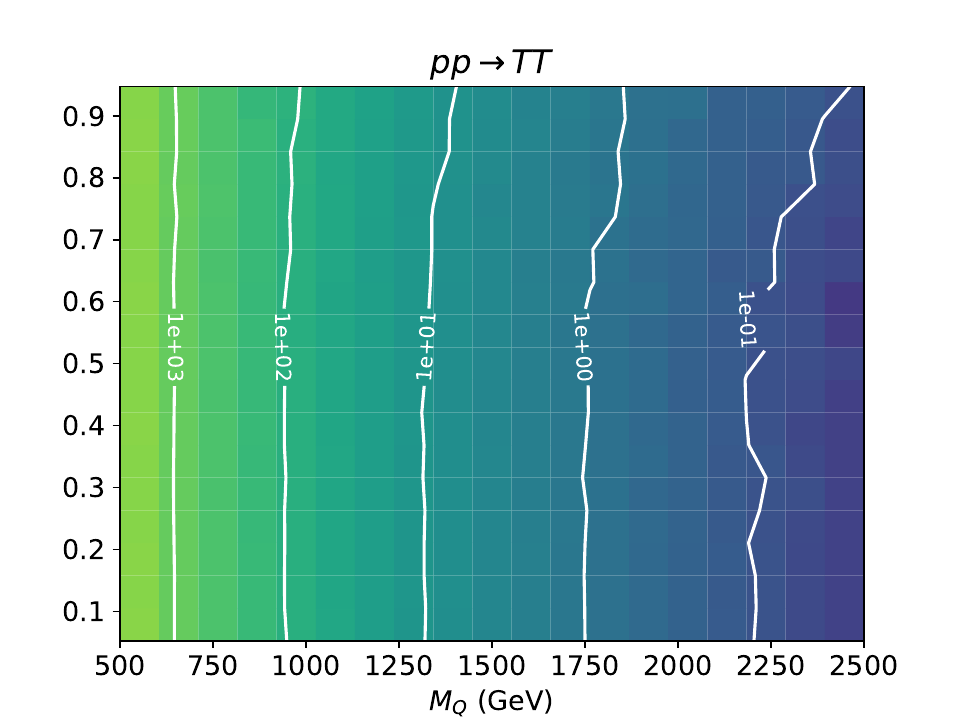}}
\includegraphics[height=3.5cm]{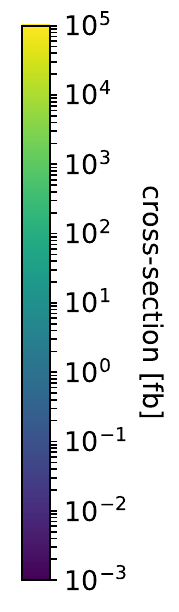} 
\caption{Leading-order cross-sections extracted from Herwig for production of a $TT$ pair as a
  function of $M_Q$ and $\kappa$, for \SI{13}{\TeV} $pp$ collisions, in the
  $\WZH = \WZHtoo$ scenario, assuming couplings to individual generations of
  quarks.  The white lines indicate the contours for production cross-sections
  in multiples of 10. The first-generation cross-sections acquire a dependence
  on $\kappa$ since pair-production initiated by proton valence quarks becomes
  possible.  The situation is analogous for other VLQ flavours, although
  somewhat attenuated for $X$ and $Y$ since they still require at least one
  antiquark from the sea to be produced via $W$ exchange in the $t$-channel.}
\label{fig:TTproduction}
\end{figure}

Pair-production of VLQs typically occurs via SM-like interactions, as shown in
Figures~\ref{fig:feyndiags}(\subref*{fig:feyndiags:QQ1}--\subref*{fig:feyndiags:QQ3}).
These vertices do not depend on $\kappa$, $\xi$ or $\zeta$, and this feature has
been exploited in several LHC analyses to reduce the model-dependence of
searches for VLQs~\cite{Aaboud:2018pii}. For a VLQ of mass
$\sim \SI{1.3}{\TeV}$ at the LHC, the cross-section is of the order of
$\SI{10}{\femto\barn}$. It is not, however, strictly accurate that VLQ
pair-production is independent of $\kappa$: the diagram in
Figure~\ref{fig:feyndiags:QQ'2} shows that VLQs may be pair-produced in diagrams
which involve two $QqV$ vertices, and which therefore have a strong dependence
on $\kappa$. Furthermore, it is not even guaranteed that this $QqV$-mediated
pair-production be negligible: indeed, it is the only pair-production diagram
which can involve two valence quarks in $pp$ collisions. If VLQs couple to
first-generation quarks, then Figure~\ref{fig:feyndiags:QQ'2} can be the
dominant production diagram at the LHC, in particular for high $M_Q$, since all
others require at least one antiquark or gluon from the proton sea. 
This effect has already been pointed out, for instance in Ref.~\cite{Cacciapaglia:2009cu}.
This
advantage disappears if the VLQs only couple to second- or third-generation SM
quarks. This is illustrated in Figure~\ref{fig:TTproduction}, which shows how
the pair-production of $T$ gains a dependence on $\kappa$ if VLQs are allowed to
couple to first-generation SM quarks. The effect is analogous for $B$, but
differs for $X$ and $Y$ since they can only be produced via
Figure~\ref{fig:feyndiags:QQ'2} if $V$ is a $W$-boson, meaning $XY$
pairs are the only option, and like-flavour valence pairs $uu$ or $dd$
do not contribute.
Another interesting feature of the diagrams in
Figures~\ref{fig:feyndiags}(\subref*{fig:feyndiags:QQ'1}--\subref*{fig:feyndiags:QQ'2})
is that the VLQs will be produced with different flavours if $V$ is the
$W$-boson, something which is not possible in the other pair-production
diagrams. It should be noted that the diagram in Figure~\ref{fig:feyndiags}\subref*{fig:feyndiags:QQ'1}
contains a model-dependent coupling of a vector boson to pairs of VLQs, the strength of which
depends on the electroweak quantum numbers of the multiplet which the VLQs belong to.
This diagram is subdominant in the analysis which we perform in this paper, so the particular
assumptions we make for this couploing do not affect the overall conclusions.

\begin{figure}[t]
\vspace{-0.4cm}
\subfloat[$T$ coupling to $u,d$ only]{\includegraphics[width=0.33\textwidth]{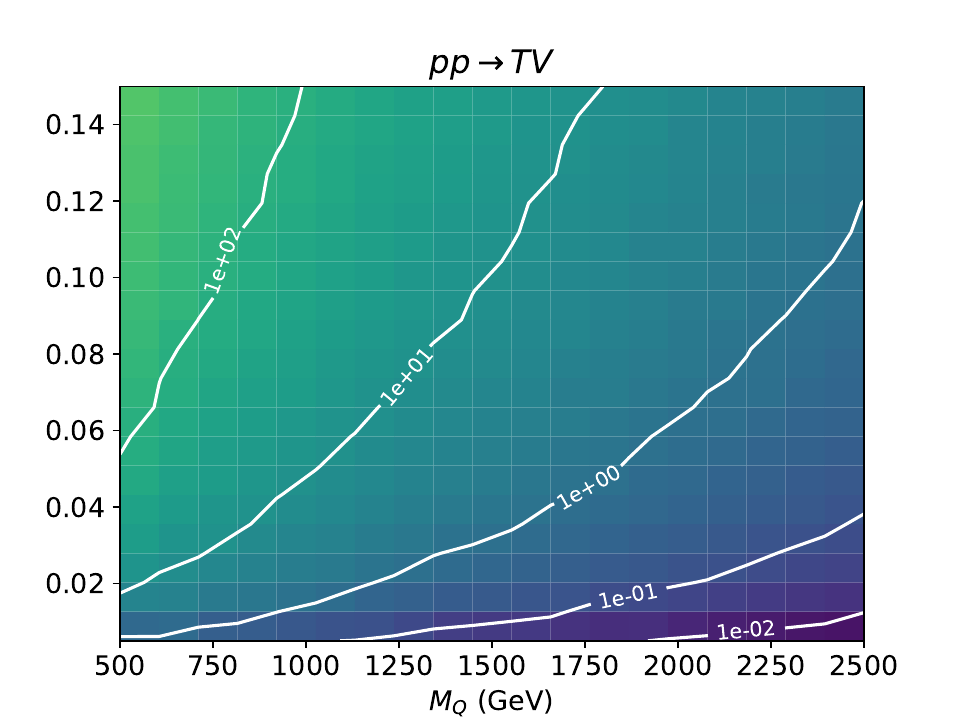}} 
\subfloat[$T$ coupling to $c,s$ only]{\includegraphics[width=0.33\textwidth]{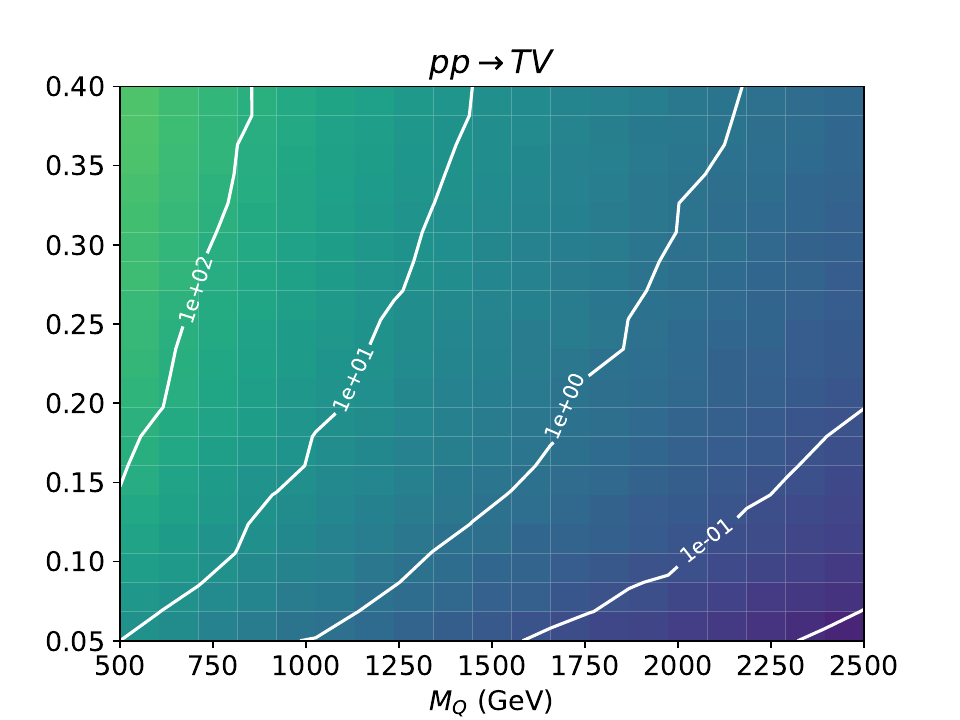}} 
\subfloat[$T$ coupling to $t,b$ only]{\includegraphics[width=0.33\textwidth]{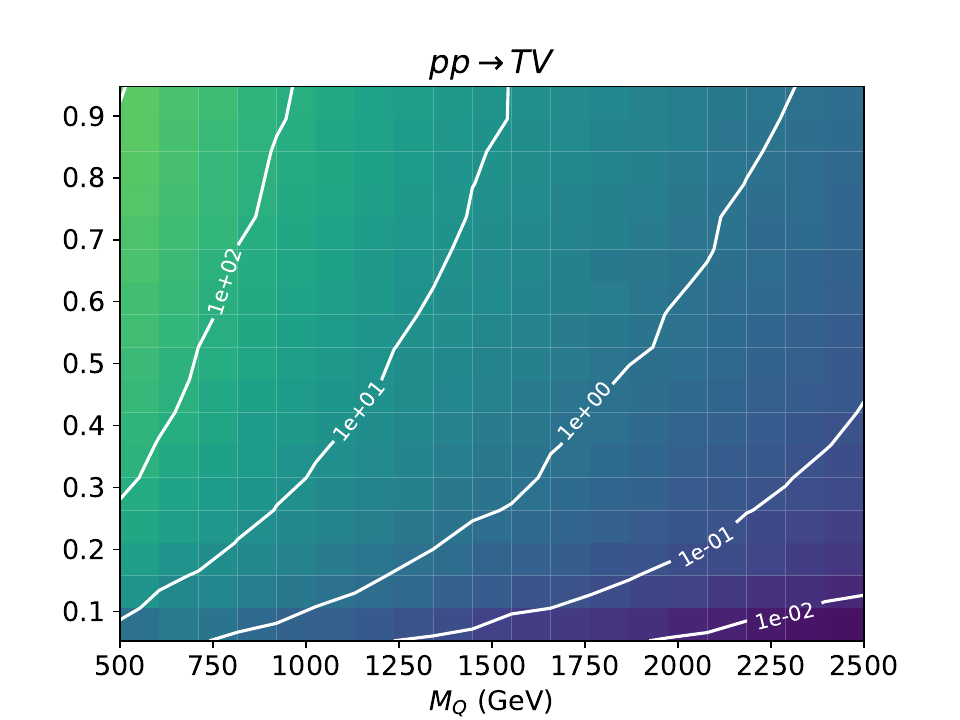}} 
\includegraphics[height=3.5cm]{xsScans/3rdGen/cbar.pdf} 
\subfloat[$Y$ coupling to $u,d$ only]{\includegraphics[width=0.33\textwidth]{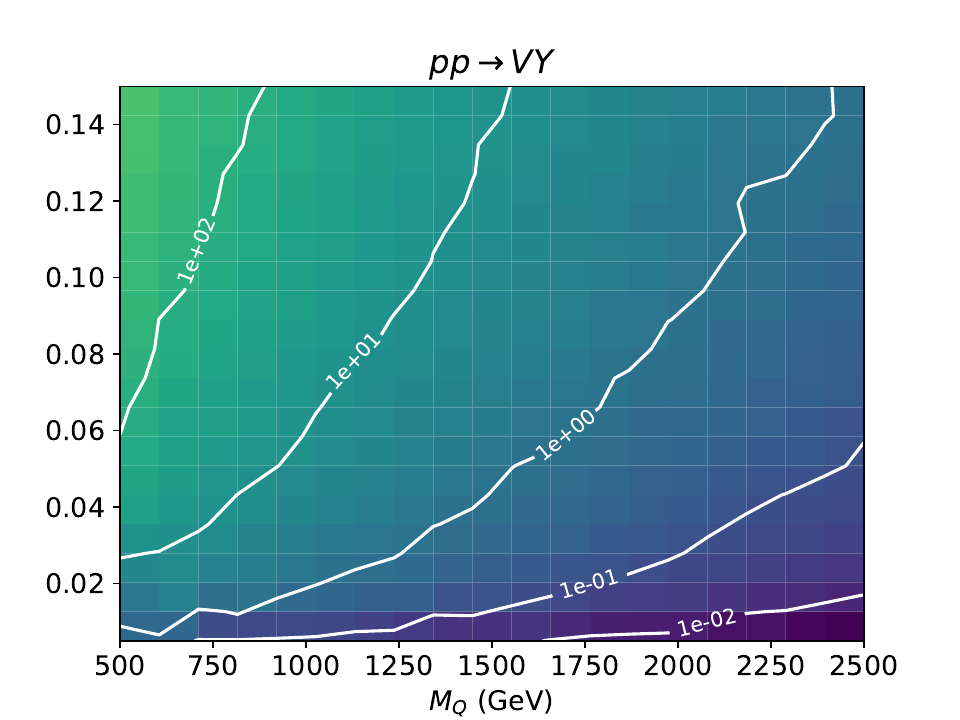}} 
\subfloat[$Y$ coupling to $c,s$ only]{\includegraphics[width=0.33\textwidth]{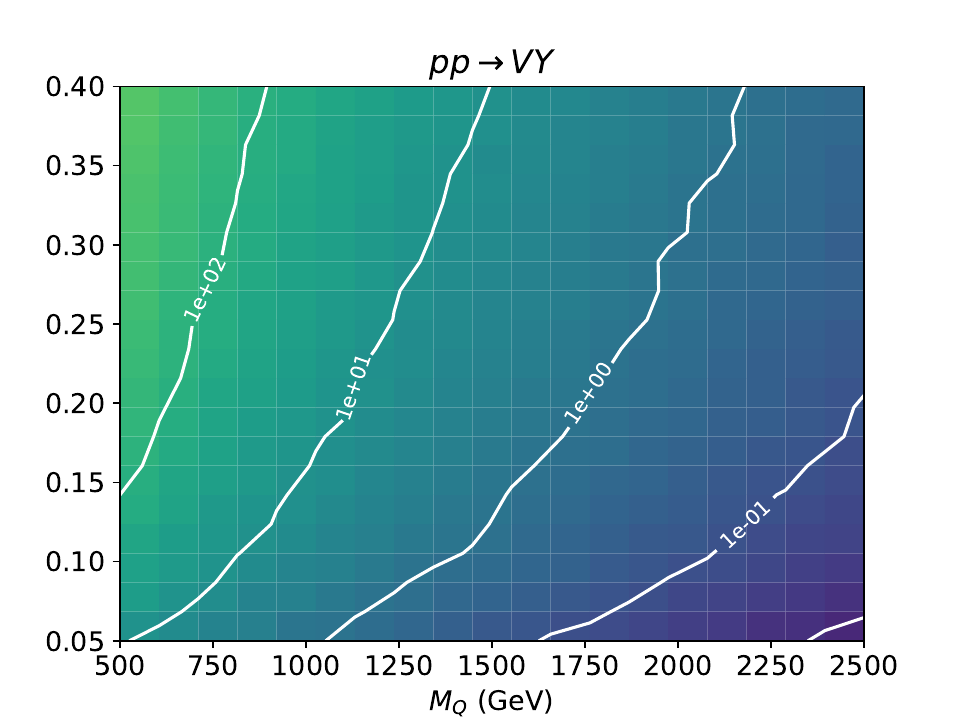}} 
\subfloat[$Y$ coupling to $t,b$ only]{\includegraphics[width=0.33\textwidth]{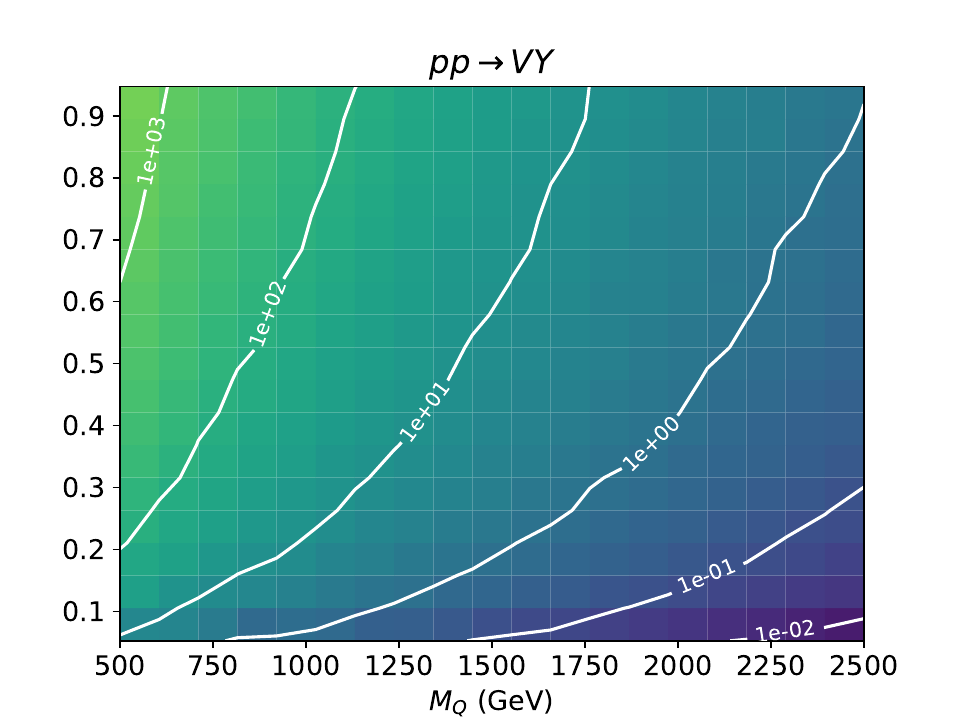}} 
\includegraphics[height=3.5cm]{xsScans/3rdGen/cbar.pdf} 
\caption{Leading-order cross-sections extracted from Herwig for production of a $T$ and $Y$
  with a weak boson as a function of $M_Q$ and $\kappa$, for \SI{13}{\TeV} $pp$
  collisions, in the $\WZH = \WZHtoo$ scenario, assuming couplings to individual
  generations of quarks.  The white lines indicate the contours for production
  cross-sections in multiples of 10. First-generation couplings lead to higher
  production cross-sections, as a result of valence-quark-induced diagrams,
  while second and third-generation couplings lead to suppressed production
  rates, according to the relevant quark PDFs.}
\label{fig:QVproduction}
\end{figure}

Single-production of VLQs also has a rich phenomenology. In this case, the
production cross-section always has a dependence on $\kappa$ since the $QqV$
vertex must always be involved. Let us consider first the case of VLQ production
in association with a weak boson $V$, as shown in
Figures~\ref{fig:feyndiags}(\subref*{fig:feyndiags:QV1}--\subref*{fig:feyndiags:QV2}). In
both diagrams, the process is initiated by a quark and a gluon, and the vector
boson is radiated from the quark. The $QqV$ vertex differs depending on the
flavour of the VLQ, as described above. The cross-section is also dependent on
$\zeta$ however, since the cross-section of
Figures~\ref{fig:feyndiags}(\subref*{fig:feyndiags:QV1}--\subref*{fig:feyndiags:QV2})
depend strongly on the incoming quark. If the VLQ couples to first-generation
quarks, then diagrams where $u$ is incoming will dominate over diagrams with $d$
by a factor of about two. 
For example, $T+H/Z$
production will be roughly twice as frequent as $B+H/Z$ production, but the
situation is reversed for $T+W$, which will be roughly half as frequent as $B+W$
production. The same argument goes for $X+W$ production, which will occur at
roughly twice the rate of $Y+W$ production. Overall, in a $\WZH = \WZHtoo$
scenario with only first-generation quark couplings $X+V$ would be the dominant
process, with $T+V$ or $B+V$ and $Y+V$ occurring $\sim25\%$ and
$\sim50\%$ less frequently respectively, driven simply by the valence quark
populations in the proton. If the VLQs only couple to second-generation quarks,
this dependence on the valence quarks disappears, as the diagram can only occur
with an incoming $c$ or $s$ from the proton sea. As a result (and leaving aside the
effects of available phase space), the production of any
flavour of VLQ in association with a $V$ becomes suppressed with respect to
pair-production, with the relative cross-sections depending chiefly on the quark
PDFs and the $\zeta$ parameters for $B$ and $T$. Finally, for third-generation
couplings, $Q+V$ production is further suppressed, and $t$-induced diagrams
disappear almost entirely. As a result, $X+V$ production is largely impossible,
while $T+V$ may only occur with a $W$.  These effects are illustrated in
Figure~\ref{fig:QVproduction}, which shows the production cross-sections for
$T+V$ and $Y+V$ depending on whether the coupling is to first-, second- or
third-generation quarks.

\begin{figure}[tbp]
\vspace{-0.4cm}
\subfloat[$T$ coupling to $u,d$ only]{\includegraphics[width=0.33\textwidth]{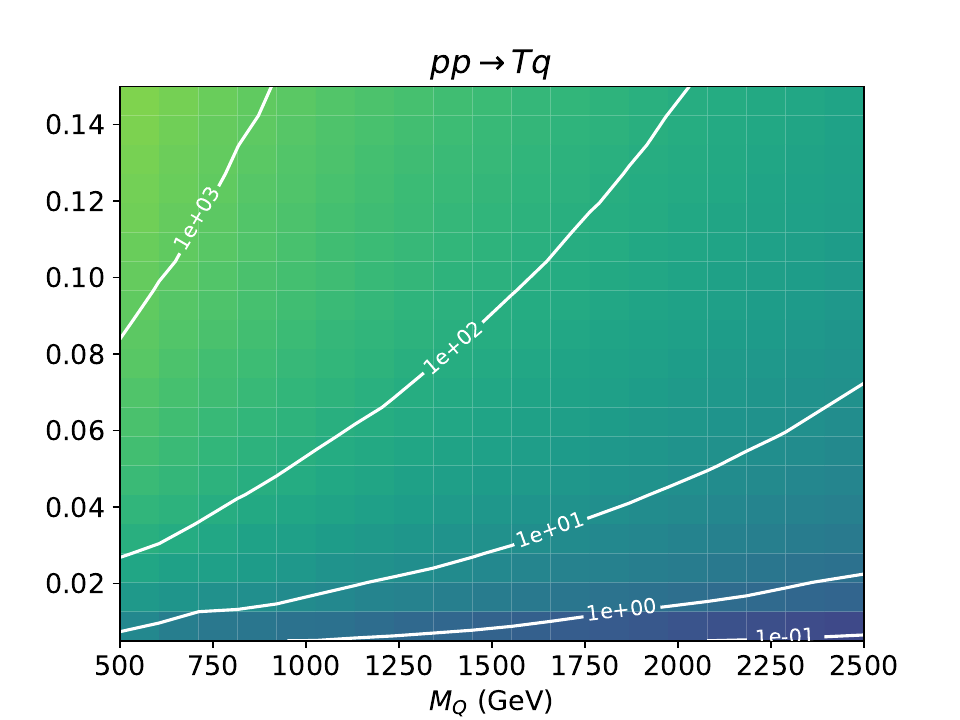}} 
\subfloat[$T$ coupling to $c,s$ only]{\includegraphics[width=0.33\textwidth]{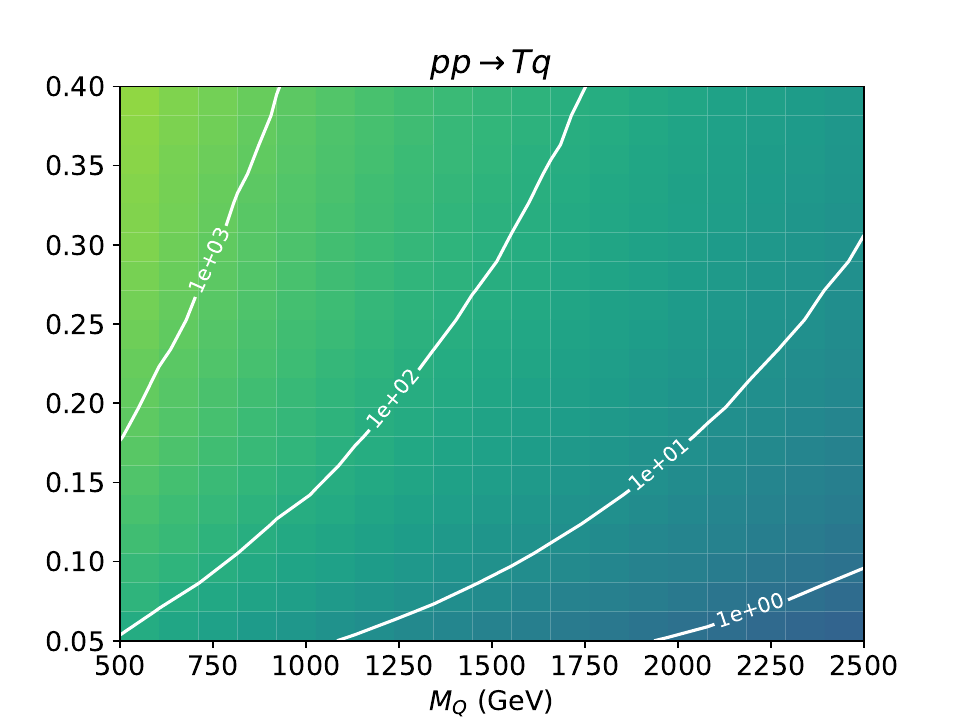}} 
\subfloat[$T$ coupling to $t,b$ only]{\includegraphics[width=0.33\textwidth]{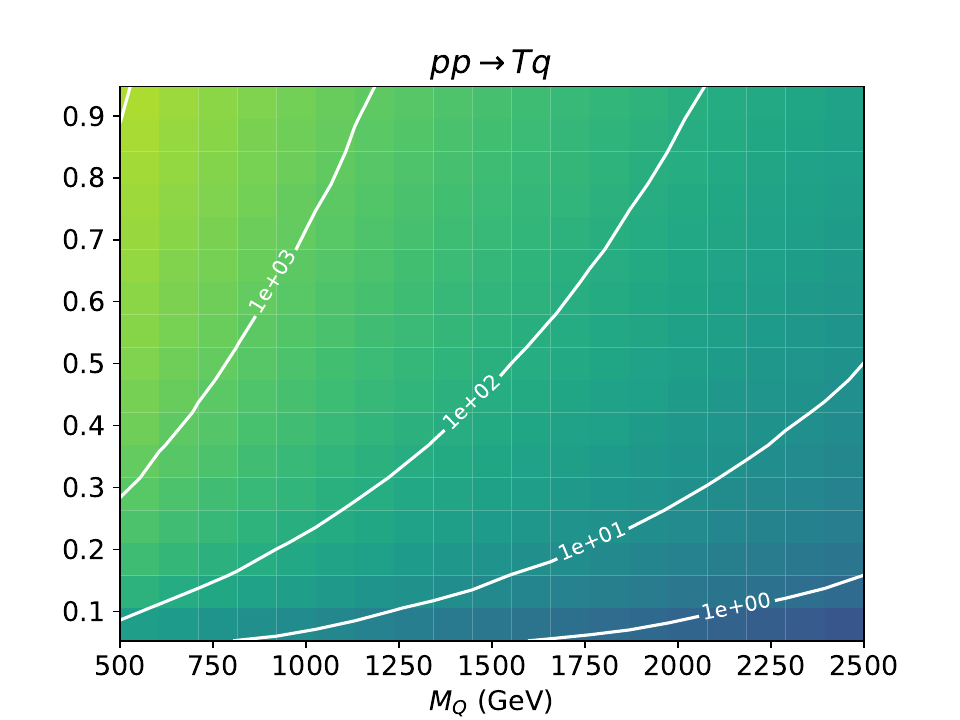}} 
\includegraphics[height=3.5cm]{xsScans/3rdGen/cbar.pdf} 
\subfloat[$X$ coupling to $u,d$ only]{\includegraphics[width=0.33\textwidth]{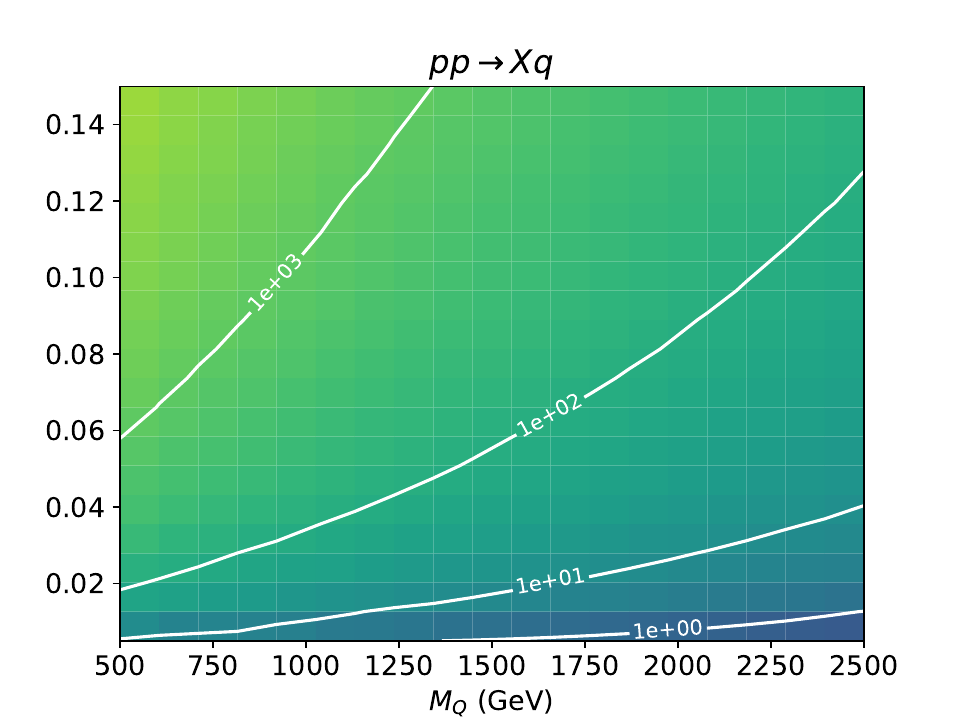}} 
\subfloat[$X$ coupling to $c,s$ only]{\includegraphics[width=0.33\textwidth]{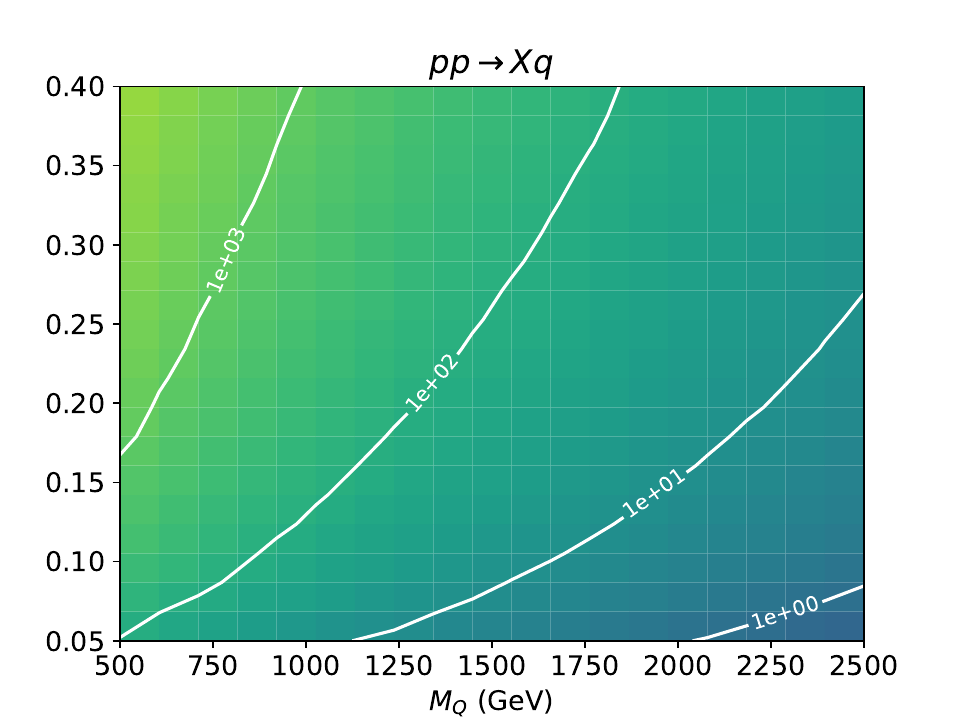}} 
\subfloat[$X$ coupling to $t,b$ only]{\includegraphics[width=0.33\textwidth]{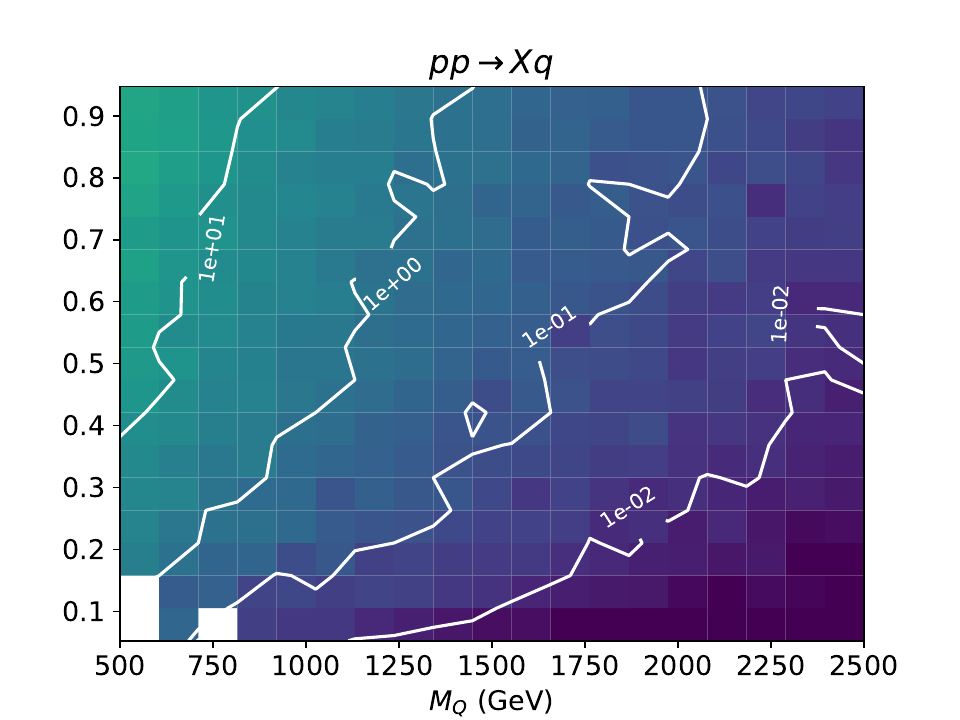}} 
\includegraphics[height=3.5cm]{xsScans/3rdGen/cbar.pdf} 
\caption{Leading-order cross-sections extracted from Herwig for production of a $T$ and $X$
  with a SM quark as a function of $M_Q$ and $\kappa$, for \SI{13}{\TeV} $pp$
  collisions, in the $\WZH = \WZHtoo$ scenario, assuming couplings to individual
  generations of quarks.  The white lines indicate the contours for production
  cross-sections in multiples of 10. First-generation couplings lead to higher
  production cross-sections, as a result of valence-quark-induced diagrams,
  while second and third-generation couplings lead to suppressed production
  rates, according to the relevant quark PDFs. $X+q$ production goes from being
  the dominant production process at the LHC if $X$ couples to first-generation
  quarks only, to vanishing if $X$ couples to third-generation quarks only.
  White cells indicate corners of phase-space where the process in question is highly subdominant,
  and therefore where the cross-section was not sampled during the Herwig run.}
\label{fig:Qqproduction}
\end{figure}

Finally, let us consider production of VLQs in association with quarks. The
diagrams for such processes are shown in
Figures~\ref{fig:feyndiags}(\subref*{fig:feyndiags:Qq1}--\subref*{fig:feyndiags:Qq2}),
and are mediated by a weak boson, either in the $s$-channel or the
$t$-channel. Once again, the importance of this production mechanism at the LHC
will depend on the flavours of the incoming quarks, since diagrams involving one
or more valence quarks will dominate. The only diagram where both quarks can be
valence quarks is the $t$-channel diagram in the scenario where VLQs may couple
to first-generation quarks. In this case diagrams involving $uu$ or $ud$ will
have the highest cross-sections, while $dd$-induced processes will acquire a
suppression factor of about four compared to $uu$ or $ud$, in addition to other
considerations such as couplings of the exchanged weak bosons.
The dominant process
will be $X+q$ production which is induced by $uu$, and which will occur roughly
four times more frequently than $Y+q$, which is induced by $dd$. $T+q$ and $B+q$
production cross-sections are also reduced by the fact that the
$H$-mediated diagram is suppressed by the very small SM coupling to $u$ or $d$
If the VLQs only couple to second- or third-generation quarks, then the valence-quark-induced
processes
cease to be available, and $Q+q$ becomes dependent on the quark PDFs, with some
production modes becoming almost entirely inaccessible if an initial $t$ quark
is involved. This effect is illustrated in Figure~\ref{fig:Qqproduction}, where
the production cross-sections for $T$ and $X$ in association with a quark are
compared.

The overall dominant production processes are unsurprisingly also extremely
dependent on the generation(s) which the VLQs are allowed to couple to.  If VLQs
can couple to first-generation quarks, then by far the dominant process,
assuming $\WZH = \WZHtoo$, is $X+q$ production, reaching up to $\sim \SI{400}{\femto\barn}$ for
$M_Q\sim \SI{1}{\TeV}$ and $\kappa \sim 0.07$.  This is followed by production
of other VLQs with quarks, which are about a factor three to four lower in
cross-section due to the relative proportion of valence quarks. Gluon-induced or
quark-induced VLQ pair-production is typically the next most common process, at
$\sim \SI{50}{\femto\barn}$ for the same parameter choices, roughly twice the rate of VLQ
production in association with a weak boson. Since pair-production does not
depend too strongly on $\kappa$ at low $M_Q$, there are regions of parameter space,
particularly at low $\kappa$, where pair-production may dominate over $Qq$
production. For higher $M_Q$, pair-production from valence quarks involving the
$QqV$ vertex may be dominant.  If first-generation couplings are forbidden, then
pair-production becomes the dominant production process, still at $\sim \SI{50}{\femto\barn}$
for the parameter values chosen above. Single-production of VLQs with $V$ or $q$
may still occur, but roughly an order of magnitude less frequently than for
first-generation couplings.

The most common experimental signature for VLQ decays is likely to be large
numbers of jets: this is unsurprising, given that VLQs decay to quarks and
weak bosons (which have their largest branching fractions to quarks). However, such
signatures would be swamped by the large QCD background from LHC $pp$
collisions. In many decay chains of the VLQs, $W$-bosons are involved, either
directly from the VLQ decay or as a result of the decay of a $t$ quark. This is
particularly the case since $X$ and $Y$ VLQs, which are often produced with the
highest cross-section, can only decay via $W$. A large fraction of VLQ decays
would therefore be expected to produce large missing transverse energy (\MET) and one
lepton, in addition to multiple jets, some of which may originate from
$b$-quarks.
In some cases, two leptons may be expected with the missing energy, which can originate from
the decay of pair-produced VLQs, with two $W$ bosons decaying
leptonically.
This signature would exhibit less dependence on $\kappa$ as a
result. Based on this analysis, one can expect LHC analyses targeting $V$+jets
or $WW$ final states to be most sensitive to VLQ models.

\section{Comparison to LHC $B$ and $T$ searches}
\label{sec:BT}

Assuming all other VLQs decouple from the SM, and that VLQs couple only to the third generation of SM
quarks, a $B$ quark with a mass of a few \SI{100}{\GeV} or more may decay to $Z+b$, $H+b$ or $W+t$, depending upon the relative couplings to
these bosons. Similarly a $T$ may decay to $Z+t, H+t$ or $W+b$.
Several dedicated searches have been made for these signatures by ATLAS and CMS, giving exclusions up to masses of around a \si{\TeV}. The
ATLAS results in particular are combined and summarised in Ref.~\cite{Aaboud:2018pii} and made available in \hepdata~\cite{Maguire:2017ypu}.
CMS results can be found in Ref.~\cite{Sirunyan:2018omb}, with roughly similar sensitivity to the ATLAS results at equivalent parameter points and integrated luminosity.
We use the ATLAS results
as exemplars for comparison since they are shown for a range of branching fraction values rather than specific points in parameter space. 
The ATLAS and CMS results each use around \SI{36}{\ifb} of \SI{13}{\TeV} data, while CONTUR uses a range of LHC measurements at \SI{7}, \SI{8} and \SI{13}~\si{\TeV}
with integrated luminosities between \SI{3} and \SI{36}{\ifb}.

Given the subsequent decays of the SM particles produced by the VLQ decays above, BSM events may enter
the fiducial phase space of a wide variety of differential cross-sections measured at the LHC, with $b$-jets, $Z,W$+jets,
dibosons and multileptons expected to be important, as discussed in the previous section.
Many of these measurements are available in \rivet, and are thus accessible to \contur. All have
been shown to agree with SM calculations, and thus the uncertainties on this agreement provide constraints on the
presence of a significant VLQ production cross-section.

\begin{figure}[tbp]
  \vspace{-0.4cm}
  \subfloat[]{\includegraphics[width=0.45\textwidth]{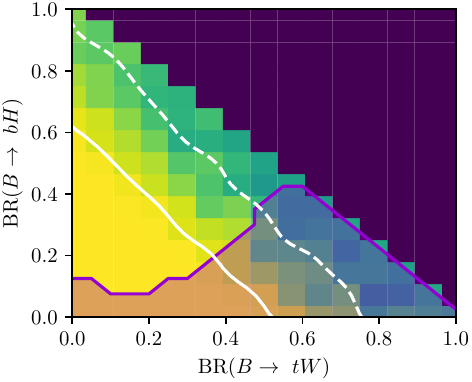}\label{fig:BTonlyB}}
  \subfloat[]{\includegraphics[width=0.45\textwidth]{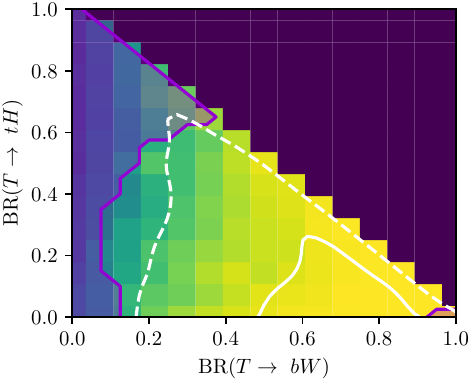}\label{fig:BTonlyT}}
  \caption{Sensitivity of LHC measurements to
    \protect\subref{fig:BTonlyB}~$B$-production for $M_B = \SI{1200}{\GeV}$ and
    \protect\subref{fig:BTonlyT}~$T$-production for $M_T = \SI{1350}{\GeV}$.
    The \contur exclusion is shown in the bins in which it is evaluated,
    graduated from yellow through green to black on a linear scale, with the 95\%~CL (solid white)
    and 68\%~CL (dashed white) exclusion contours superimposed. The mauve region
    is excluded at 95\%~CL by the ATLAS combination~\cite{Aaboud:2018pii}.}
  \label{fig:BTonly}
\end{figure}

Figure~\ref{fig:BTonlyB} shows the \contur exclusion region, in the half-plane
of different branching ratios, for $M_B = \SI{1200}{\GeV}$. This mass places it
in middle of the exclusion from ATLAS, which ranges from
\SIrange{1040}{1350}{\GeV} over the half-plane. The mauve region shows the
exclusion limit at 95\% confidence level (CL) by ATLAS. There are four main searches contributing to
this limit, one each targeting the $Z$ and $H$ decay
channels~\cite{ZllSearch,HadSearch}, and the remaining two target $B$-decay to
$Wt$~\cite{WtSearch,TriLepSearch}. This results in a high sensitivity in the bottom-right
corner of the triangle, where $BR(B\rightarrow tW)$ is high. The sensitivity in
the measurements is somewhat complementary to the searches, and comes primarily
from $Z+$jet
measurements~\cite{Aad:2015auj,Aaboud:2017hox,Aaboud:2017hbk,Aaboud:2019jcc}.

Figure~\ref{fig:BTonlyT} shows the \contur exclusion region for
$M_T = \SI{1350}{\GeV}$.  The ATLAS search exclusion for this mass value is also shown; in
this case, the ATLAS exclusion ranges from \SIrange{1310}{1420}{\GeV} over the
branching ratio half-plane. Here the difference in sensitivity between the ATLAS searches and \contur is nicely seen. From \contur, the exclusion comes primarily from measurements involving top quarks and $W$ bosons~\cite{Aaboud:2017fye,Aaboud:2018eqg,Sirunyan:2018wem,Khachatryan:2016mnb,Sirunyan:2018ptc}. The ATLAS combination limit in mauve on the other hand, has three contributions from searches sensitive to $T$-decay to $Ht$~\cite{HbbSearch,TriLepSearch,HadSearch}, two that target $T$-decay to $Zt$~\cite{ZnunuSearch,ZllSearch}, and only one sensitive in the $W$ channel~\cite{WbSearch}. Again, the measurement sensitivity is quite complementary to the searches.

\section{Four VLQ flavours}
\label{sec:all}

\begin{figure}[tbp]
\subfloat[\B]{\includegraphics[width=0.3\textwidth]{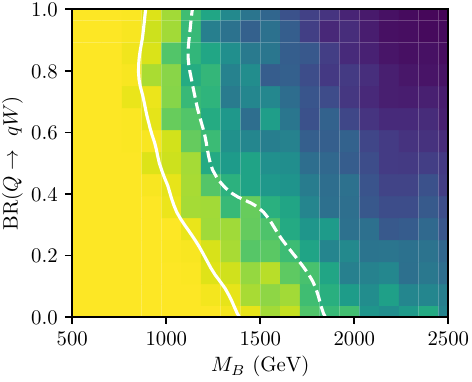}}
\subfloat[\T]{\includegraphics[width=0.3\textwidth]{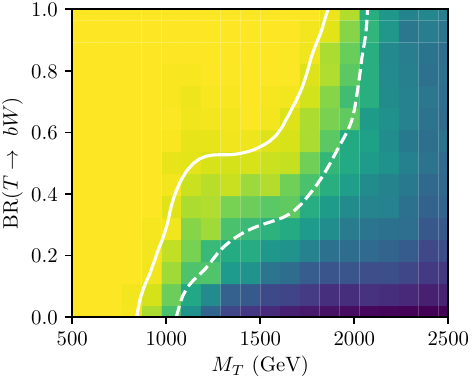}}\\
\subfloat[\BT]{\includegraphics[width=0.3\textwidth]{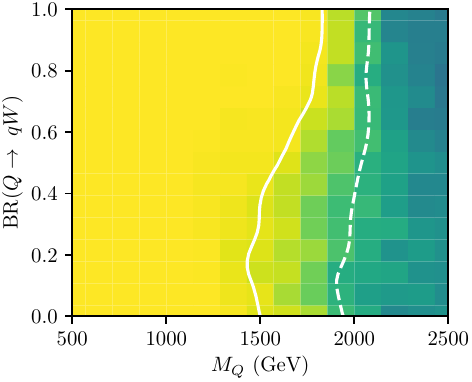}}
\subfloat[\BY]{\includegraphics[width=0.3\textwidth]{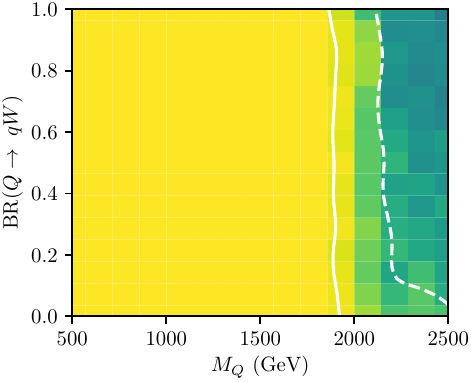}}
\subfloat[\XT]{\includegraphics[width=0.3\textwidth]{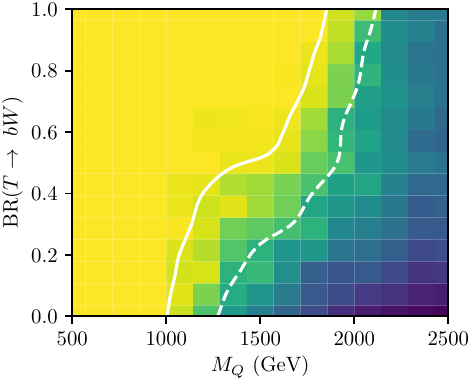}}\\
\subfloat[\BTY]{\includegraphics[width=0.3\textwidth]{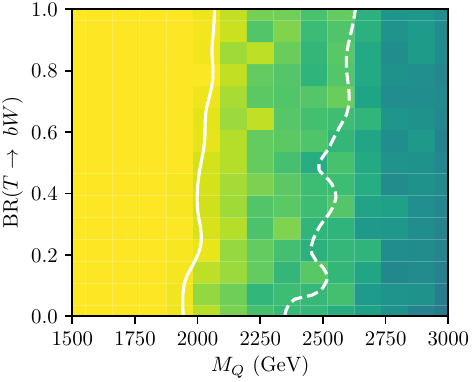}}
\subfloat[\BTX]{\includegraphics[width=0.3\textwidth]{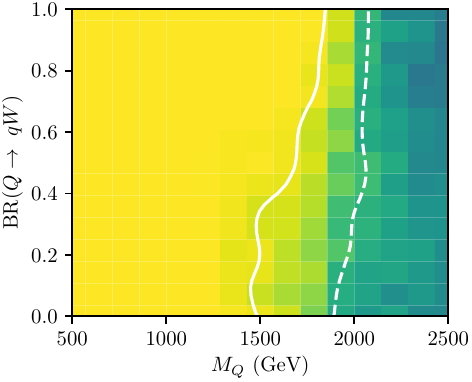}}
\unskip\qquad\vrule\quad
\subfloat[\BTXY]{\includegraphics[width=0.3\textwidth]{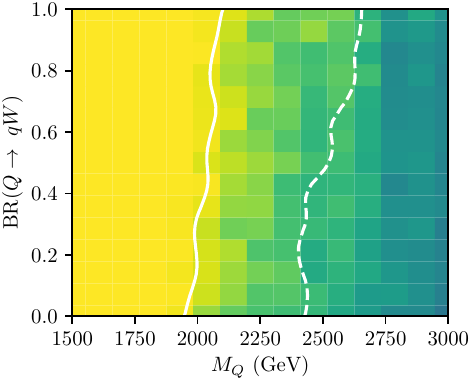}}
\caption{
Sensitivity of LHC measurements to VLQ production when all VLQs are degenerate in mass, assuming $\mathrm{BF}(Q \rightarrow Hq) = \mathrm{BF}(Q \rightarrow Zq)$.
The \contur exclusion is shown in the bins in which it is evaluated,
graduated from yellow through green to black on a linear scale, with the 95\%~CL (solid white)
and 68\%~CL (dashed white) exclusion contours superimposed.
The results for various multiplets are shown, in addition to the four-VLQ case which is unrealistic but still a useful benchmark.
In the axis labels, $Q$ may refer to $B$ or $T$, while $q$ may refer to $b$ or $t$.}
\label{fig:all}
\end{figure}

There is no particular reason, other than the desire to take simple benchmarks, why one VLQ should have lower mass
than all the others. An equally, and perhaps more, natural scenario is that they all have similar masses.
Setting the branching fractions to each gauge boson to be the same for $B$ and $T$ (with $X$ and $Y$ always decaying
to $W$ as discussed in Section~\ref{sec:pheno}), and assuming still only couplings to third-generation quarks,
we have studied the branching fraction half plane for various values of a generic VLQ mass $M_Q$.
As might be expected, for a given mass the sensitivity of the measurements is greater than for a single VLQ,
with the entire half-plane in the branching fraction space being disfavoured up to about \SI{1800}{\GeV}.
Figure~\ref{fig:all} shows the results of a scan of the branching fraction of $B$ or $T$ to $qW$ versus VLQ mass $M_Q$, for various multiplet assumptions, 
and assuming $\mathrm{BF}(Q \rightarrow Hq) = \mathrm{BF}(Q \rightarrow Zq)$.
The difference in exclusion shape between Figures~\ref{fig:all}a and~\ref{fig:all}b can be explained by the lack of top density in the proton PDF.
The lower parts of the plots only allow single-VLQ production via a neutral boson: for the $T$ this means a $t$-quark would be needed,
which is vanishingly rare, and instead QCD/EM pair-production is the only viable mechanism, while the $B$ can be singly-produced via
an incoming $b$-quark in the proton sea. In the top half of the plots, the situation is reversed, with $W$-bosons mediating
single production, and correspondingly, it is single-$B$ production which is suppressed by the lack of $t$-quarks in the proton sea.

It should be noted that the bounds from direct searches would be enhanced in the presence of multiple degenerate VLQs, so the results shown in
Figure~\ref{fig:all} should not be compared directly to search results without appropriate re-interpretation.

\section{All quark generations, and coupling strength}
\label{sec:kappa}

The results discussed in the previous sections apply to scenarios where the VLQs
only couple directly to third generation SM quarks and heavy bosons.
We now consider cases where the VLQs can couple to other generations of quarks.
All production processes described in Section~\ref{sec:pheno} and Figure~\ref{fig:feyndiags} are included.
As discussed in Section~\ref{sec:pheno}, at low VLQ masses, the dominant production process is
expected to be pair production via the strong interaction, which does
not depend on the VLQ--quark coupling, $\kappa$. 
Despite small values of this coupling
suppressing the single-production channels, production of a single VLQ is less
kinematically penalised than pair-production and so single-production can
dominate at high masses. This is particularly of interest for couplings to
first-generation quarks, where amplitudes involving VLQ couplings direct to
initial-state quarks become significant due to large high-$x$ valence-quark
densities.  The $t$-channel weak VLQ-pair production process of
Figure~\ref{fig:feyndiags:QQ'2} is a double-recipient of this PDF enhancement
for first-generation couplings.  On the other hand, the existing, non-collider
limits on the coupling between the VLQ and SM quarks are more stringent for the
first- and second-generations than for the third. In this section we study the
relative impacts of these considerations on constraints from LHC measurements.
For the single-light-VLQ representative scenarios considered in
Ref.~\cite{Buchkremer:2013bha}, the existing limits on the couplings are
estimated as being $\kappa \le 0.07$ for coupling to the first generation only,
$\kappa \le 0.2$ for the second only, and an order of magnitude smaller when
more than one generation is coupled. (For third generation only, $\kappa = 1$ is
allowed, as used in the previous section.) These values are indicated in the studies
below, although we note that it has been pointed out in Ref.~\cite{Atre:2008iu} that additional symmteries
in specific models may lead to natural cancellations which mean these limits
do not apply.

\subsection{First quark generation}
\label{sec:kappa:1gen}

\begin{figure}[tbp]
  \centering
  \subfloat[]{
    \includegraphics[width=0.45\textwidth]{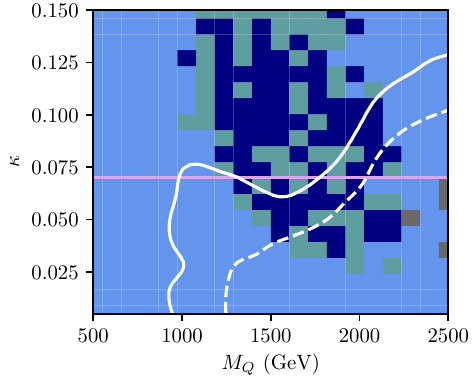}\label{fig:gen1:001:dom}}
  \subfloat[]{
    \includegraphics[width=0.45\textwidth]{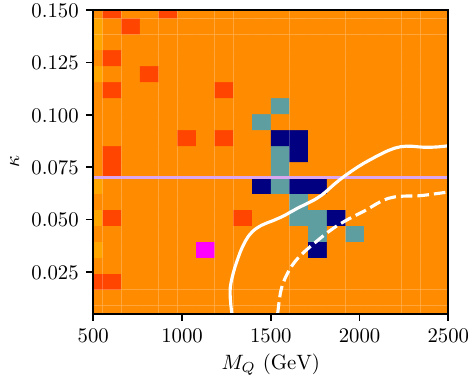}\label{fig:gen1:010:dom}} \\
  \subfloat[]{
    \includegraphics[width=0.45\textwidth]{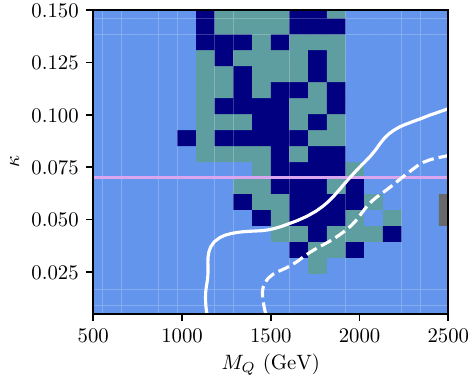}\label{fig:gen1:100:dom}}
  \subfloat[]{
    \includegraphics[width=0.45\textwidth]{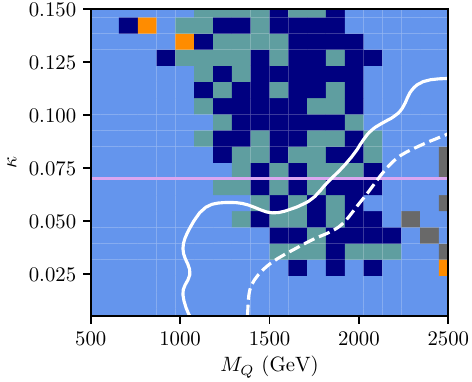}\label{fig:gen1:121:dom}}
  \vspace*{2ex} \\
  \begin{tabular}{lll}
    \swatch{cornflowerblue}~ATLAS $WW$ & \swatch{navy}~ATLAS $\mu$+\MET{}+jet & \swatch{cadetblue}~ATLAS $e$+\MET{}+jet\\
    \swatch{darkorange}~ATLAS $\mu\mu$+jet & \swatch{orangered}~ATLAS $ee$+jet & \swatch{orange!60}~ATLAS $\ell\ell$+jet\\
    \swatch{silver}~ATLAS jets & \swatch{dimgrey}~CMS jets & \\
    \swatch{deepskyblue} CMS~$e$+\MET{}+jet & \swatch{magenta} ATLAS~4$\ell$ &
  \end{tabular}
  \vspace*{2ex}
  \caption{Dominant LHC analysis pools contributing to VLQ limit-setting in the
    $\kappa$ vs VLQ mass plane, where $\kappa$ is the coupling to
    first-generation SM quarks.  All VLQ ($B, T, X, Y$) masses are set to be
    degenerate. The disfavoured regions are located above and to the left of the
    dashed (68\%~CL) and solid (95\%~CL) white contours respectively. The lower
    bounds in $\kappa$ from non-LHC flavour physics are indicated with the pink
    horizontal contour. The VLQ branching fractions to \WZH are
    \protect\subref{fig:gen1:001:dom}~\WZHzzo
    \protect\subref{fig:gen1:010:dom}~\WZHzoz
    \protect\subref{fig:gen1:100:dom}~\WZHozz and
    \protect\subref{fig:gen1:121:dom}~\WZHtoo.}
  \label{fig:1gen:dom}
\end{figure}

In Figure~\ref{fig:1gen:dom}, we show the current LHC measurement sensitivity in
the plane of the coupling $\kappa$ to the first generation SM quarks and the VLQ
mass $M_Q$, overlaid on a map of the experimental analyses dominating each
$(\kappa, M_Q)$ point's \CLs value. These maps are shown for the three extreme
VLQ--boson branching-fraction configurations
($\WZH = \WZHzzo, \WZHzoz, \WZHozz$), and an example admixture of all three
bosons ($\WZH = \WZHtoo$). In the following sections we will use this same set
of $\xi$ configurations to exemplify the LHC $\kappa$--$M_Q$ sensitivities for
couplings to the second and third quark generations. The detailed \CLs maps in
$\kappa$--$M_Q$, from which these limit contours are constructed, are presented
in Appendix~\ref{app:clsmaps} for couplings to each quark generation.

First, we note from the white contour lines that the majority of the
first-generation $\kappa$--$M_Q$ plane is excluded at 95\%~CL, meaning that ---
despite the lack of dedicated searches --- LHC measurements set stringent limits
on first-generation VLQs. The tightest limits are set for the $Z$-only $\xi$
configuration in Figure~\ref{fig:gen1:010:dom}, and the least stringent for the
Higgs-only in Figure~\ref{fig:gen1:001:dom}.  At low VLQ masses, below
\SI{1}{\TeV} for VLQ decay via a Higgs and below $\sim\SI{1.3}{\TeV}$ for decay
via a $Z$, QCD and EM pair-production dominates and the exclusion is insensitive
to $\kappa$. Above this threshold, a strong $\kappa$ dependence enters via weak
$Qq$ production, with the allowed regions of all \WZH configurations extending
to the same maximum $\kappa \sim 0.07$ at high mass. This value is similar to
the non-collider limits conservatively estimated by
Ref.~\cite{Buchkremer:2013bha} for a mass scale of the order of a \si{\TeV} from
atomic parity violation measurements~\cite{Deandrea:1997wk,Wood:1997zq}.

The sets of independent measurement analyses which provide the dominant
contributions to each $\kappa$-$M_Q$ point in the scans are identified by the
background colouring in the plot. This is highly sensitive to the $\xi$
configuration governing VLQ--boson couplings, with the most striking feature
being the difference between the $Z$ corner of the $\xi$ triangle (shown in
Figure~\ref{fig:gen1:010:dom}) as compared to the $W$ and $H$ corners
(Figures~\ref{fig:gen1:100:dom} and~\ref{fig:gen1:001:dom} respectively). The
former is unsurprisingly dominated by dilepton (+ jet) analyses, which also make
a significant appearance in the excluded $\kappa$-independent region and the
un-excluded high-mass region of the \WZHtoo mixture
(Figure~\ref{fig:gen1:121:dom}).

Exclusions in the low-mass regions of the $H$ and $W$ corners of the $\xi^V$
triangle (Figures~\ref{fig:gen1:001:dom} and~\ref{fig:gen1:100:dom}) are
dominated by $WW$ measurements (either direct or via Higgs
decays). Interestingly, the dominant contribution to the exclusion power from the
$WW$ analysis pool is the measurement of detector-corrected distributions in the
\emph{control regions} of the \SI{13}{\TeV} ATLAS leptoquark
search~\cite{Aaboud:2019jcc}, an addition to that search study made to enable
testing of MC generator models as well as BSM re-interpretation studies.  This
analysis has exclusion power across the whole mass range, but in the
$M_Q = \SIrange{1}{2}{\TeV}$ region it becomes subdominant to
lepton+\MET (+ jet) analyses sensitive to single-VLQ production, which drive the
exclusion contour downward to enclose lower $\kappa$ values in that mass range.
These analyses also dominate the \SIrange{1}{2}{\TeV} model exclusions in the
mixed $\WZH = \WZHtoo$ configuration, and make an appearance at
$\kappa \sim \numrange{0.05}{0.10}$ in the $\xi^Z = 1$ configuration. Further
investigation of this intrusion of lepton+\MET{}+jet analyses into the sea of
$WW$-based sensitivity reveals interesting phenomenology: it is dominated by
measurements of leading jet \pT in the detector-corrected control regions of the
\SI{8}{\TeV} ATLAS vector-boson fusion (VBF) $Wjj$ analysis~\cite{Aaboud:2017fye}.

\begin{figure}[tb]
  \centering
  \hspace*{-3.5em}%
  \mbox{%
    \subfloat[]{
      \includegraphics[width=0.36\textwidth]{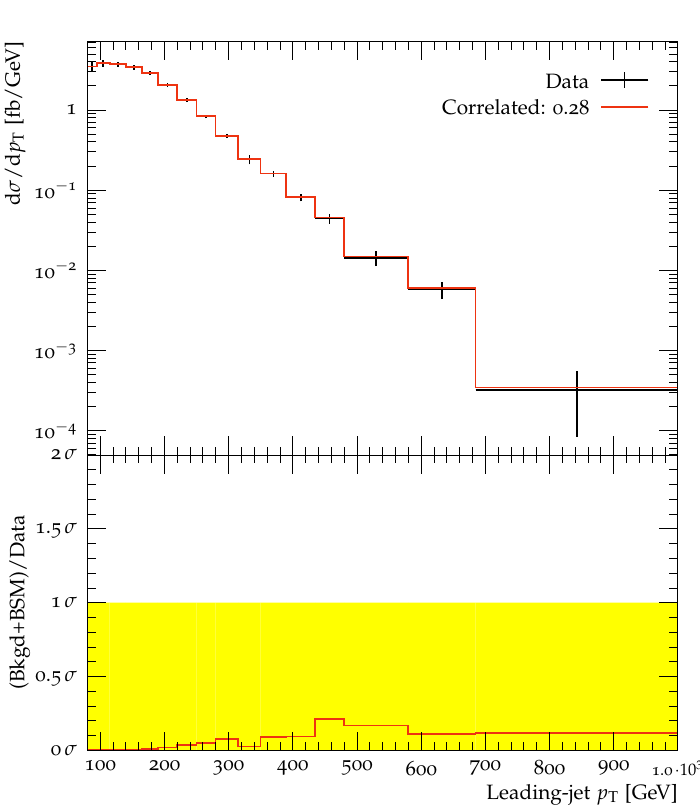}\label{fig:gen1:wjjscan:a}}
    \quad
    \subfloat[]{
      \includegraphics[width=0.36\textwidth]{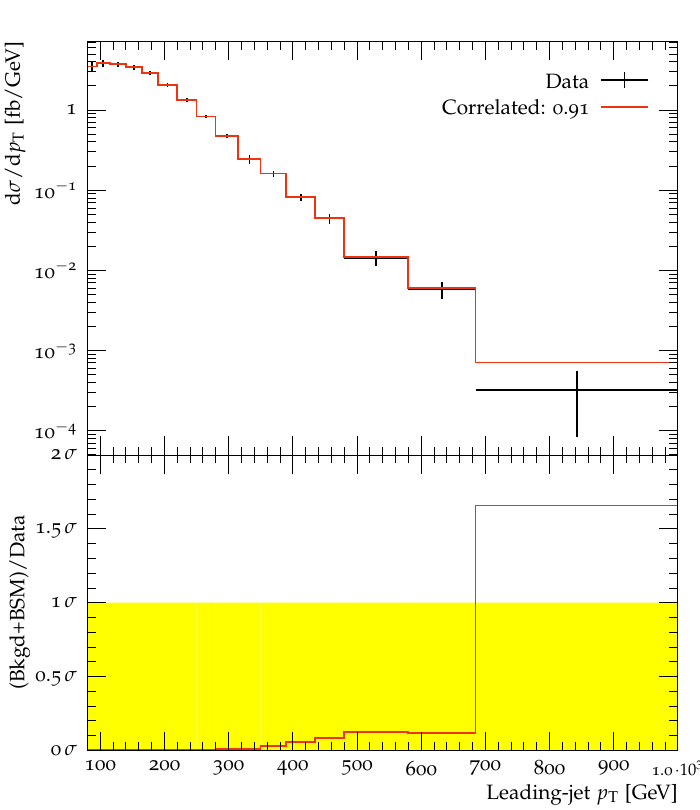}\label{fig:gen1:wjjscan:b}}
    \quad
    \subfloat[]{
      \includegraphics[width=0.36\textwidth]{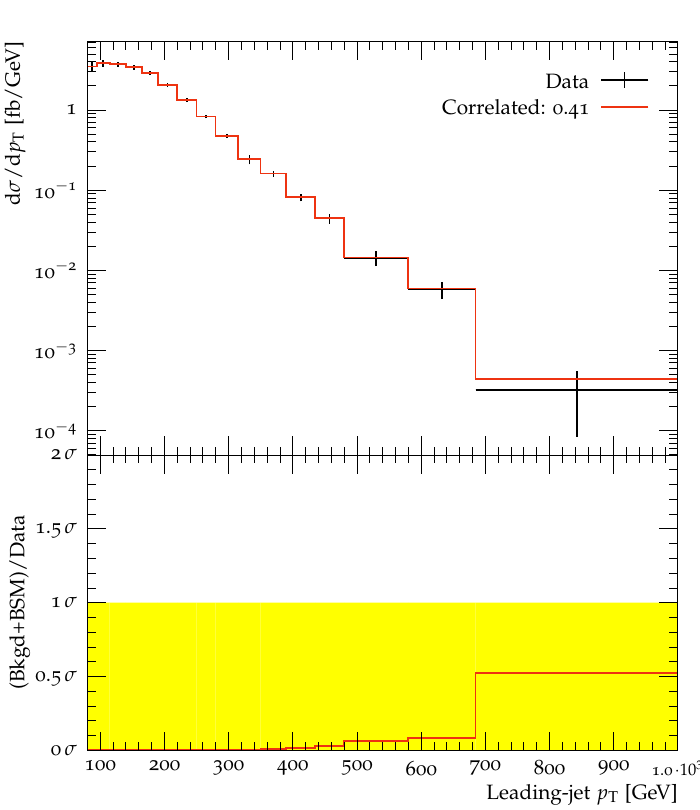}\label{fig:gen1:wjjscan:c}}
    }
    \caption{ATLAS \SI{8}{\TeV} $Wjj$ forward-lepton control region leading-jet
      \pT distributions at three points on the 95\% exclusion contour for
      $\WZH=\WZHozz$, respectively at $M_Q$ values of
      \protect\subref{fig:gen1:wjjscan:a} \SI{1000}{\GeV},
      \protect\subref{fig:gen1:wjjscan:b} \SI{1750}{\GeV}, and
      \protect\subref{fig:gen1:wjjscan:c} \SI{2250}{\GeV}. The rise and
      subsidence of a 90\% \CLs exclusion from a single $Wjj$ bin is seen as the
      contour passes from below \SI{1}{\TeV} to above \SI{2}{\TeV}.
      The black points are data, the red histogram is the VLQ contribution stacked on top of the data.
      In the lower insets, the ratio is shown and the yellow band indicates the significance, taking into
      account the statistical and systematic uncertainties on the data. The legend gives the exclusion
      (i.e. one minus the $p$-value) for that histogram, using all the bins in the distribution and treating systematic uncertainties as uncorrelated between bins.
}
  \label{fig:gen1:wjjscan}
\end{figure}

Figure~\ref{fig:gen1:wjjscan} shows the $Wjj$ leading-jet \pT distribution for
three $\WZH=\WZHozz$ points near the 95\% exclusion contour
respectively below, in, and above the
single-VLQ-dominated exclusion region. The highest bins of the \pT distribution
are seen to be significantly enhanced for non-$Z$-coupling VLQs with masses in
the \SIrange{1}{2}{\TeV} range, leading to the increased exclusion sensitivity
in that region, but for higher masses this excess subsides, due to both the
falling cross-section and the excess localising to out-of-range bins.


\subsection{Second quark generation}
\label{sec:kappa:2gen}

\begin{figure}[tbp]
  \centering
  \subfloat[]{
    \includegraphics[width=0.45\textwidth]{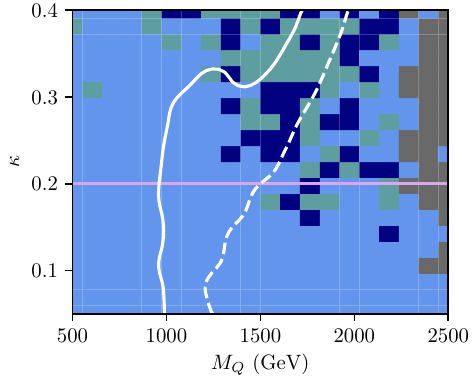}\label{fig:gen2:001:dom}}
  \subfloat[]{
    \includegraphics[width=0.45\textwidth]{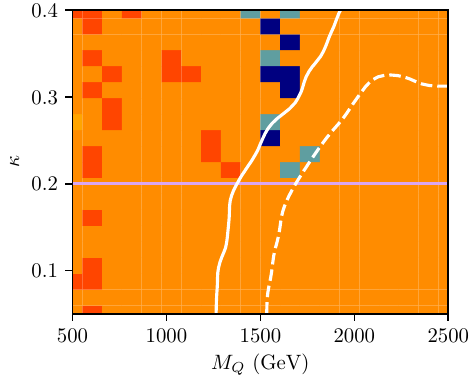}\label{fig:gen2:010:dom}} \\
  \subfloat[]{
    \includegraphics[width=0.45\textwidth]{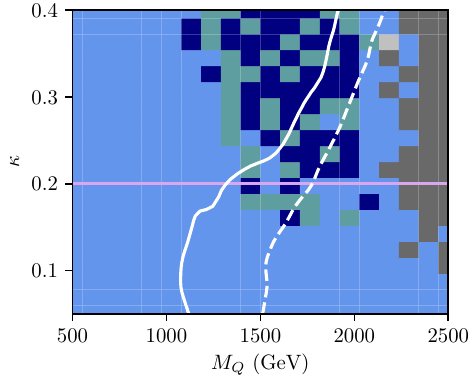}\label{fig:gen2:100:dom}}
  \subfloat[]{
    \includegraphics[width=0.45\textwidth]{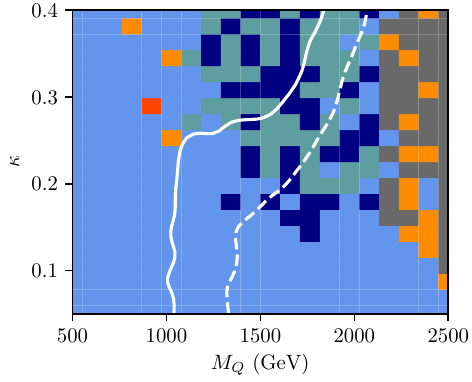}\label{fig:gen2:121:dom}}
  \vspace*{2ex}
  \begin{tabular}{llll}
    \swatch{cornflowerblue}~ATLAS $WW$ & \swatch{navy}~ATLAS $\mu$+\MET{}+jet & \swatch{cadetblue}~ATLAS $e$+\MET{}+jet\\
    \swatch{orangered}~ATLAS $ee$+jet & \swatch{darkorange}~ATLAS $\mu\mu$+jet & \swatch{orange!60}~ATLAS $\ell\ell$+jet\\
    \swatch{magenta} ATLAS~4$\ell$ & \swatch{silver}~ATLAS jets & \swatch{dimgrey}~CMS jets\\
  \end{tabular}
  \vspace*{2ex}
  \caption{Dominant LHC analysis pools contributing to VLQ limit-setting in the
    $\kappa$ vs VLQ mass plane, where $\kappa$ is the coupling to
    second-generation SM quarks.  All VLQ ($B, T, X, Y$) masses are set to be
    degenerate. The disfavoured regions are located above and to the left of the
    dashed (68\%~CL) and solid (95\%~CL) white contours respectively. The lower
    bounds in $\kappa$ from non-LHC flavour physics are indicated with the pink
    horizontal contour. The VLQ branching fractions to \WZH are
    \protect\subref{fig:gen2:001:dom}~\WZHzzo
    \protect\subref{fig:gen2:010:dom}~\WZHzoz
    \protect\subref{fig:gen2:100:dom}~\WZHozz and
    \protect\subref{fig:gen2:121:dom}~\WZHtoo.}
  \label{fig:2gen:dom}
\end{figure}

In Figure~\ref{fig:2gen:dom}, we perform the equivalent scan and analysis for
weak VLQ couplings only to the second quark generation.  Again, at low VLQ
masses, pair production dominates and the exclusion is insensitive to
$\kappa$. The same pattern of mass thresholds as a function of $\xi$
configuration is seen as for the first-generation scan. However, the $\kappa$-dependent
exclusion which dominated the parameter space for the first generation
is here absent, with only a hint of $\kappa$-dependence in the $Z$-only coupling
configuration (Figure~\ref{fig:gen2:010:dom}). This is in keeping with the
expectation summarised in Section~\ref{sec:pheno}, since the weak
pair-production and single production processes are highly suppressed if the
VLQs cannot couple to the proton's valence quarks (nor the first-generation
sea).
In this case, the estimated limit ($\kappa <0.2$) from non-LHC measurements
comes from measurements of the $Z \rightarrow q\bar{q}$ couplings at
LEP~\cite{Buchkremer:2013bha,ALEPH:2005ab}

The $WW$ analysis group dominates for the second generation
$H$ and $W$ $\xi$ corners (Figures~\ref{fig:gen2:001:dom}
and~\ref{fig:gen2:100:dom}) and for most of the mixed-boson configuration, at
least for all $\kappa$ values allowed by the flavour constraints.  At higher
$\kappa$ values, weak VLQ--boson production mechanisms again contribute, giving
some weak dependence on the coupling.  The $Z$-compatible dilepton+jets analyses
dominate all of the $Z$-only configuration in Figure~\ref{fig:gen2:010:dom}.
Thus for the most interesting $\kappa < 0.2$ region, second-generation limits
are driven by VLQ pair production, with decays into the permitted bosons and SM
quarks.  This is consistent with the discussion in Section~\ref{sec:pheno}:
production of VLQs in association to quarks is suppressed by an order of
magnitude compared to the case where VLQs can couple to first-generation quarks,
while pair-production via EM/QCD processes occurs at roughly the same rate. This
explains the weaker dependence on $\kappa$ of the exclusion reported above.

We again see the intrusion of single-VLQ exclusion by the ATLAS VBF $Wjj$
analysis for second-generation VLQs between \SIrange{1}{2}{\TeV}, explained by
the same mechanism. Compared to the first generation, however, the impact of QCD
jet analyses is also seen at the highest masses, for all but the pure-$Z$
configuration of Figure~\ref{fig:gen2:010:dom}. This is primarily driven by the
CMS \SI{13}{\TeV} jet mass analysis~\cite{Sirunyan:2018xdh}, and to a lesser
extent the ATLAS \SI{13}{\TeV} dijet \& inclusive jet analysis~\cite{Aaboud:2017wsi}.

\subsection{Third quark generation}
\label{sec:kappa:3gen}

\begin{figure}[tbp]
  \centering
  \subfloat[]{
    \includegraphics[width=0.45\textwidth]{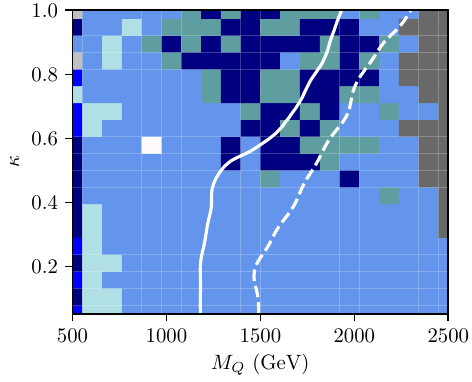}\label{fig:gen3:001:dom}}
  \subfloat[]{
    \includegraphics[width=0.45\textwidth]{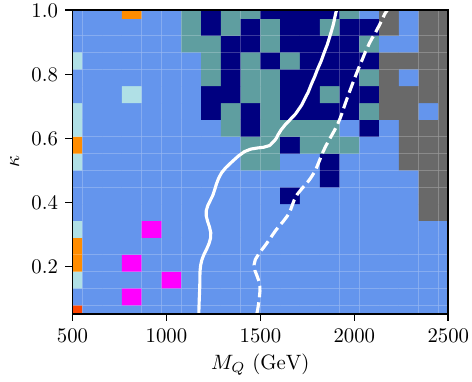}\label{fig:gen3:010:dom}} \\
  \subfloat[]{
    \includegraphics[width=0.45\textwidth]{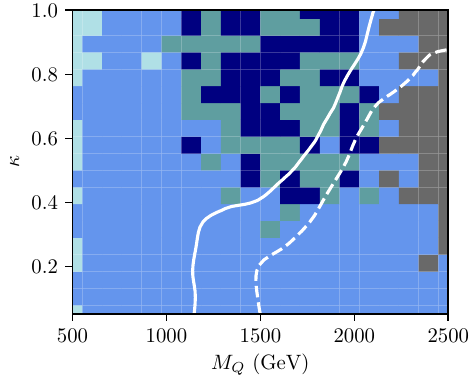}\label{fig:gen3:100:dom}}
  \subfloat[]{
    \includegraphics[width=0.45\textwidth]{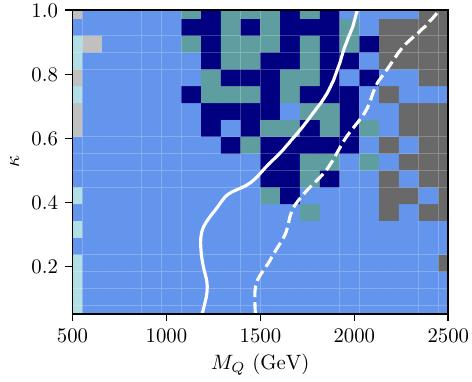}\label{fig:gen3:121:dom}}
  \vspace*{2ex}
  \begin{tabular}{lll}
    \swatch{cornflowerblue}~ATLAS $WW$ & \swatch{navy}~ATLAS $\mu$+\MET{}+jet & \swatch{cadetblue}~ATLAS $e$+\MET{}+jet \\
    \swatch{powderblue} CMS $\ell$+\MET{}+jet & \swatch{darkorange}~ATLAS $\mu\mu$+jet & \swatch{magenta} ATLAS~4$\ell$ \\
    \swatch{silver}~ATLAS jets & \swatch{dimgrey}~CMS jets & \swatch{snow}~ATLAS $t\bar{t}$ hadronic
  \end{tabular}
  \vspace*{2ex}
  \caption{Dominant LHC analysis pools contributing to VLQ limit-setting in the $\kappa$ vs
    VLQ mass plane, where $\kappa$ is the coupling to third-generation SM quarks.
    All VLQ ($B, T, X, Y$) masses are set to be degenerate. The disfavoured regions
    are located above and to the left of the dashed (68\%~CL)
    and solid (95\%~CL) white contours respectively. The VLQ branching
    fractions to \WZH are \protect\subref{fig:gen3:001:dom}~\WZHzzo
    \protect\subref{fig:gen3:010:dom}~\WZHzoz \protect\subref{fig:gen3:100:dom}~\WZHozz
    and \protect\subref{fig:gen3:121:dom}~\WZHtoo. %
  }
  \label{fig:3gen:dom}
\end{figure}

In Figure~\ref{fig:3gen:dom}, we show the sensitivity in the plane of the
coupling $\kappa$ to the third-generation SM quarks and VLQ mass.  Again, as
discussed in Section~\ref{sec:pheno}, at low VLQ masses, pair production
dominates and the exclusion is insensitive to $\kappa$. Since there is no top
parton density, and the $b$ content of the proton is suppressed relative to
light quarks, single production cross-sections are lower than for the lighter
generations. Single production does still bring some additional sensitivity to
higher masses (around \SI{2}{\TeV}) when $W+q$ decays dominate, which is
consistent with the fact that in this region \contur does better than the
searches (which focus on pair production) in Figure~\ref{fig:BTonlyB} (for $B$
VLQs).

Comparing Figure~\ref{fig:gen3:010:dom} with the equivalent figures for the
first- and second-generation interactions (Figures~\ref{fig:gen1:010:dom}
and~\ref{fig:gen2:010:dom} respectively), it is interesting to note that
the type of measurement which provides the best sensitivity changes for the $Z$-corner
in third-generation couplings. In the first- and second-generation cases,
the sensitivity is dominated by dilepton-plus-jet measurements, while the third-generation
case becomes dominated by measurements involving leptons and missing energy with jets,
or $WW$-like measurements --- only a few points in the scan are dominated by the
ATLAS \SI{13}{\TeV} dilepton~\cite{Aaboud:2017hbk,Aaboud:2019jcc} or
four-lepton~\cite{Aaboud:2019lxo,Aaboud:2017rwm} measurements.

This change is due to the fact that when third-generation couplings are the only
ones allowed, $X+q$ and $T+q$ processes are suppressed due to the lack of top
quarks in the proton sea. $Y+q$ dominates the high-mass region, with
pair-production dominant for lower VLQ masses. The $T$ and $Y$ VLQs, produced
singly or in pairs, will decay to top quarks, resulting in the production of at
least one $W$-boson. This in turn leads to the missing energy signatures which
were not present for first- or second-generation couplings.

The $WW$ dominance across all $\xi$ configurations also reflects the impact of
exploiting multiple bins in the distribution: without this information, the \SI{13}{\TeV}
CMS $\ell$+\MET{}+jet
analyses~\cite{Aaboud:2018uzf,Aaboud:2017fha,Aaboud:2018eki,Sirunyan:2018wem,Khachatryan:2016mnb,Sirunyan:2018ptc}
would dominate at low mass in the non-$Z$ configurations, and the statistically
limited ATLAS four-lepton analyses would dominate a larger region of
the $\xi^Z = 1$ $\kappa$--$M_Q$ plane.

\subsection{Singlets, doublets and triplets}
\label{sec:kappa:multiplets}

The results in this section assumed four VLQs, $B,T,X,Y$. This was a good didactic scenario, as it contains the most rich phenomenology, but is not the most favoured scenario.
In Appendix~\ref{app:multiplets} we present equivalent results for singlets $(B)$, $(T)$, doublets $(B,T)$, $(B,Y)$, $(X,T)$, and triplets $(B,T,X)$,
$(B,T,Y)$, which have stronger theoretical motivation. 
The $\WZH=\WZHzoo$ case is considered for doublets instead of $\WZH=\WZHtoo$, as motivated by Ref.~\cite{Buchkremer:2013bha}. 
The results are broadly similar to those shown in the previous sections, with a few important differences which we summarise here:
\begin{itemize}
\item The constraints are typically slightly weaker the fewer particles there
  are in the multiplet: more new particles means more opportunities for
  divergences from the SM.
\item Scenarios without $X$ or $Y$ particles and where $\WZH=\WZHzzo$ have
  substantially weaker constraints than other scenarios, since the lack of $W$
  decays means that only Higgs measurements would be sensitive (manifested as
  $\gamma+X$ analyses dominating low-mass sensitivity via $H \to \gamma\gamma$),
  and low single-production cross-sections through the Higgs mean that only
  comparatively high values of $\kappa$ can be excluded beyond the
  pair-production region.
\item Continuing this focus on the $\WZH=\WZHzzo$ configuration, if there are
  only first-generation couplings, the triplets and doublets containing only one
  of $X$ or $Y$ have differing degrees of $\kappa$-sensitive exclusion power,
  with stronger constraints if $X$ is present. This is due to the larger valence
  up- versus down-quark PDFs, for $u \to W^- X$ and $d \to W^+ Y$ respectively.
  This effect disappears (along with the valence PDFs) for second-generation couplings.
\item Assuming only third-generation couplings, the $\WZH=\WZHzoz$ and
  $\WZH=\WZHzzo$ corners for multiplets containing only $T$ and $Y$ VLQs have no
  $\kappa$-dependence in their constraints, since single-VLQ production is not
  allowed: the top proton density is vanishingly small for $t \to T \{Z,H\}$ and
  $t \to X W^-$ production. Only pair-production is a viable mechanism.
\item The converse would apply for only $B$ and $X$ VLQs in the $\WZH=\WZHozz$
  case, but the only such natural multiplet is the $B$ singlet. In this we do
  indeed see $\kappa$ independence again.
\end{itemize}

\afterpage{\clearpage}

\section{Discussion and conclusions}

We have presented studies of a generic class of Vector-Like Quark models,
comparing the predictions from all $2\rightarrow2$ production diagrams for VLQs
to a large bank of LHC differential cross-section measurements.
Despite the measurements not being explicitly designed for this purpose,
we find that they can exclude significant regions of VLQ parameter space,
in a wider range of model parameters than those
typically considered in dedicated searches.

This approach, using inclusive event generation and
model-independent measurements to study BSM signals, is hence not only
competitive with dedicated searches, but indeed can outperform them in
generic model spaces for which specific searches have not been optimised.
This
is due to the fact that
searches often make simplifying assumptions, or focus their efforts on
detecting the most spectacular signature from a model.
But
studying all the implications of a given model simultaneously
may instead reveal
moderate changes to many SM distributions,
an effect most evident through combination of multiple precision measurements.
This is very much the case in VLQ models. Firstly, LHC searches for such
particles have often focused on pair-produced VLQs as a $\kappa$-dependence is
technically challenging for interpretations made using detector-level physics
objects. Secondly, it is typical at the LHC for only VLQs coupling to
third-generation quarks to be considered, as they give the most striking
experimental signatures and are most weakly constrained by previous measurements.
Thirdly we note efforts to reinterpret searches in terms of models
that include additonal exotic VLQ production mechanisms~\cite{Araque:2015cna}; our approach could
be easily extended to cover such scenarios.

Reviewing the phenomenology of VLQ models at the LHC, we see 
that the way
in which VLQs are produced in proton--proton collisions depends intimately on
interplays between the composition of the proton, and the strengths of the
coupling of VLQs to different SM bosons and the different generations of SM
quarks. For example, VLQs coupled to first-generation quarks are more
likely to be singly-produced than if coupled to second- or third-generation
quarks, where pair production is typically more likely. Furthermore, production
mechanisms such as $X+W$ become almost forbidden if only third-generation
couplings are allowed, leading to an asymmetry between $X$ and $Y$ production
which is not present for second-generation couplings.  Finally, the
phenomenology of VLQ decays when $B,T$ only couple to $Z$ bosons changes
drastically if VLQs are allowed to couple to third-generation quarks:
lepton-plus-missing-energy signatures then dominate over the expected
dilepton-plus-jet signatures, due to the production of top quarks and subsequent
decays involving missing energy.

These effects are neither small nor trivial, and suggest that the
richly intertwined phenomenology of VLQ production and decay at the LHC could
lead to novel analysis strategies and potentially new routes to discovery at
hadron colliders. Indeed, if a discovery were made, one could use these
considerations, along with the latest understanding of proton PDFs, to constrain
both the overall scale of weak VLQ interactions and their relative couplings to
different SM quark generations.
These insights from our inclusive approach
herald a very interesting era for VLQ searches at the LHC.

Given that a dedicated search can take a large team several years to prepare and
publish, while running a \contur scan takes less than a day, it is arguable that
checking compatibility of the search targets with the current canon of
model-independent measurements constitutes an important ``due diligence'' step
in analysis design.  The increased volume of data, and pressures on computing
resources in future LHC runs, make this argument ever more compelling in our
view.
Conversely, actively considering the contributions of inclusive-production
studies when designing future analyses enables analysis teams to focus on
those model regions which are not already covered.

This would naturally focus search attention on more exotic signatures, for
example of long-lived particles or other
anomalous, intrinsically non-SM-like features. In the precision era of the LHC there
is an increasing motivation to make detailed measurements, rather than
searches, in regions of phase-space with a significant SM cross-section.
Search analyses can also contribute to the resource of model-independent results
by making
unfolded measurements in their control regions\footnote{And in signal regions if
  no excess is observed, both adding re-interpretation value and enhancing the
  impact of an otherwise null-result paper.}: as discussed in this paper, a
prototype of control-region measurements in a leptoquark search has proven to
have significant exclusion power in regions of VLQ parameter space.
One search's control region can be anothers
signal region: model-independent measurements may help avoid the danger of
accidentally fitting away a signal.

Finally, this approach relies on careful preservation of measurements in
\hepdata and \rivet. Although \rivet was designed as a way to compare MC
generators, it turns out to be perfectly suited to re-interpretation, especially
if the reference data are published with a full breakdown of uncertainties and
state-of-the-art background predictions to aid rigorous statistical
interpretation.  Most LHC measurements do provide \rivet routines, but
unfortunately some very powerful measurements are still missing from the \rivet
database, or only come out many years after the associated publication. Furthermore, CMS has historically
published fewer routines than ATLAS, a fact manifest in the dominance by ATLAS
measurements of the \CLs limits shown in this paper. We hope that our results
add further encouragement for all collider physics experiments to make public
re-interpretation routines and bin-correlation data a core feature of their publication and
data-preservation processes~\cite{Abdallah:2020pec}.

\section*{Acknowledgements}
We thank Jack Burton, Khadeeja Bepari, Ben Waugh and David Yallup for helpful discussions,
Martin Habedank for implementing the presentation of dominant \CLs pools in \contur,
and Peter Richardson for improvements and enhancements of the Herwig UFO interface.
Our thanks to the many contributors to the \rivet analysis collection.

\paragraph{Funding information}
AB, JMB, LC and DH have received funding from the European Union's Horizon 2020
research and innovation programme as part of the Marie Skłodowska-Curie
Innovative Training Network MCnetITN3 (grant agreement no.~722104). AB and PS
acknowledge Royal Society funding under grants UF160548 and
RGF\textbackslash{}EA\textbackslash{}180252. AB, JMB and LC have received
funding from the UKRI Science and Technology Facilities Council (STFC)
consolidated grants for experimental particle physics.


\clearpage
\appendix

\section{Additional VLQ multiplets}
\label{app:multiplets}

\subsection{First generation}

\begin{figure}[h]
  \centering
  \subfloat[\BTX~\WZHzzo]{
    \includegraphics[width=0.24\textwidth]{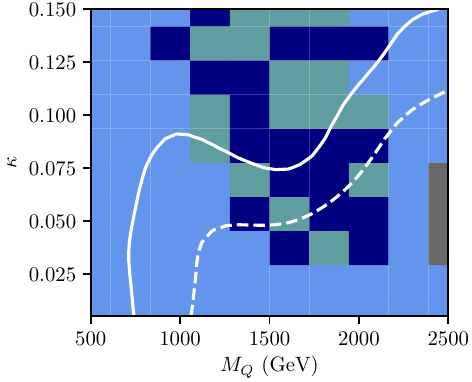}\label{fig:3gen1:BTX001}}
  \subfloat[\BTX~\WZHzoz]{
    \includegraphics[width=0.24\textwidth]{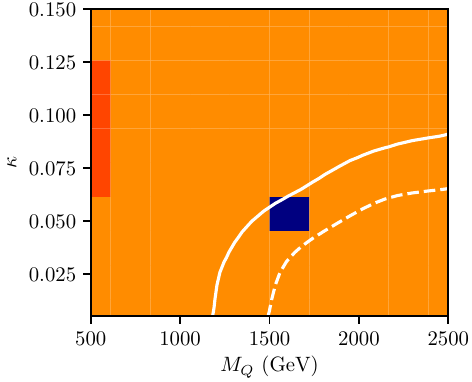}\label{fig:3gen1:BTX010}}
  \subfloat[\BTX~\WZHozz]{
    \includegraphics[width=0.24\textwidth]{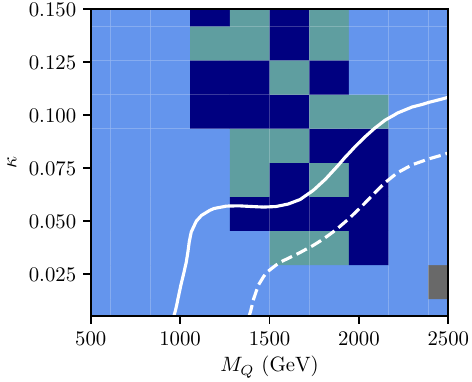}\label{fig:3gen1:BTX100}}
  \subfloat[\BTX~\WZHtoo]{
    \includegraphics[width=0.24\textwidth]{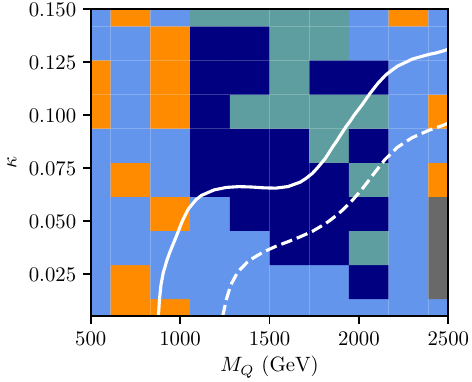}\label{fig:3gen1:BTX211}}\\
  \subfloat[\BTY~\WZHzzo]{
    \includegraphics[width=0.24\textwidth]{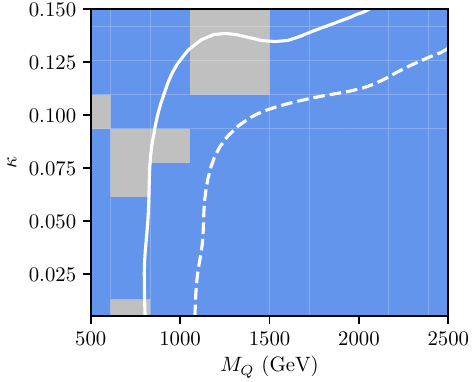}\label{fig:3gen1:BTY001}}
  \subfloat[\BTY~\WZHzoz]{
    \includegraphics[width=0.24\textwidth]{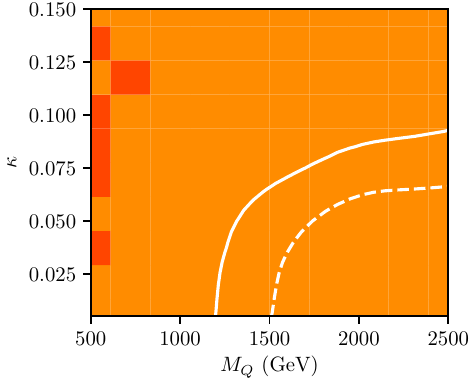}\label{fig:3gen1:BTY010}}
  \subfloat[\BTY~\WZHozz]{
    \includegraphics[width=0.24\textwidth]{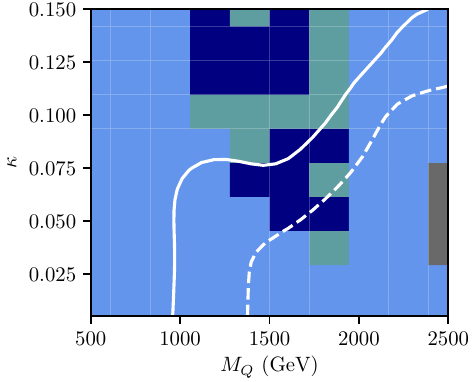}\label{fig:3gen1:BTY100}}
  \subfloat[\BTY~\WZHtoo]{
    \includegraphics[width=0.24\textwidth]{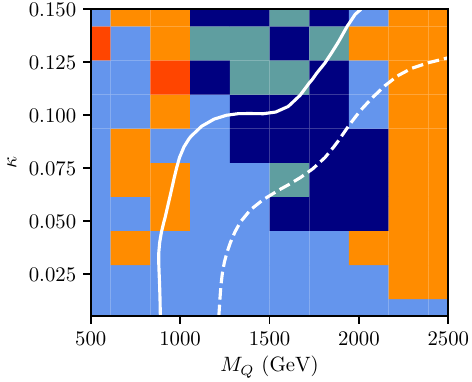}\label{fig:3gen1:BTY211}}\\[1em]
  \hrule
  \subfloat[\BTXY~\WZHzzo]{
    \includegraphics[width=0.24\textwidth]{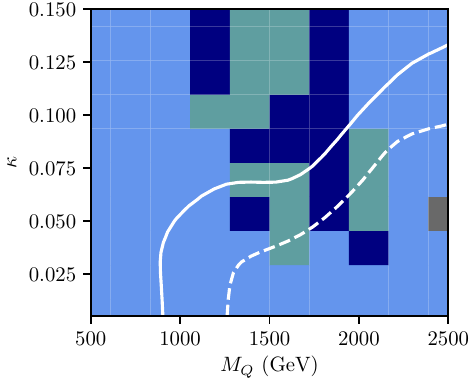}\label{fig:3gen1:BTXY001}}
  \subfloat[\BTXY~\WZHzoz]{
    \includegraphics[width=0.24\textwidth]{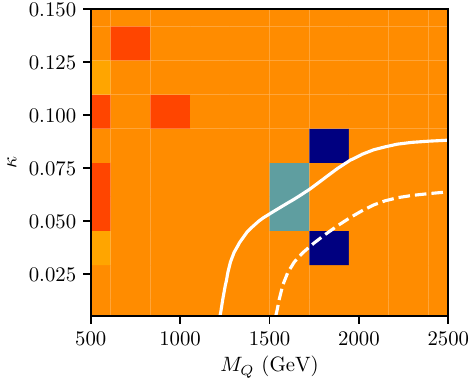}\label{fig:3gen1:BTXY010}}
  \subfloat[\BTXY~\WZHozz]{
    \includegraphics[width=0.24\textwidth]{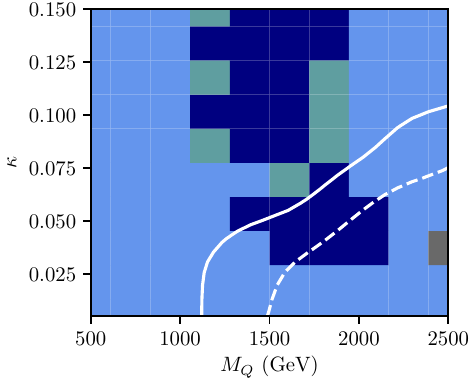}\label{fig:3gen1:BTXY100}}
  \subfloat[\BTXY~\WZHtoo]{
    \includegraphics[width=0.24\textwidth]{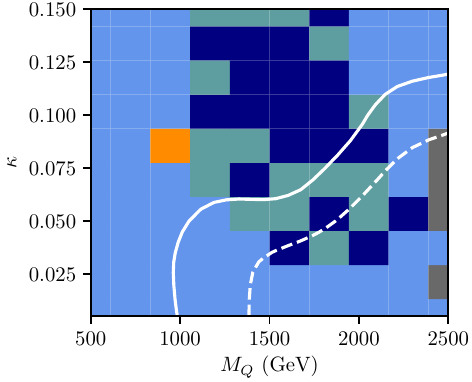}\label{fig:3gen1:BTXY211}}
  \vspace*{2ex}
  \begin{tabular}{llll}
    \swatch{cornflowerblue}~ATLAS $WW$ & \swatch{navy}~ATLAS $\mu$+\MET{}+jet & \swatch{cadetblue}~ATLAS $e$+\MET{}+jet\\
    \swatch{orangered}~ATLAS $ee$+jet & \swatch{darkorange}~ATLAS $\mu\mu$+jet & \swatch{orange!60}~ATLAS $\ell\ell$+jet\\
                                      \swatch{silver}~ATLAS jets & \swatch{dimgrey}~CMS jets &
  \end{tabular}
  \vspace*{2ex}
  \caption{ Sensitivity of LHC measurements to VLQ triplet production in the $\kappa$ vs
    VLQ mass plane, where $\kappa$ is the coupling to first-generation SM quarks.
    All VLQ masses are set to be degenerate.
    The multiplets are given as rows:
    \protect\subref{fig:3gen1:BTX001}--\protect\subref{fig:3gen1:BTX211}($B,T,X$) triplet,
    \protect\subref{fig:3gen1:BTY001}--\protect\subref{fig:3gen1:BTY211}($B,T,Y$) triplet, and for comparison to the main text, the
    \protect\subref{fig:3gen1:BTXY001}--\protect\subref{fig:3gen1:BTXY211}($B,T,X,Y$) quadruplet.
    The VLQ branching fractions to \WZH are arranged in columns of \WZHzzo, \WZHzoz, \WZHozz, and \WZHtoo from left to right.
  }%
  \label{fig:1gen_triplets}
\end{figure}

\begin{figure}[h]
  \centering
  \subfloat[\BT~\WZHzzo]{
    \includegraphics[width=0.24\textwidth]{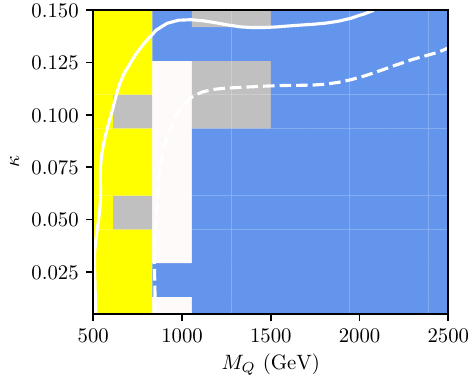}\label{fig:2gen1:BT001}}
  \subfloat[\BT~\WZHzoz]{
    \includegraphics[width=0.24\textwidth]{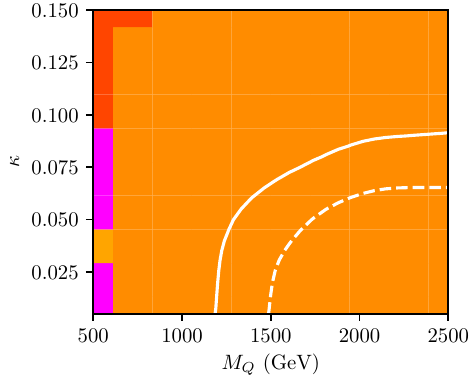}\label{fig:2gen1:BT010}}
  \subfloat[\BT~\WZHozz]{
    \includegraphics[width=0.24\textwidth]{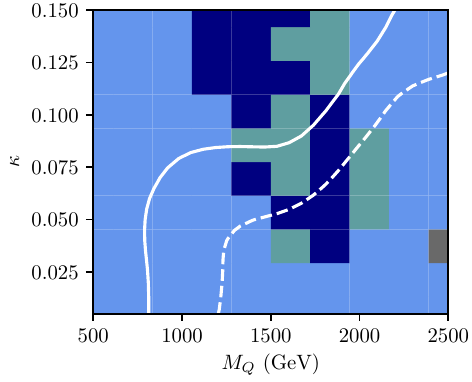}\label{fig:2gen1:BT100}}
  \subfloat[\BT~\WZHzoo]{
    \includegraphics[width=0.24\textwidth]{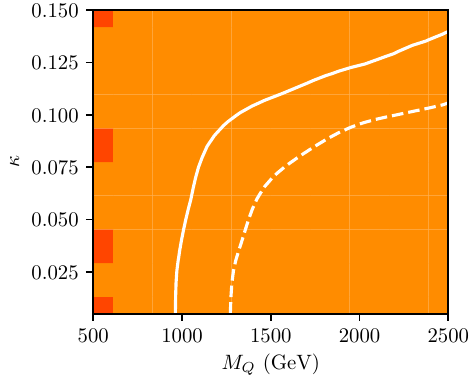}\label{fig:2gen1:BT011}}\\
  \subfloat[\XT~\WZHzzo]{
    \includegraphics[width=0.24\textwidth]{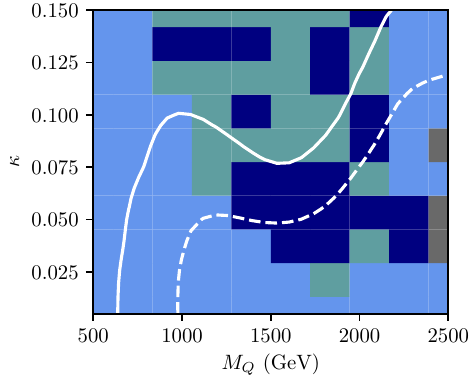}\label{fig:2gen1:XT001}}
  \subfloat[\XT~\WZHzoz]{
    \includegraphics[width=0.24\textwidth]{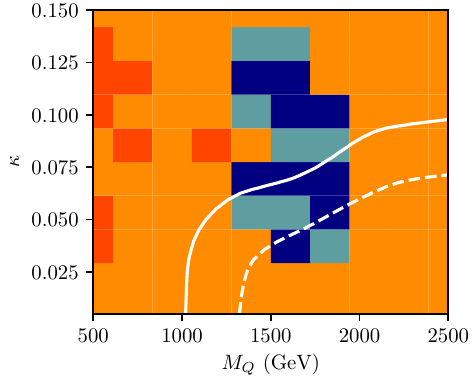}\label{fig:2gen1:XT010}}
  \subfloat[\XT~\WZHozz]{
    \includegraphics[width=0.24\textwidth]{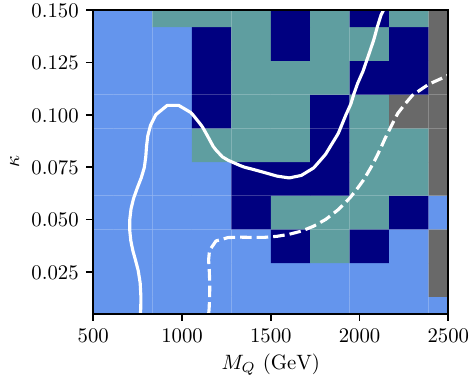}\label{fig:2gen1:XT100}}
  \subfloat[\XT~\WZHzoo]{
    \includegraphics[width=0.24\textwidth]{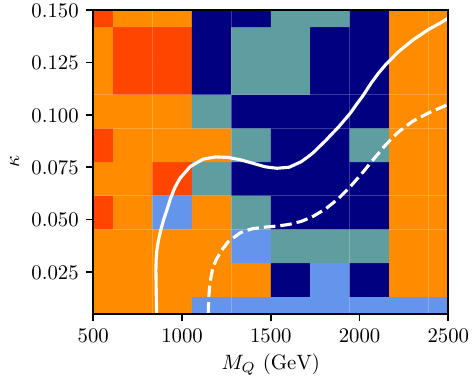}\label{fig:2gen1:XT011}}\\
  \subfloat[\BY~\WZHzzo]{
    \includegraphics[width=0.24\textwidth]{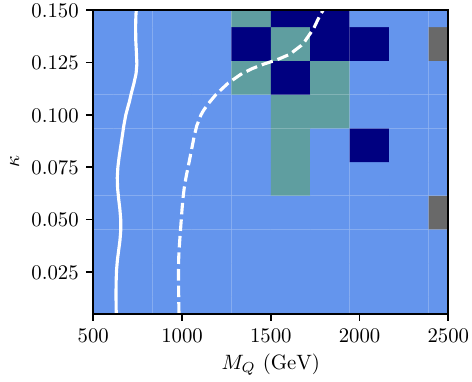}\label{fig:2gen1:BY001}}
  \subfloat[\BY~\WZHzoz]{
    \includegraphics[width=0.24\textwidth]{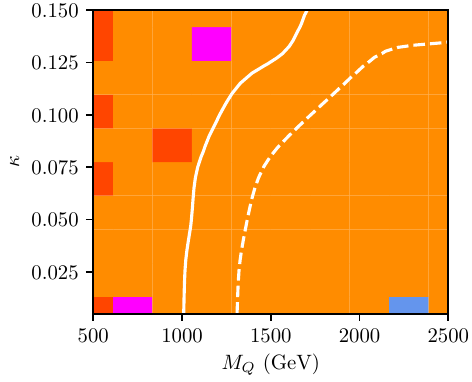}\label{fig:2gen1:BY010}}
  \subfloat[\BY~\WZHozz]{
    \includegraphics[width=0.24\textwidth]{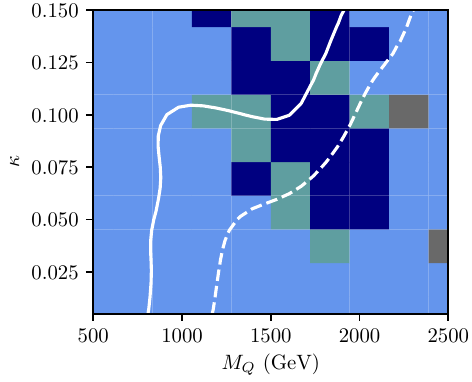}\label{fig:2gen1:BY100}}
  \subfloat[\BY~\WZHzoo]{
    \includegraphics[width=0.24\textwidth]{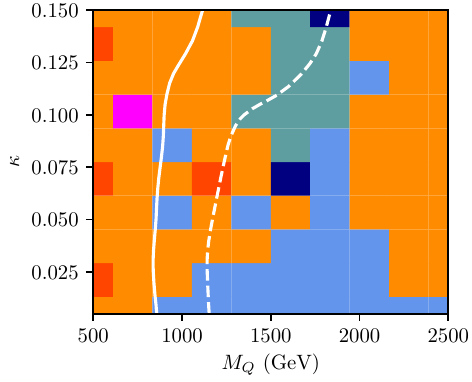}\label{fig:2gen1:BY011}}\\[1em]
  \hrule
  \subfloat[\BTXY~\WZHzzo]{
    \includegraphics[width=0.24\textwidth]{BTYX/gen1_hwz100/dominantPools0}\label{fig:2gen1:BTXY001}}
  \subfloat[\BTXY~\WZHzoz]{
    \includegraphics[width=0.24\textwidth]{BTYX/gen1_hwz001/dominantPools0}\label{fig:2gen1:BTXY010}}
  \subfloat[\BTXY~\WZHozz]{
    \includegraphics[width=0.24\textwidth]{BTYX/gen1_hwz010/dominantPools0}\label{fig:2gen1:BTXY100}}
  \subfloat[\BTXY~\WZHzoo]{
    \includegraphics[width=0.24\textwidth]{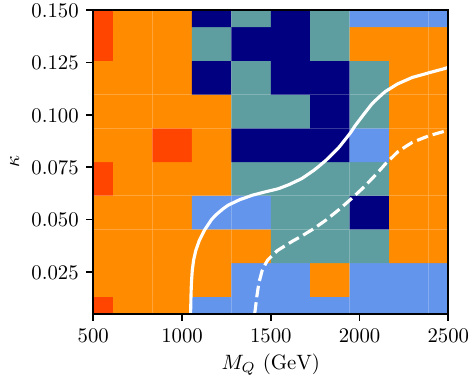}\label{fig:2gen1:BTXY011}}
  \vspace*{2ex}
  \begin{tabular}{llll}
    \swatch{snow}~ATLAS $t\bar{t}$ hadr & \swatch{yellow!60!white} ATLAS $\gamma\gamma$ \& $\gamma\!+\!X$ \\
    \swatch{cornflowerblue}~ATLAS $WW$ & \swatch{navy}~ATLAS $\mu$+\MET{}+jet & \swatch{cadetblue}~ATLAS $e$+\MET{}+jet\\
    \swatch{orangered}~ATLAS $ee$+jet & \swatch{darkorange}~ATLAS $\mu\mu$+jet & \swatch{orange!60}~ATLAS $\ell\ell$+jet\\
    \swatch{magenta} ATLAS~4$\ell$ & \swatch{silver}~ATLAS jets & \swatch{dimgrey}~CMS jets\\
  \end{tabular}
  \vspace*{2ex}
  \caption{ Sensitivity of LHC measurements to VLQ doublet production in the $\kappa$ vs
    VLQ mass plane, where $\kappa$ is the coupling to first-generation SM quarks.
    All VLQ masses are set to be degenerate.
    The multiplets are given as rows:
    \protect\subref{fig:2gen1:BT001}--\protect\subref{fig:2gen1:BT011}($B,T$) doublet,
    \protect\subref{fig:2gen1:XT001}--\protect\subref{fig:2gen1:XT011}($X,T$) doublet,
    \protect\subref{fig:2gen1:BY001}--\protect\subref{fig:2gen1:BY011}($B,Y$) doublet, and for comparison to the main text, the
    \protect\subref{fig:2gen1:BTXY001}--\protect\subref{fig:2gen1:BTXY100}($B,T,X,Y$) quadruplet.
    The VLQ branching fractions to \WZH are arranged in columns of \WZHzzo, \WZHzoz, \WZHozz, and \WZHzoo from left to right.
    The \WZHzoo case is considered for doublets instead of \WZHtoo, as motivated by Ref.~\cite{Buchkremer:2013bha}. 
  }%
  \label{fig:1gen_doublets}
\end{figure}

\begin{figure}[h]
  \centering
   \subfloat[$B$~\WZHzzo]{
    \includegraphics[width=0.24\textwidth]{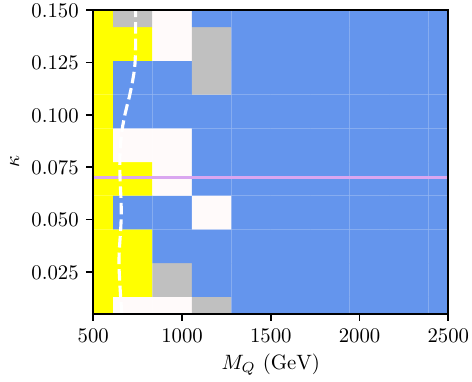}\label{fig:1gen1:B001}}
  \subfloat[$B$~\WZHzoz]{
    \includegraphics[width=0.24\textwidth]{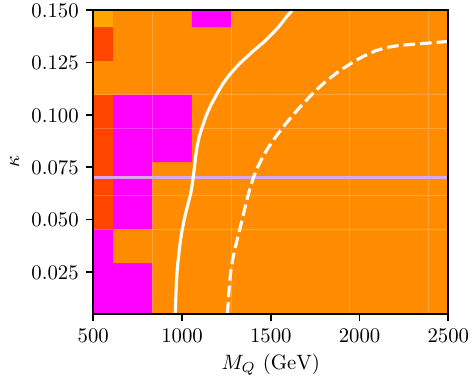}\label{fig:1gen1:B010}}
  \subfloat[$B$~\WZHozz]{
    \includegraphics[width=0.24\textwidth]{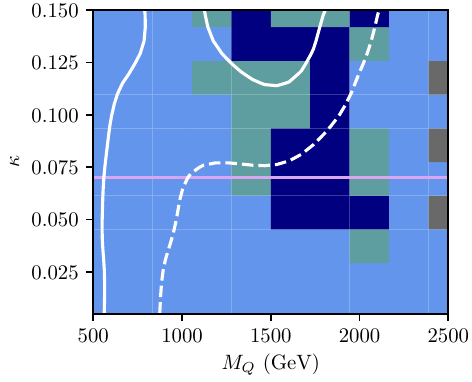}\label{fig:1gen1:B100}}
  \subfloat[$B$~\WZHtoo]{
    \includegraphics[width=0.24\textwidth]{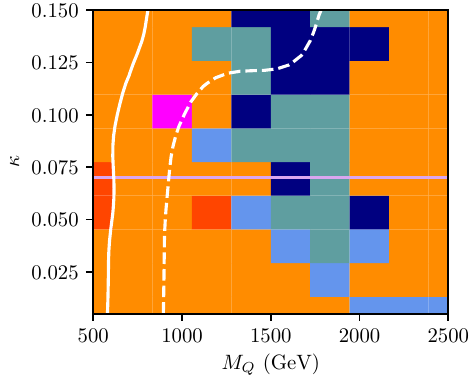}\label{fig:1gen1:B211}}\\
  \subfloat[$T$~\WZHzzo]{
    \includegraphics[width=0.24\textwidth]{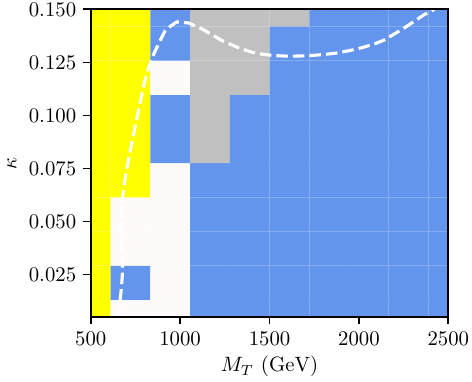}\label{fig:1gen1:T001}}
  \subfloat[$T$~\WZHzoz]{
    \includegraphics[width=0.24\textwidth]{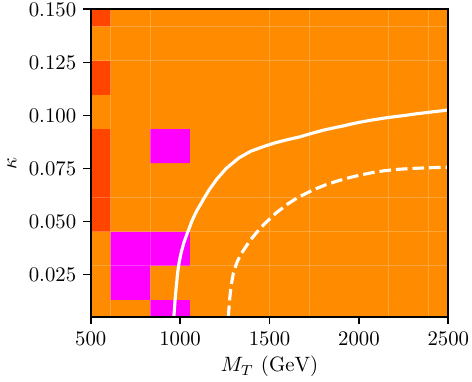}\label{fig:1gen1:T010}}
  \subfloat[$T$~\WZHozz]{
    \includegraphics[width=0.24\textwidth]{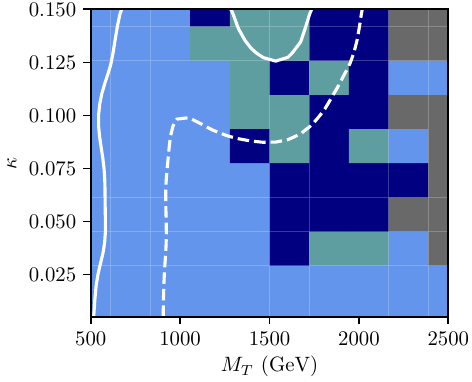}\label{fig:1gen1:T100}}
  \subfloat[$T$~\WZHtoo]{
    \includegraphics[width=0.24\textwidth]{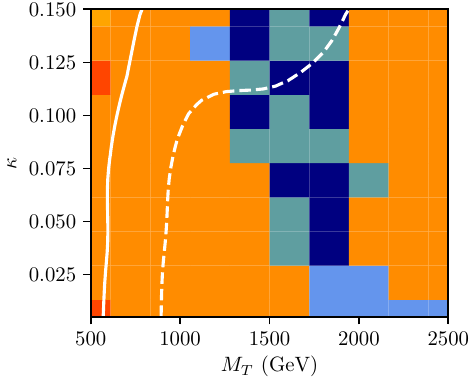}\label{fig:1gen1:T211}}\\[1em]
  \hrule
  \subfloat[\BTXY~\WZHzzo]{
    \includegraphics[width=0.24\textwidth]{BTYX/gen1_hwz100/dominantPools0}\label{fig:1gen1:BTXY001}}
  \subfloat[\BTXY~\WZHzoz]{
    \includegraphics[width=0.24\textwidth]{BTYX/gen1_hwz001/dominantPools0}\label{fig:1gen1:BTXY010}}
  \subfloat[\BTXY~\WZHozz]{
    \includegraphics[width=0.24\textwidth]{BTYX/gen1_hwz010/dominantPools0}\label{fig:1gen1:BTXY100}}
  \subfloat[\BTXY~\WZHtoo]{
    \includegraphics[width=0.24\textwidth]{BTYX/gen1_hwz121/dominantPools0}\label{fig:1gen1:BTXY211}}
  \vspace*{2ex}
  \begin{tabular}{llll}
    \swatch{snow}~ATLAS $t\bar{t}$ hadr & \swatch{yellow!60!white} ATLAS $\gamma\gamma$ \& $\gamma\!+\!X$ \\
    \swatch{cornflowerblue}~ATLAS $WW$ & \swatch{navy}~ATLAS $\mu$+\MET{}+jet & \swatch{cadetblue}~ATLAS $e$+\MET{}+jet\\
    \swatch{orangered}~ATLAS $ee$+jet & \swatch{darkorange}~ATLAS $\mu\mu$+jet & \swatch{orange!60}~ATLAS $\ell\ell$+jet\\
    \swatch{magenta} ATLAS~4$\ell$ & \swatch{silver}~ATLAS jets & \swatch{dimgrey}~CMS jets\\
  \end{tabular}
  \vspace*{2ex}
  \caption{ Sensitivity of LHC measurements to VLQ singlet production in the $\kappa$ vs
    VLQ mass plane, where $\kappa$ is the coupling to first-generation SM quarks.
    All VLQ masses are set to be degenerate.
    The multiplets are given as rows:
    \protect\subref{fig:1gen1:B001}--\protect\subref{fig:1gen1:B211} $B$ singlet,
    \protect\subref{fig:1gen1:T001}--\protect\subref{fig:1gen1:T211} $T$ singlet, and for comparison to the main text, the
    \protect\subref{fig:1gen1:BTXY001}--\protect\subref{fig:1gen1:BTXY211}($B,T,X,Y$) quadruplet.
    The VLQ branching fractions to \WZH are arranged in columns of \WZHzzo, \WZHzoz, \WZHozz, and \WZHtoo from left to right.
  }%
  \label{fig:1gen_singlets}
\end{figure}

\clearpage

\subsection{Second generation}

\begin{figure}[h]
  \centering
  \subfloat[\BTX~\WZHzzo]{
    \includegraphics[width=0.24\textwidth]{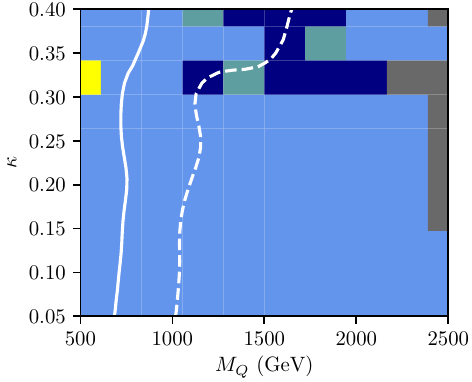}\label{fig:3gen2:BTX001}}
  \subfloat[\BTX~\WZHzoz]{
    \includegraphics[width=0.24\textwidth]{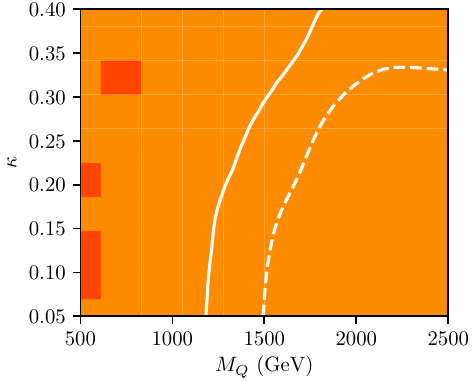}\label{fig:3gen2:BTX010}}
  \subfloat[\BTX~\WZHozz]{
    \includegraphics[width=0.24\textwidth]{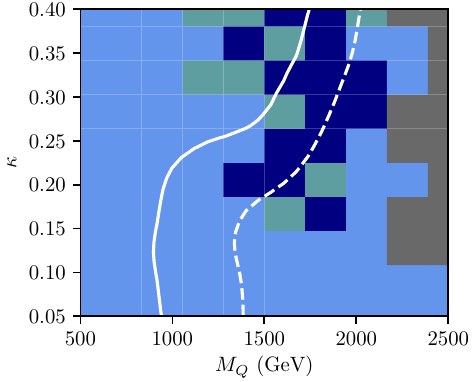}\label{fig:3gen2:BTX100}}
  \subfloat[\BTX~\WZHtoo]{
    \includegraphics[width=0.24\textwidth]{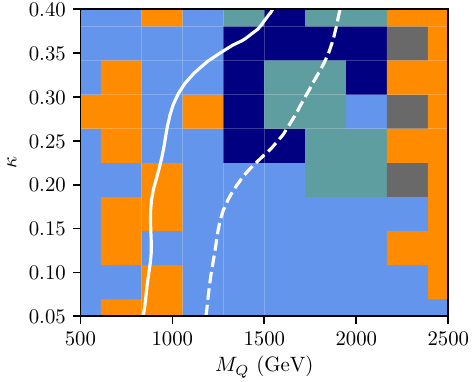}\label{fig:3gen2:BTX211}}\\
  \subfloat[\BTY~\WZHzzo]{
    \includegraphics[width=0.24\textwidth]{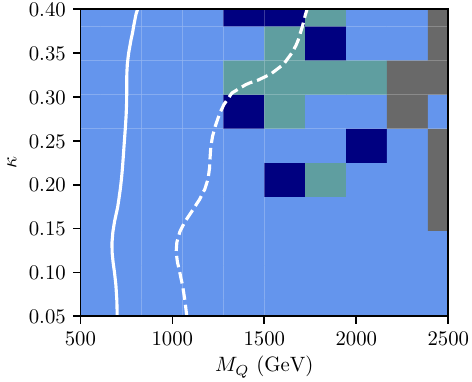}\label{fig:3gen2:BTY001}}
  \subfloat[\BTY~\WZHzoz]{
    \includegraphics[width=0.24\textwidth]{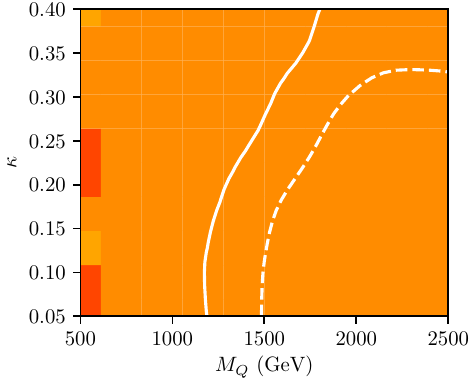}\label{fig:3gen2:BTY010}}
  \subfloat[\BTY~\WZHozz]{
    \includegraphics[width=0.24\textwidth]{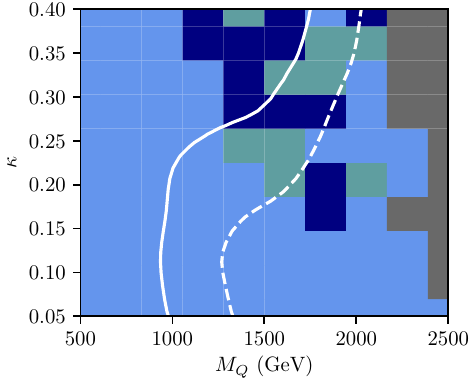}\label{fig:3gen2:BTY100}}
  \subfloat[\BTY~\WZHtoo]{
    \includegraphics[width=0.24\textwidth]{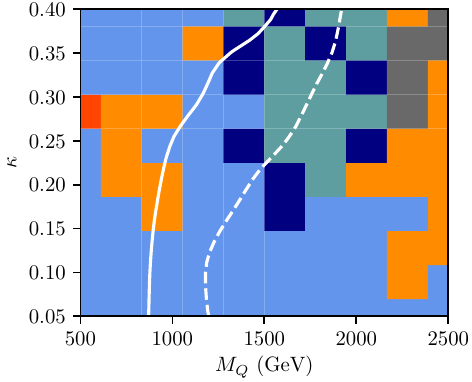}\label{fig:3gen2:BTY211}}\\[1em]
  \hrule
  \subfloat[\BTXY~\WZHzzo]{
    \includegraphics[width=0.24\textwidth]{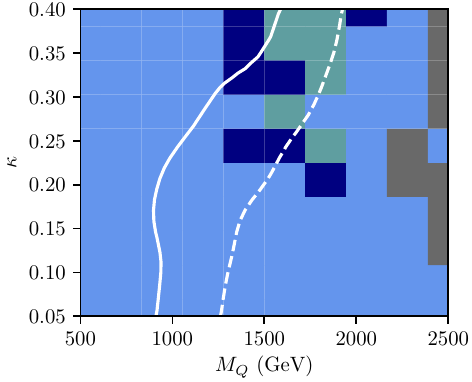}\label{fig:3gen2:BTXY001}}
  \subfloat[\BTXY~\WZHzoz]{
    \includegraphics[width=0.24\textwidth]{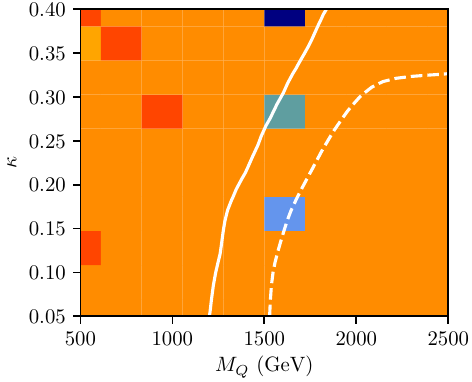}\label{fig:3gen2:BTXY010}}
  \subfloat[\BTXY~\WZHozz]{
    \includegraphics[width=0.24\textwidth]{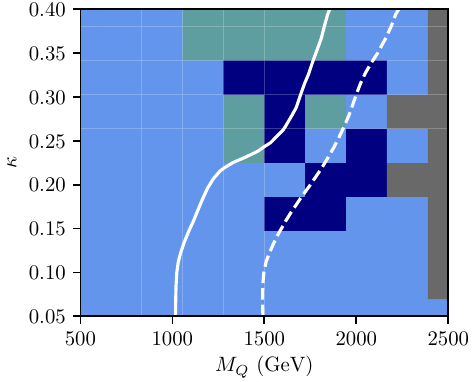}\label{fig:3gen2:BTXY100}}
  \subfloat[\BTXY~\WZHtoo]{
    \includegraphics[width=0.24\textwidth]{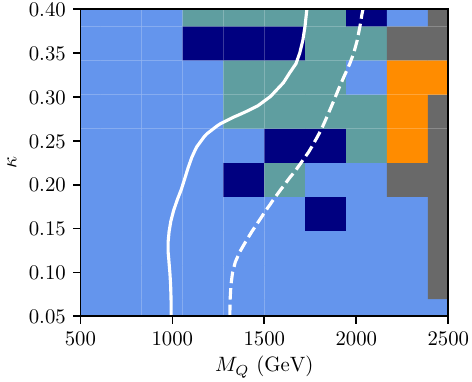}\label{fig:3gen2:BTXY211}}
  \vspace*{2ex}
  \begin{tabular}{llll}
    \swatch{cornflowerblue}~ATLAS $WW$ & \swatch{navy}~ATLAS $\mu$+\MET{}+jet & \swatch{cadetblue}~ATLAS $e$+\MET{}+jet\\
    \swatch{orangered}~ATLAS $ee$+jet & \swatch{darkorange}~ATLAS $\mu\mu$+jet & \swatch{orange!60}~ATLAS $\ell\ell$+jet\\
                                       \swatch{silver}~ATLAS jets & \swatch{dimgrey}~CMS jets
                                                                  & \swatch{yellow!60!white} ATLAS $\gamma\gamma$ \& $\gamma\!+\!X$
  \end{tabular}
  \vspace*{2ex}
  \caption{ Sensitivity of LHC measurements to VLQ triplet production in the $\kappa$ vs
    VLQ mass plane, where $\kappa$ is the coupling to second-generation SM quarks.
    All VLQ masses are set to be degenerate.
    The multiplets are given as rows:
    \protect\subref{fig:3gen2:BTX001}--\protect\subref{fig:3gen2:BTX211}($B,T,X$) triplet,
    \protect\subref{fig:3gen2:BTY001}--\protect\subref{fig:3gen2:BTY211}($B,T,Y$) triplet, and for comparison to the main text, the
    \protect\subref{fig:3gen2:BTXY001}--\protect\subref{fig:3gen2:BTXY211}($B,T,X,Y$) quadruplet.
    The VLQ branching fractions to \WZH are arranged in columns of \WZHzzo, \WZHzoz, \WZHozz, and \WZHtoo from left to right.
  }%
  \label{fig:2gen_triplets}
\end{figure}

\begin{figure}[h]
  \centering
  \subfloat[\BT~\WZHzzo]{
    \includegraphics[width=0.24\textwidth]{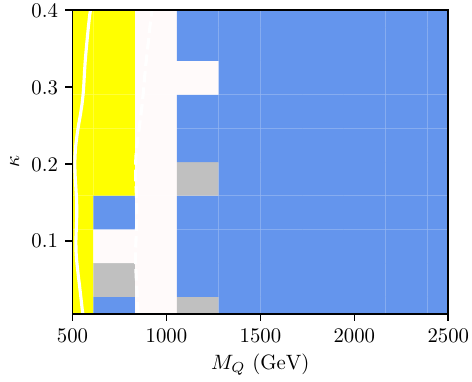}\label{fig:2gen2:BT001}}
  \subfloat[\BT~\WZHzoz]{
    \includegraphics[width=0.24\textwidth]{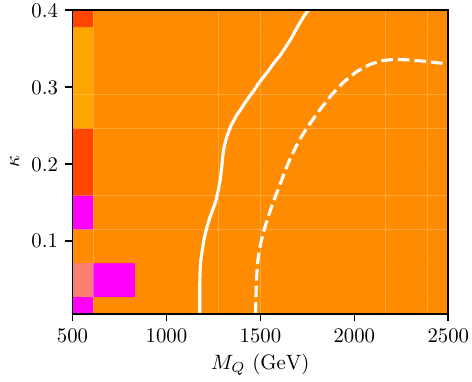}\label{fig:2gen2:BT010}}
  \subfloat[\BT~\WZHozz]{
    \includegraphics[width=0.24\textwidth]{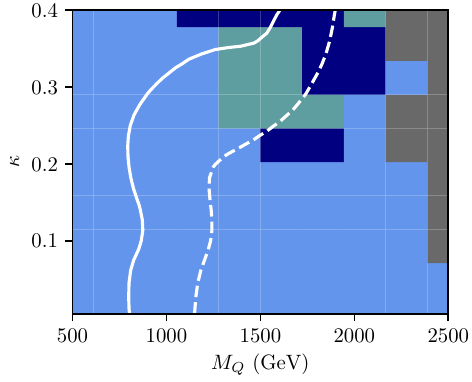}\label{fig:2gen2:BT100}}
  \subfloat[\BT~\WZHzoo]{
    \includegraphics[width=0.24\textwidth]{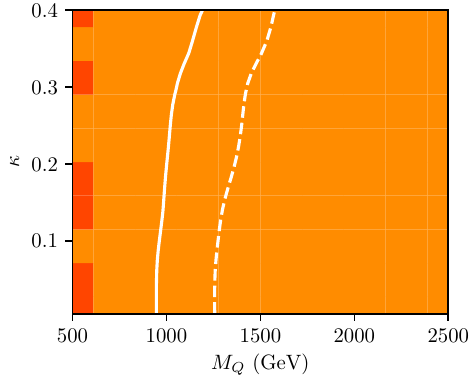}\label{fig:2gen2:BT011}}\\
  \subfloat[\XT~\WZHzzo]{
    \includegraphics[width=0.24\textwidth]{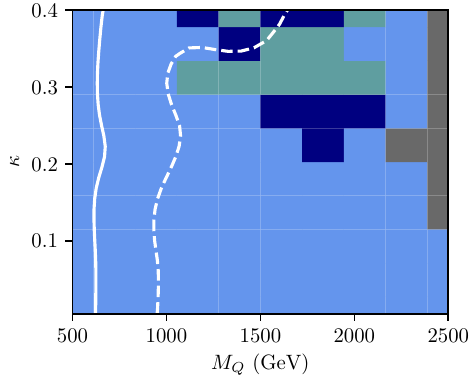}\label{fig:2gen2:XT001}}
  \subfloat[\XT~\WZHzoz]{
    \includegraphics[width=0.24\textwidth]{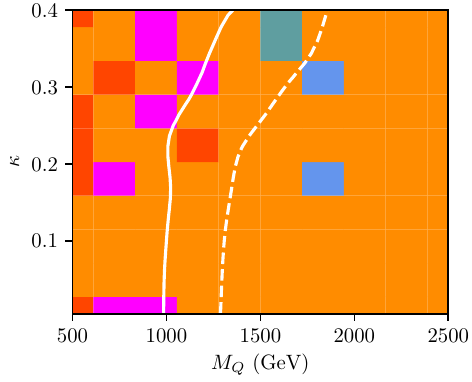}\label{fig:2gen2:XT010}}
  \subfloat[\XT~\WZHozz]{
    \includegraphics[width=0.24\textwidth]{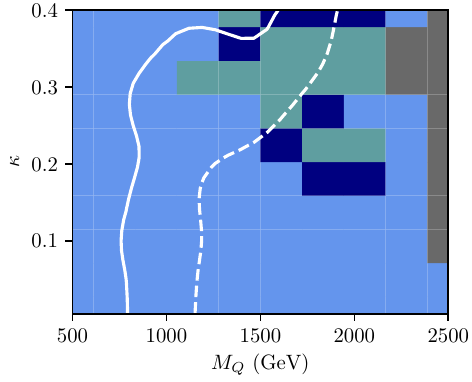}\label{fig:2gen2:XT100}}
  \subfloat[\XT~\WZHzoo]{
    \includegraphics[width=0.24\textwidth]{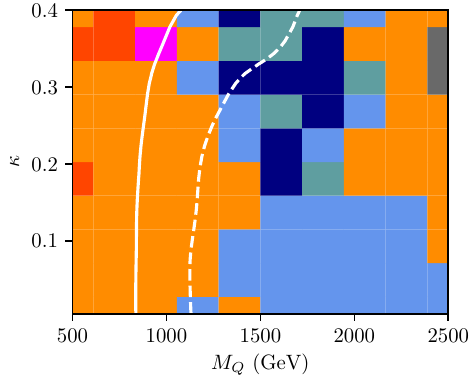}\label{fig:2gen2:XT011}}\\
  \subfloat[\BY~\WZHzzo]{
    \includegraphics[width=0.24\textwidth]{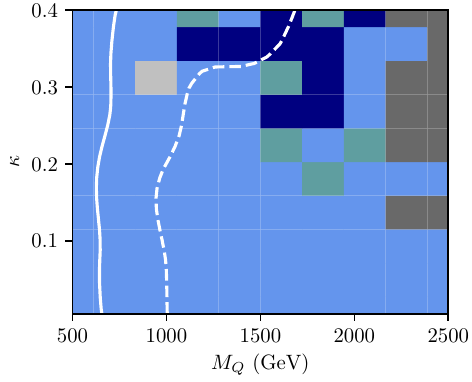}\label{fig:2gen2:BY001}}
  \subfloat[\BY~\WZHzoz]{
    \includegraphics[width=0.24\textwidth]{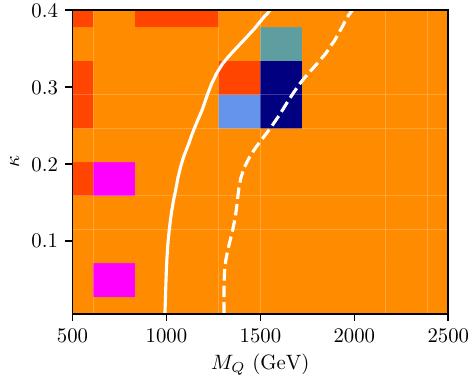}\label{fig:2gen2:BY010}}
  \subfloat[\BY~\WZHozz]{
    \includegraphics[width=0.24\textwidth]{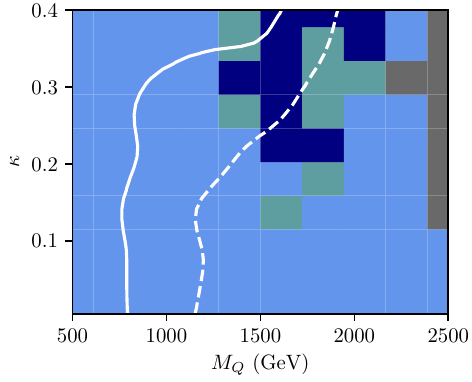}\label{fig:2gen2:BY100}}
  \subfloat[\BY~\WZHzoo]{
    \includegraphics[width=0.24\textwidth]{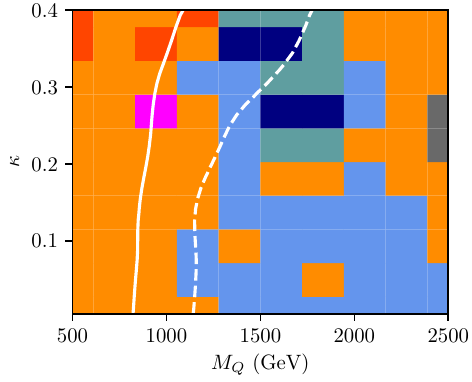}\label{fig:2gen2:BY011}}\\[1em]
  \hrule
  \subfloat[\BTXY~\WZHzzo]{
    \includegraphics[width=0.24\textwidth]{BTYX/gen2_hwz100/dominantPools0}\label{fig:2gen2:BTXY001}}
  \subfloat[\BTXY~\WZHzoz]{
    \includegraphics[width=0.24\textwidth]{BTYX/gen2_hwz001/dominantPools0}\label{fig:2gen2:BTXY010}}
  \subfloat[\BTXY~\WZHozz]{
    \includegraphics[width=0.24\textwidth]{BTYX/gen2_hwz010/dominantPools0}\label{fig:2gen2:BTXY100}}
  \subfloat[\BTXY~\WZHzoo]{
    \includegraphics[width=0.24\textwidth]{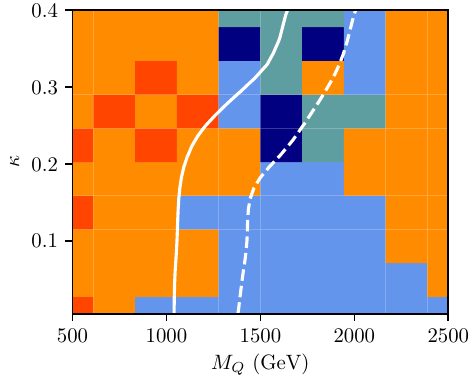}\label{fig:2gen2:BTXY011}}
  \vspace*{2ex}
  \begin{tabular}{llll}
    \swatch{snow}~ATLAS $t\bar{t}$ hadr & \swatch{yellow!60!white} ATLAS $\gamma\gamma$ \& $\gamma\!+\!X$ \\
    \swatch{cornflowerblue}~ATLAS $WW$ & \swatch{navy}~ATLAS $\mu$+\MET{}+jet & \swatch{cadetblue}~ATLAS $e$+\MET{}+jet\\
    \swatch{orangered}~ATLAS $ee$+jet & \swatch{darkorange}~ATLAS $\mu\mu$+jet & \swatch{orange!60}~ATLAS $\ell\ell$+jet\\
    \swatch{magenta} ATLAS~4$\ell$ & \swatch{silver}~ATLAS jets & \swatch{dimgrey}~CMS jets\\
  \end{tabular}
  \vspace*{2ex}
  \caption{ Sensitivity of LHC measurements to VLQ doublet production in the $\kappa$ vs
    VLQ mass plane, where $\kappa$ is the coupling to second-generation SM quarks.
    All VLQ masses are set to be degenerate.
    The multiplets are given as rows:
    \protect\subref{fig:2gen2:BT001}--\protect\subref{fig:2gen2:BT011}($B,T$) doublet,
    \protect\subref{fig:2gen2:XT001}--\protect\subref{fig:2gen2:XT011}($X,T$) doublet,
    \protect\subref{fig:2gen2:BY001}--\protect\subref{fig:2gen2:BY011}($B,Y$) doublet, and for comparison to the main text, the
    \protect\subref{fig:2gen2:BTXY001}--\protect\subref{fig:2gen2:BTXY100}($B,T,X,Y$) quadruplet.
    The VLQ branching fractions to \WZH are arranged in columns of \WZHzzo, \WZHzoz, \WZHozz, and \WZHzoo from left to right.
    The \WZHzoo case is considered for doublets instead of \WZHtoo, as motivated by Ref.~\cite{Buchkremer:2013bha}. 
  }%
  \label{fig:2gen_doublets}
\end{figure}

\begin{figure}[h]
  \centering
   \subfloat[$B$~\WZHzzo]{
    \includegraphics[width=0.24\textwidth]{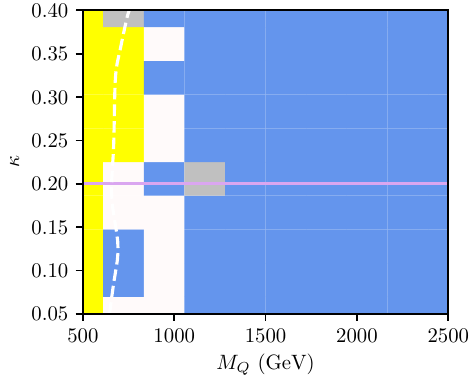}\label{fig:1gen2:B001}}
  \subfloat[$B$~\WZHzoz]{
    \includegraphics[width=0.24\textwidth]{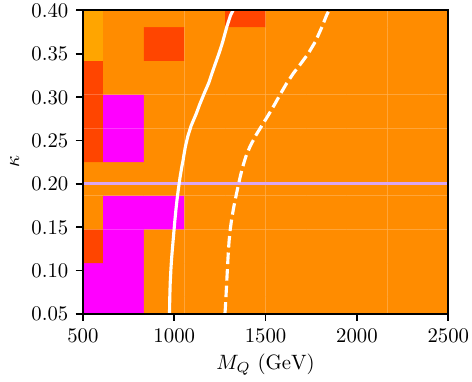}\label{fig:1gen2:B010}}
  \subfloat[$B$~\WZHozz]{
    \includegraphics[width=0.24\textwidth]{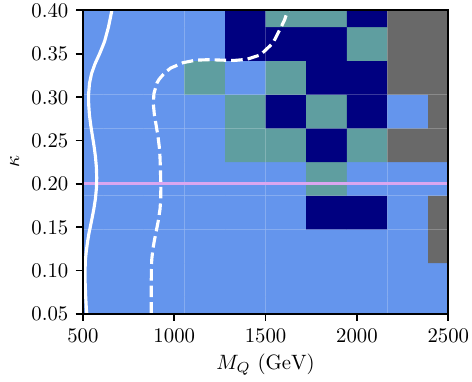}\label{fig:1gen2:B100}}
  \subfloat[$B$~\WZHtoo]{
    \includegraphics[width=0.24\textwidth]{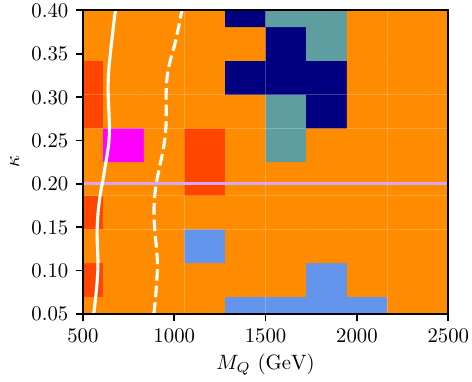}\label{fig:1gen2:B211}}\\
  \subfloat[$T$~\WZHzzo]{
    \includegraphics[width=0.24\textwidth]{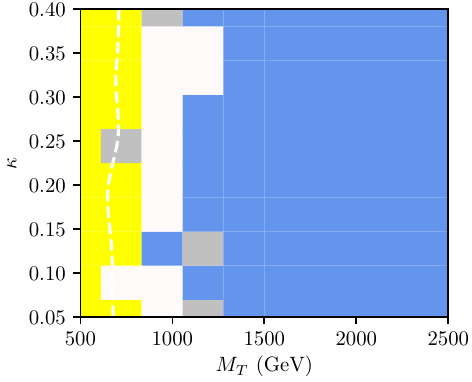}\label{fig:1gen2:T001}}
  \subfloat[$T$~\WZHzoz]{
    \includegraphics[width=0.24\textwidth]{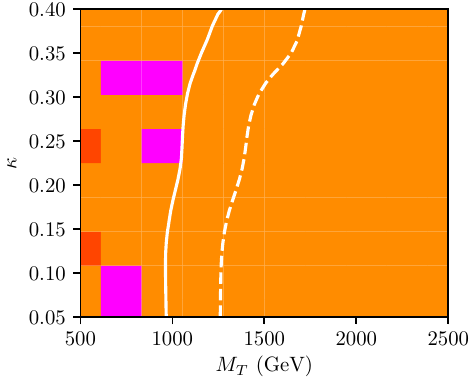}\label{fig:1gen2:T010}}
  \subfloat[$T$~\WZHozz]{
    \includegraphics[width=0.24\textwidth]{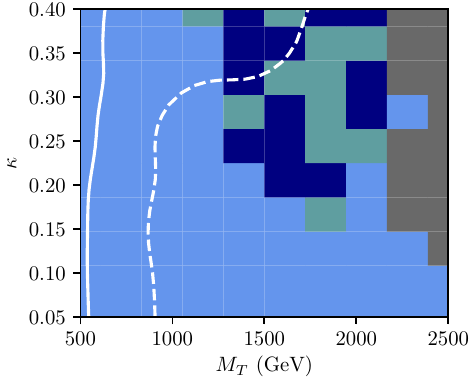}\label{fig:1gen2:T100}}
  \subfloat[$T$~\WZHtoo]{
    \includegraphics[width=0.24\textwidth]{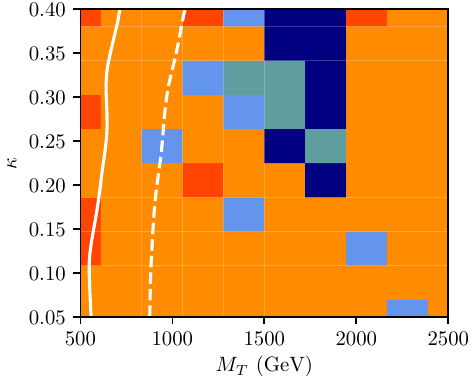}\label{fig:1gen2:T211}}\\[1em]
  \hrule
  \subfloat[\BTXY~\WZHzzo]{
    \includegraphics[width=0.24\textwidth]{BTYX/gen2_hwz100/dominantPools0}\label{fig:1gen2:BTXY001}}
  \subfloat[\BTXY~\WZHzoz]{
    \includegraphics[width=0.24\textwidth]{BTYX/gen2_hwz001/dominantPools0}\label{fig:1gen2:BTXY010}}
  \subfloat[\BTXY~\WZHozz]{
    \includegraphics[width=0.24\textwidth]{BTYX/gen2_hwz010/dominantPools0}\label{fig:1gen2:BTXY100}}
  \subfloat[\BTXY~\WZHtoo]{
    \includegraphics[width=0.24\textwidth]{BTYX/gen2_hwz121/dominantPools0}\label{fig:1gen2:BTXY211}}
  \vspace*{2ex}
  \begin{tabular}{llll}
    \swatch{snow}~ATLAS $t\bar{t}$ hadr & \swatch{yellow!60!white} ATLAS $\gamma\gamma$ \& $\gamma\!+\!X$ \\
    \swatch{cornflowerblue}~ATLAS $WW$ & \swatch{navy}~ATLAS $\mu$+\MET{}+jet & \swatch{cadetblue}~ATLAS $e$+\MET{}+jet\\
    \swatch{orangered}~ATLAS $ee$+jet & \swatch{darkorange}~ATLAS $\mu\mu$+jet & \swatch{orange!60}~ATLAS $\ell\ell$+jet\\
    \swatch{magenta} ATLAS~4$\ell$ & \swatch{silver}~ATLAS jets & \swatch{dimgrey}~CMS jets\\
  \end{tabular}
  \vspace*{2ex}
  \caption{ Sensitivity of LHC measurements to VLQ singlet production in the $\kappa$ vs
    VLQ mass plane, where $\kappa$ is the coupling to second-generation SM quarks.
    All VLQ masses are set to be degenerate.
    The multiplets are given as rows:
    \protect\subref{fig:1gen2:B001}--\protect\subref{fig:1gen2:B211} $B$ singlet,
    \protect\subref{fig:1gen2:T001}--\protect\subref{fig:1gen2:T211} $T$ singlet, and for comparison to the main text, the
    \protect\subref{fig:1gen2:BTXY001}--\protect\subref{fig:1gen2:BTXY211}($B,T,X,Y$) quadruplet.
    The VLQ branching fractions to \WZH are arranged in columns of \WZHzzo, \WZHzoz, \WZHozz, and \WZHtoo from left to right.
  }%
  \label{fig:2gen_singlets}
\end{figure}

\clearpage

\subsection{Third generation}

\begin{figure}[h]
  \centering
  \subfloat[\BTX~\WZHzzo]{
    \includegraphics[width=0.24\textwidth]{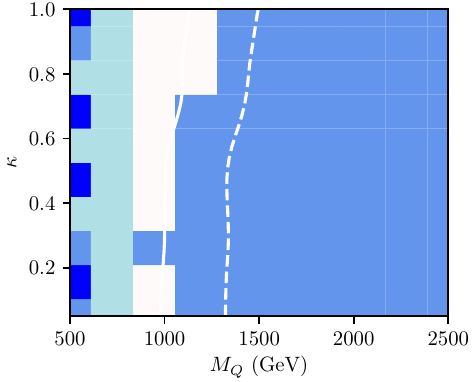}\label{fig:3gen3:BTX001}}
  \subfloat[\BTX~\WZHzoz]{
    \includegraphics[width=0.24\textwidth]{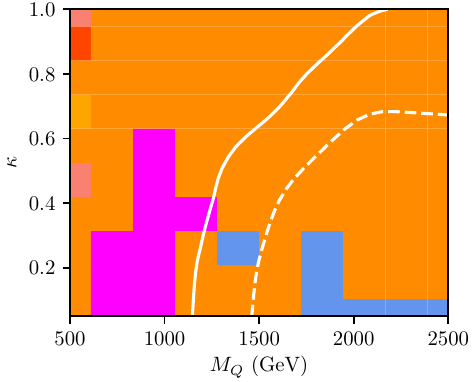}\label{fig:3gen3:BTX010}}
  \subfloat[\BTX~\WZHozz]{
    \includegraphics[width=0.24\textwidth]{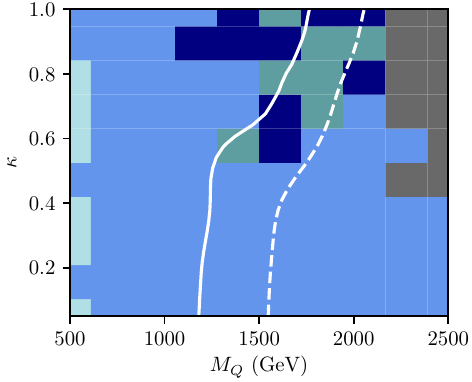}\label{fig:3gen3:BTX100}}
  \subfloat[\BTX~\WZHtoo]{
    \includegraphics[width=0.24\textwidth]{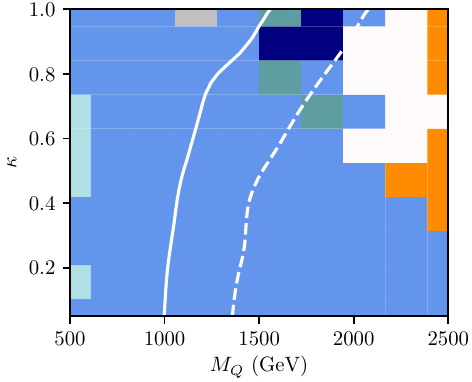}\label{fig:3gen3:BTX211}}\\
  \subfloat[\BTY~\WZHzzo]{
    \includegraphics[width=0.24\textwidth]{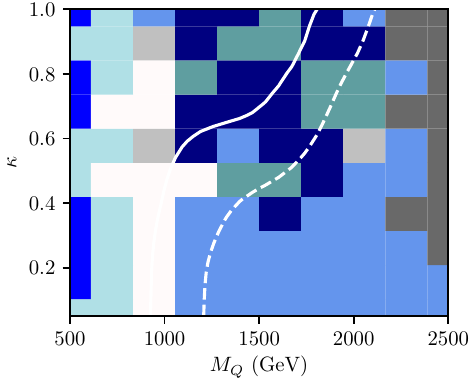}\label{fig:3gen3:BTY001}}
  \subfloat[\BTY~\WZHzoz]{
    \includegraphics[width=0.24\textwidth]{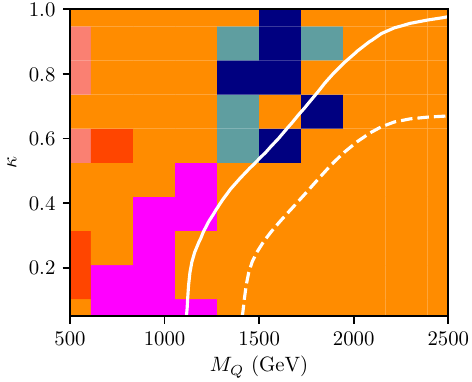}\label{fig:3gen3:BTY010}}
  \subfloat[\BTY~\WZHozz]{
    \includegraphics[width=0.24\textwidth]{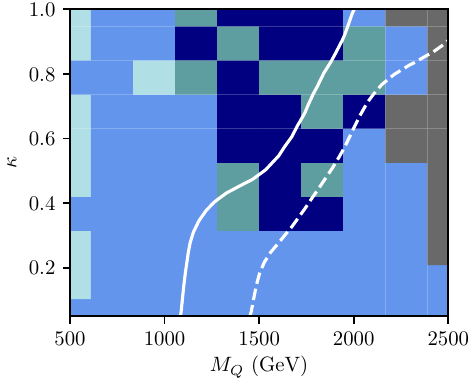}\label{fig:3gen3:BTY100}}
  \subfloat[\BTY~\WZHtoo]{
    \includegraphics[width=0.24\textwidth]{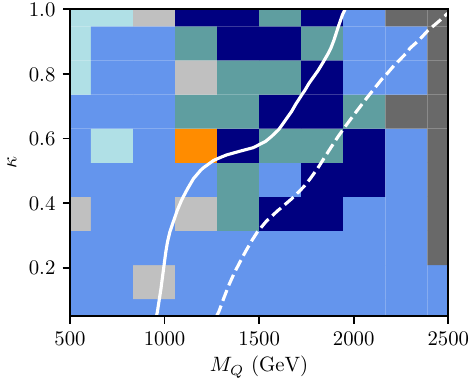}\label{fig:3gen3:BTY211}}\\[1em]
  \hrule
  \subfloat[\BTXY~\WZHzzo]{
    \includegraphics[width=0.24\textwidth]{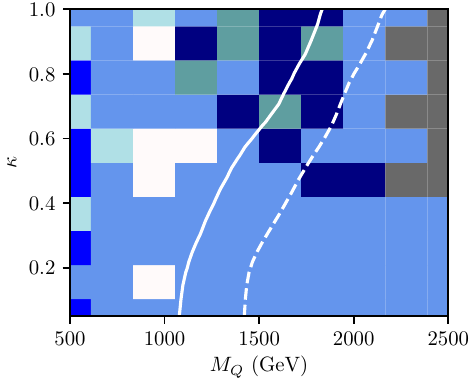}\label{fig:3gen3:BTXY001}}
  \subfloat[\BTXY~\WZHzoz]{
    \includegraphics[width=0.24\textwidth]{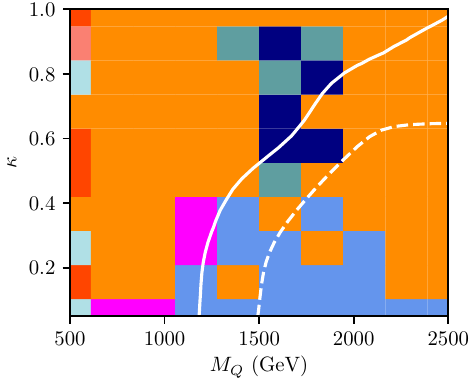}\label{fig:3gen3:BTXY010}}
  \subfloat[\BTXY~\WZHozz]{
    \includegraphics[width=0.24\textwidth]{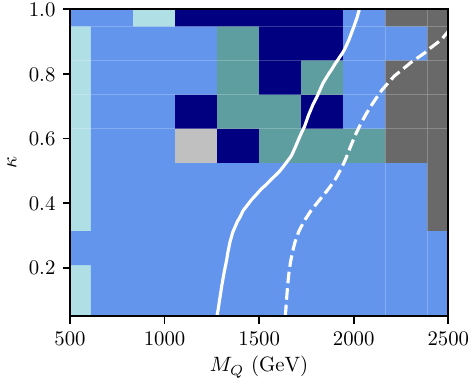}\label{fig:3gen3:BTXY100}}
  \subfloat[\BTXY~\WZHtoo]{
    \includegraphics[width=0.24\textwidth]{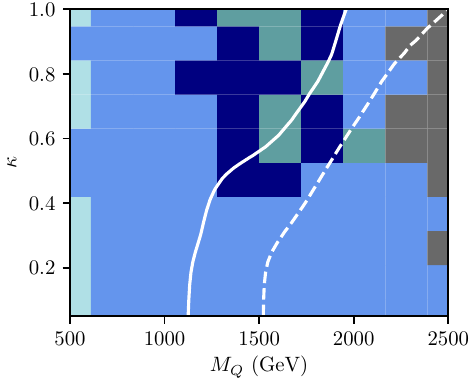}\label{fig:3gen3:BTXY211}}
  \vspace*{2ex}
  \begin{tabular}{llll}
    \swatch{blue}~ATLAS $\ell$+MET+jet & \swatch{powderblue}~ATLAS $\ell$+MET+jet & \swatch{snow}~ATLAS $t\bar{t}$ hadr\\
    \swatch{cornflowerblue}~ATLAS $WW$ & \swatch{navy}~ATLAS $\mu$+\MET{}+jet & \swatch{cadetblue}~ATLAS $e$+\MET{}+jet\\
    \swatch{orangered}~ATLAS $ee$+jet & \swatch{darkorange}~ATLAS $\mu\mu$+jet & \swatch{orange!60}~ATLAS $\ell\ell$+jet\\
    \swatch{magenta} ATLAS~4$\ell$ & \swatch{silver}~ATLAS jets & \swatch{dimgrey}~CMS jets
  \end{tabular}
  \vspace*{2ex}
  \caption{ Sensitivity of LHC measurements to VLQ triplet production in the $\kappa$ vs
    VLQ mass plane, where $\kappa$ is the coupling to third-generation SM quarks.
    All VLQ masses are set to be degenerate.
    The multiplets are given as rows:
    \protect\subref{fig:3gen3:BTX001}--\protect\subref{fig:3gen3:BTX211}($B,T,X$) triplet,
    \protect\subref{fig:3gen3:BTY001}--\protect\subref{fig:3gen3:BTY211}($B,T,Y$) triplet, and for comparison to the main text, the
    \protect\subref{fig:3gen3:BTXY001}--\protect\subref{fig:3gen3:BTXY211}($B,T,X,Y$) quadruplet.
    The VLQ branching fractions to \WZH are arranged in columns of \WZHzzo, \WZHzoz, \WZHozz, and \WZHtoo from left to right.
  }%
  \label{fig:3gen_triplets}
\end{figure}

\begin{figure}[h]
  \centering
  \subfloat[\BT~\WZHzzo]{
    \includegraphics[width=0.24\textwidth]{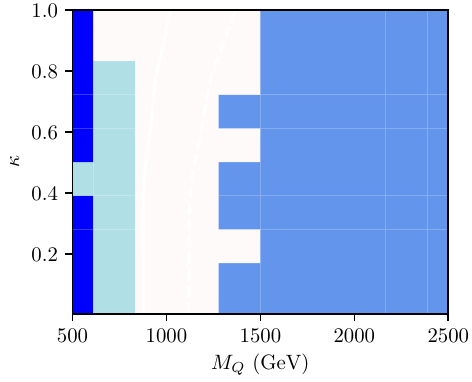}\label{fig:2gen3:BT001}}
  \subfloat[\BT~\WZHzoz]{
    \includegraphics[width=0.24\textwidth]{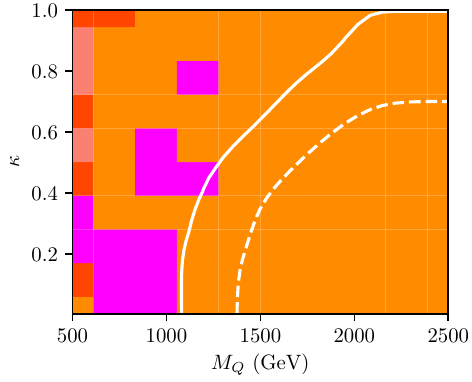}\label{fig:2gen3:BT010}}
  \subfloat[\BT~\WZHozz]{
    \includegraphics[width=0.24\textwidth]{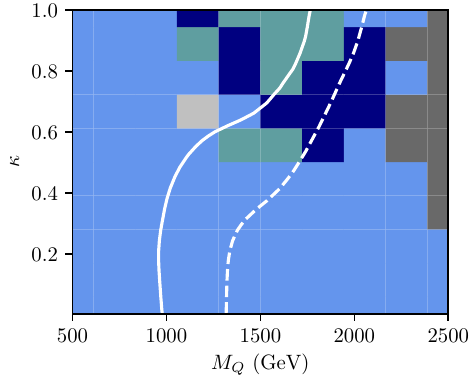}\label{fig:2gen3:BT100}}
  \subfloat[\BT~\WZHzoo]{
    \includegraphics[width=0.24\textwidth]{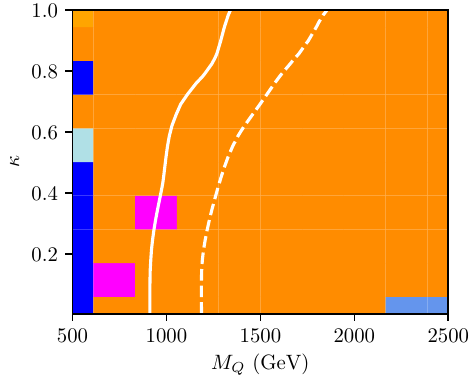}\label{fig:2gen3:BT011}}\\
  \subfloat[\XT~\WZHzzo]{
    \includegraphics[width=0.24\textwidth]{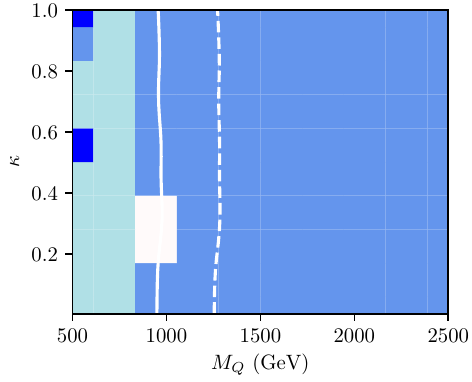}\label{fig:2gen3:XT001}}
  \subfloat[\XT~\WZHzoz]{
    \includegraphics[width=0.24\textwidth]{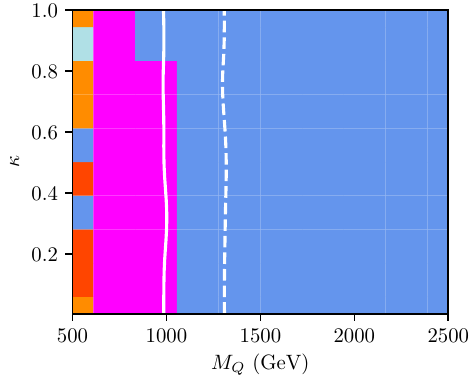}\label{fig:2gen3:XT010}}
  \subfloat[\XT~\WZHozz]{
    \includegraphics[width=0.24\textwidth]{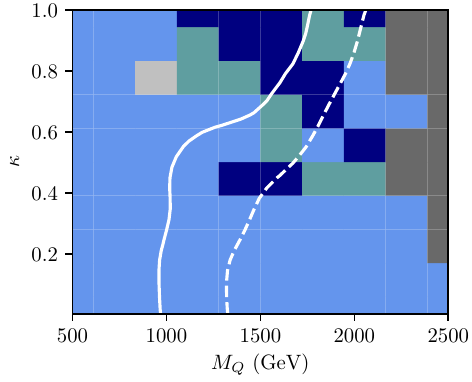}\label{fig:2gen3:XT100}}
  \subfloat[\XT~\WZHzoo]{
    \includegraphics[width=0.24\textwidth]{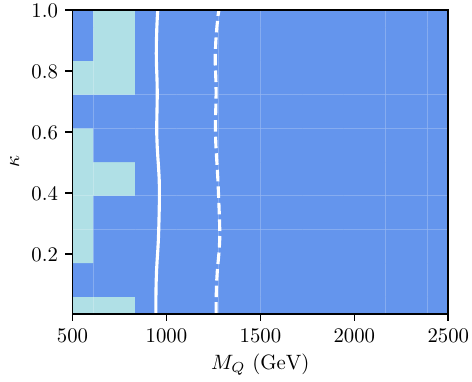}\label{fig:2gen3:XT011}}\\
  \subfloat[\BY~\WZHzzo]{
    \includegraphics[width=0.24\textwidth]{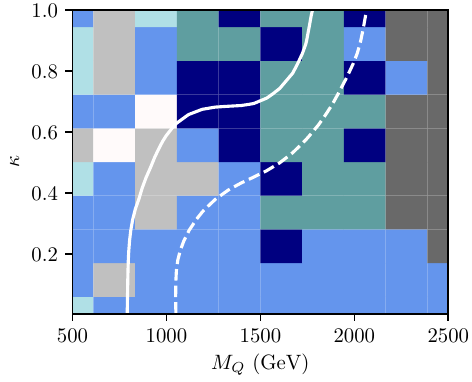}\label{fig:2gen3:BY001}}
  \subfloat[\BY~\WZHzoz]{
    \includegraphics[width=0.24\textwidth]{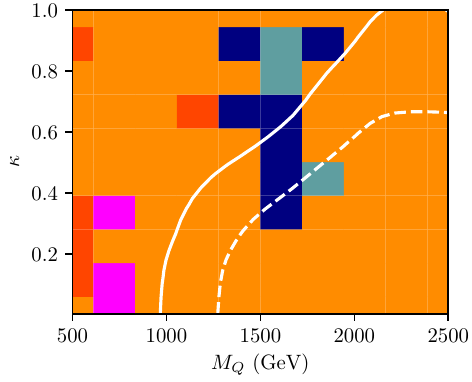}\label{fig:2gen3:BY010}}
  \subfloat[\BY~\WZHozz]{
    \includegraphics[width=0.24\textwidth]{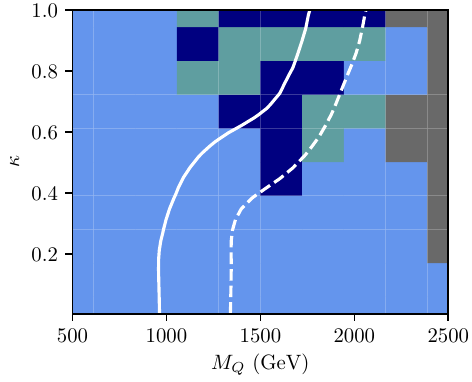}\label{fig:2gen3:BY100}}
  \subfloat[\BY~\WZHzoo]{
    \includegraphics[width=0.24\textwidth]{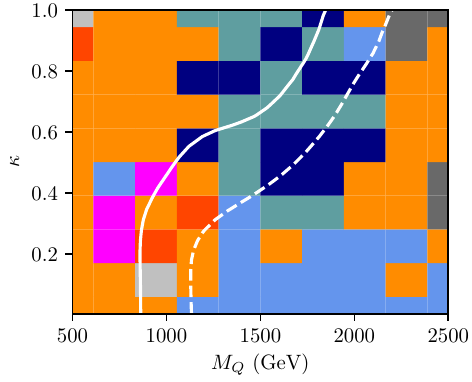}\label{fig:2gen3:BY011}}\\[1em]
  \hrule
  \subfloat[\BTXY~\WZHzzo]{
    \includegraphics[width=0.24\textwidth]{BTYX/gen3_hwz100/dominantPools0}\label{fig:2gen3:BTXY001}}
  \subfloat[\BTXY~\WZHzoz]{
    \includegraphics[width=0.24\textwidth]{BTYX/gen3_hwz001/dominantPools0}\label{fig:2gen3:BTXY010}}
  \subfloat[\BTXY~\WZHozz]{
    \includegraphics[width=0.24\textwidth]{BTYX/gen3_hwz010/dominantPools0}\label{fig:2gen3:BTXY100}}
  \subfloat[\BTXY~\WZHzoo]{
    \includegraphics[width=0.24\textwidth]{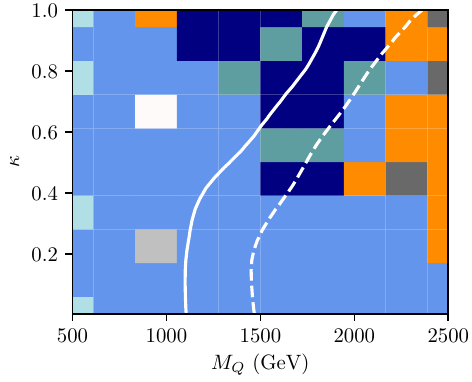}\label{fig:2gen3:BTXY011}}
  \vspace*{2ex}
  \begin{tabular}{llll}
    \swatch{blue}~ATLAS $\ell$+MET+jet & \swatch{powderblue}~ATLAS $\ell$+MET+jet & \swatch{snow}~ATLAS $t\bar{t}$ hadr\\
    \swatch{cornflowerblue}~ATLAS $WW$ & \swatch{navy}~ATLAS $\mu$+\MET{}+jet & \swatch{cadetblue}~ATLAS $e$+\MET{}+jet\\
    \swatch{orangered}~ATLAS $ee$+jet & \swatch{darkorange}~ATLAS $\mu\mu$+jet & \swatch{orange!60}~ATLAS $\ell\ell$+jet\\
    \swatch{magenta} ATLAS~4$\ell$ & \swatch{silver}~ATLAS jets & \swatch{dimgrey}~CMS jets
  \end{tabular}
  \vspace*{2ex}
  \caption{ Sensitivity of LHC measurements to VLQ doublet production in the $\kappa$ vs
    VLQ mass plane, where $\kappa$ is the coupling to third-generation SM quarks.
    All VLQ masses are set to be degenerate.
    The multiplets are given as rows:
    \protect\subref{fig:2gen3:BT001}--\protect\subref{fig:2gen3:BT011}($B,T$) doublet,
    \protect\subref{fig:2gen3:XT001}--\protect\subref{fig:2gen3:XT011}($X,T$) doublet,
    \protect\subref{fig:2gen3:BY001}--\protect\subref{fig:2gen3:BY011}($B,Y$) doublet, and for comparison to the main text, the
    \protect\subref{fig:2gen3:BTXY001}--\protect\subref{fig:2gen3:BTXY100}($B,T,X,Y$) quadruplet.
    The VLQ branching fractions to \WZH are arranged in columns of \WZHzzo, \WZHzoz, \WZHozz, and \WZHzoo from left to right.
    The \WZHzoo case is considered for doublets instead of \WZHtoo, as motivated by Ref.~\cite{Buchkremer:2013bha}. 
  }%
  \label{fig:3gen_doublets}
\end{figure}

\begin{figure}[h]
  \centering
  \subfloat[$B$~\WZHzzo]{
    \includegraphics[width=0.24\textwidth]{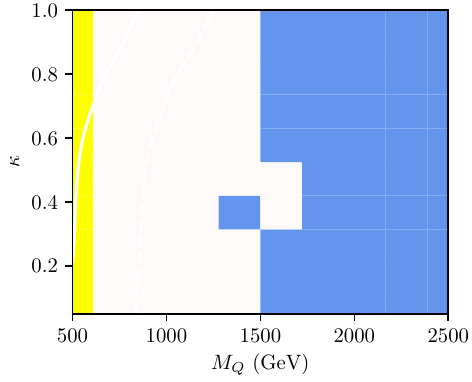}\label{fig:1gen3:B001}}
  \subfloat[$B$~\WZHzoz]{
    \includegraphics[width=0.24\textwidth]{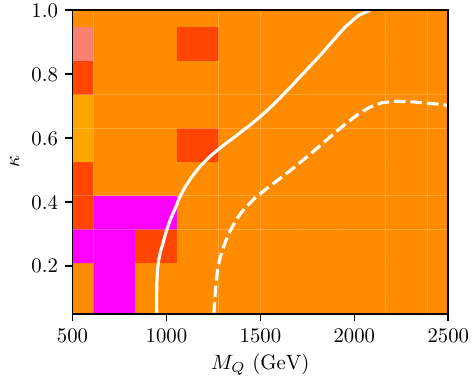}\label{fig:1gen3:B010}}
  \subfloat[$B$~\WZHozz]{
    \includegraphics[width=0.24\textwidth]{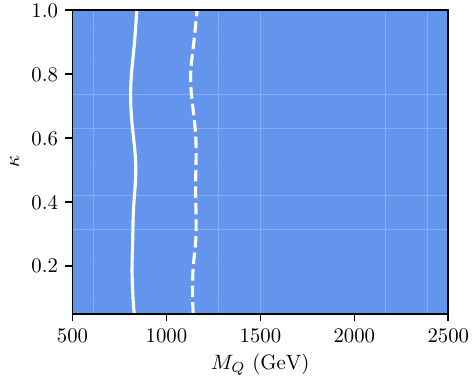}\label{fig:1gen3:B100}}
  \subfloat[$B$~\WZHtoo]{
    \includegraphics[width=0.24\textwidth]{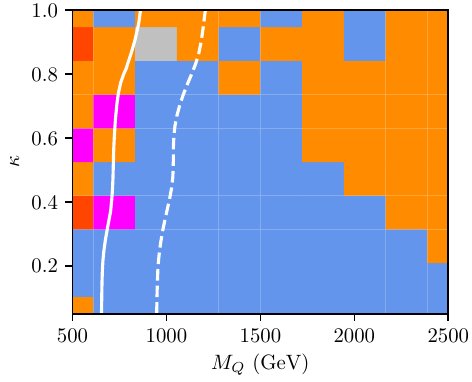}\label{fig:1gen3:B211}}\\[1ex]
  \subfloat[$T$~\WZHzzo]{
    \includegraphics[width=0.24\textwidth]{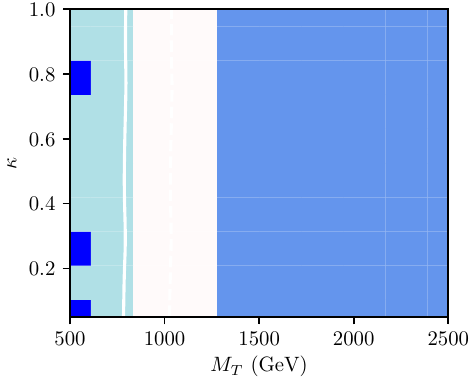}\label{fig:1gen3:T001}}
  \subfloat[$T$~\WZHzoz]{
    \includegraphics[width=0.24\textwidth]{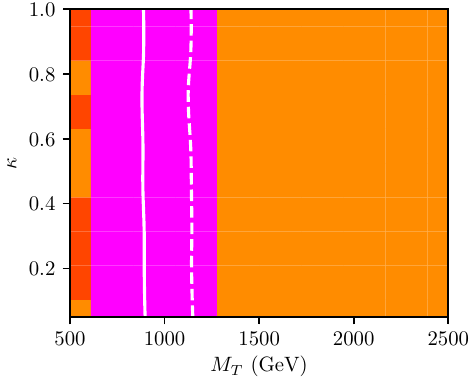}\label{fig:1gen3:T010}}
  \subfloat[$T$~\WZHozz]{
    \includegraphics[width=0.24\textwidth]{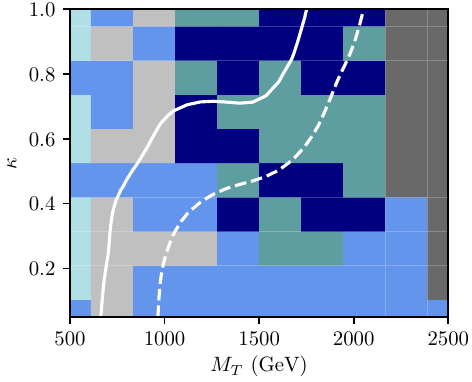}\label{fig:1gen3:T100}}
  \subfloat[$T$~\WZHtoo]{
    \includegraphics[width=0.24\textwidth]{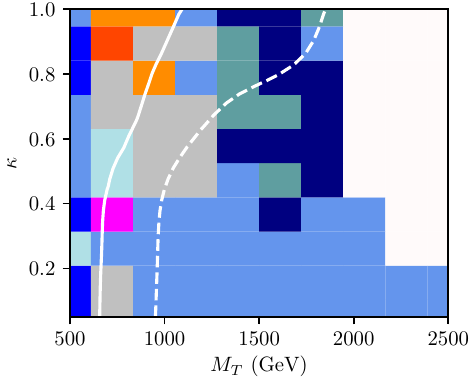}\label{fig:1gen3:T211}}\\[1em]
  \hrule
  \subfloat[\BTXY~\WZHzzo]{
    \includegraphics[width=0.24\textwidth]{BTYX/gen3_hwz100/dominantPools0}\label{fig:1gen3:BTXY001}}
  \subfloat[\BTXY~\WZHzoz]{
    \includegraphics[width=0.24\textwidth]{BTYX/gen3_hwz001/dominantPools0}\label{fig:1gen3:BTXY010}}
  \subfloat[\BTXY~\WZHozz]{
    \includegraphics[width=0.24\textwidth]{BTYX/gen3_hwz010/dominantPools0}\label{fig:1gen3:BTXY100}}
  \subfloat[\BTXY~\WZHtoo]{
    \includegraphics[width=0.24\textwidth]{BTYX/gen3_hwz121/dominantPools0}\label{fig:1gen3:BTXY211}}
  \vspace*{2ex}
  \begin{tabular}{llll}
    \swatch{blue}~ATLAS $\ell$+MET+jet & \swatch{powderblue}~ATLAS $\ell$+MET+jet & \swatch{snow}~ATLAS $t\bar{t}$ hadr\\
    \swatch{cornflowerblue}~ATLAS $WW$ & \swatch{navy}~ATLAS $\mu$+\MET{}+jet & \swatch{cadetblue}~ATLAS $e$+\MET{}+jet\\
    \swatch{orangered}~ATLAS $ee$+jet & \swatch{darkorange}~ATLAS $\mu\mu$+jet & \swatch{orange!60}~ATLAS $\ell\ell$+jet\\
    \swatch{magenta} ATLAS~4$\ell$ & \swatch{silver}~ATLAS jets & \swatch{dimgrey}~CMS jets
  \end{tabular}
  \vspace*{2ex}
  \caption{ Sensitivity of LHC measurements to VLQ singlet production in the $\kappa$ vs
    VLQ mass plane, where $\kappa$ is the coupling to third-generation SM quarks.
    All VLQ masses are set to be degenerate.
    The multiplets are given as rows:
    \protect\subref{fig:1gen3:B001}--\protect\subref{fig:1gen3:B211} $B$ singlet,
    \protect\subref{fig:1gen3:T001}--\protect\subref{fig:1gen3:T211} $T$ singlet, and for comparison to the main text, the
    \protect\subref{fig:1gen3:BTXY001}--\protect\subref{fig:1gen3:BTXY211}($B,T,X,Y$) quadruplet.
    The VLQ branching fractions to \WZH are arranged in columns of \WZHzzo, \WZHzoz, \WZHozz, and \WZHtoo from left to right.
  }%
  \label{fig:3gen_singlets}
\end{figure}

\clearpage

\section{\CLs maps in $\kappa$ vs $M_Q$}
\label{app:clsmaps}

\vspace*{17ex}

\begin{figure}[h]
  \centering
  \subfloat[]{
    \includegraphics[width=0.45\textwidth]{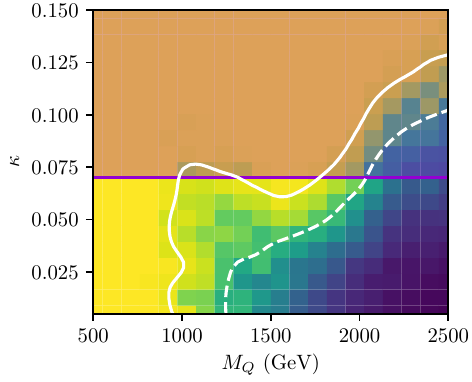}\label{fig:gen1:001}}
  \subfloat[]{
    \includegraphics[width=0.45\textwidth]{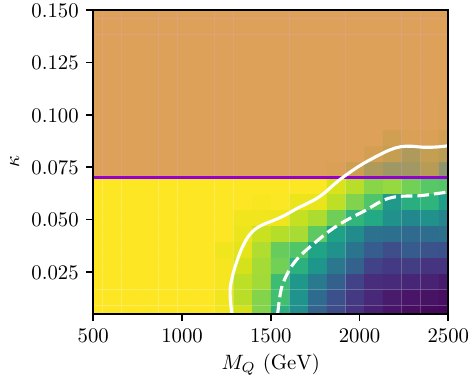}\label{fig:gen1:010}} \\
  \subfloat[]{
    \includegraphics[width=0.45\textwidth]{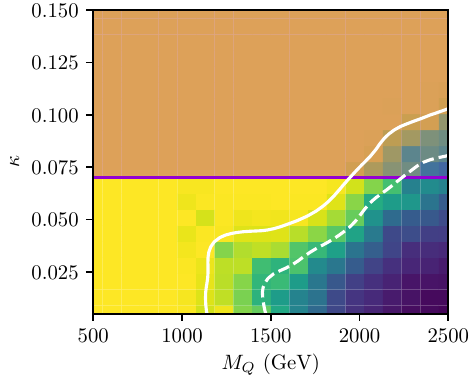}\label{fig:gen1:100}}
  \subfloat[]{
    \includegraphics[width=0.45\textwidth]{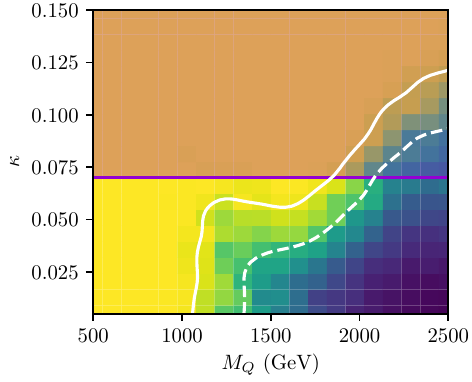}\label{fig:gen1:111}}
  \caption{ Sensitivity of LHC measurements to VLQ production in the $\kappa$ vs
    VLQ mass plane, where $\kappa$ is the coupling to first-generation SM quarks.
    All VLQ ($B, T, X, Y$) masses are set to be degenerate. The green and yellow
    regions are disfavoured at 68\%~CL and 95\%~CL respectively, with the dashed
    and solid white contours delineating the boundaries. The VLQ branching
    fractions to \WZH are \protect\subref{fig:gen1:001}~\WZHzzo
    \protect\subref{fig:gen1:010}~\WZHzoz \protect\subref{fig:gen1:100}~\WZHozz
    and \protect\subref{fig:gen1:211}~\WZHtoo. %
  }%
  \label{fig:1gen}
\end{figure}

\begin{figure}[tbp]
  \centering
  \subfloat[]{
    \includegraphics[width=0.45\textwidth]{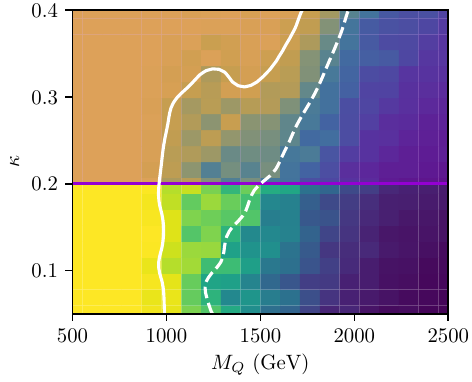}\label{fig:gen2:001}}
  \subfloat[]{
    \includegraphics[width=0.45\textwidth]{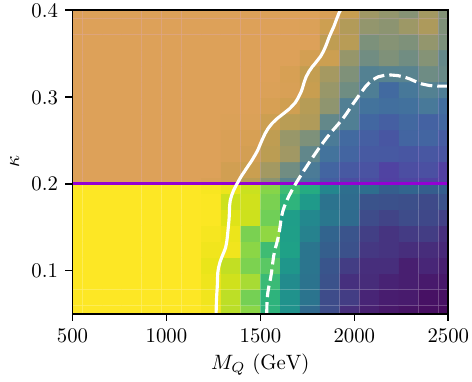}\label{fig:gen2:010}} \\
  \subfloat[]{
    \includegraphics[width=0.45\textwidth]{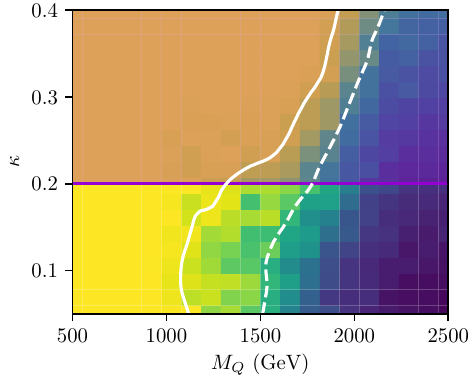}\label{fig:gen2:100}}
  \subfloat[]{
    \includegraphics[width=0.45\textwidth]{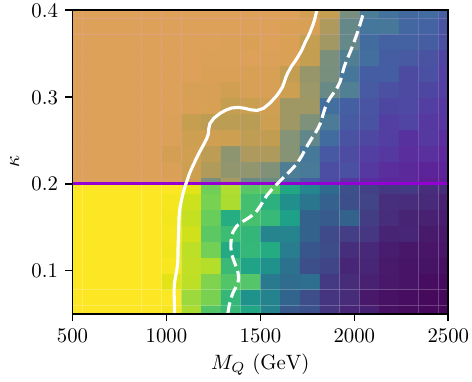}\label{fig:gen2:121}}
  \caption{ Sensitivity of LHC measurements to VLQ production in the $\kappa$ vs
    VLQ mass plane, where $\kappa$ is the coupling to second-generation SM quarks.
    All VLQ ($B, T, X, Y$) masses are set to be degenerate. The green and yellow
    regions are disfavoured at 68\%~CL and 95\%~CL respectively, with the dashed
    and solid white contours delineating the boundaries. The VLQ branching
    fractions to \WZH are \protect\subref{fig:gen2:001}~\WZHzzo
    \protect\subref{fig:gen2:010}~\WZHzoz \protect\subref{fig:gen2:100}~\WZHozz and
    \protect\subref{fig:gen2:121}~\WZHtoo.
  }%
  \label{fig:2gen}
\end{figure}

\begin{figure}[tbp]
  \centering
  \subfloat[]{
    \includegraphics[width=0.45\textwidth]{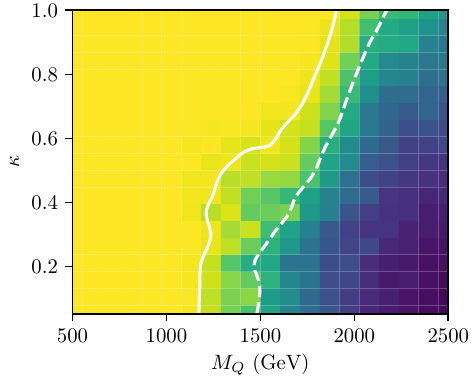}\label{fig:gen3:001}}
  \subfloat[]{
    \includegraphics[width=0.45\textwidth]{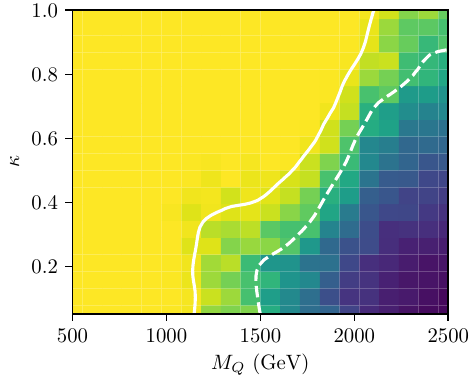}\label{fig:gen3:010}} \\
  \subfloat[]{
    \includegraphics[width=0.45\textwidth]{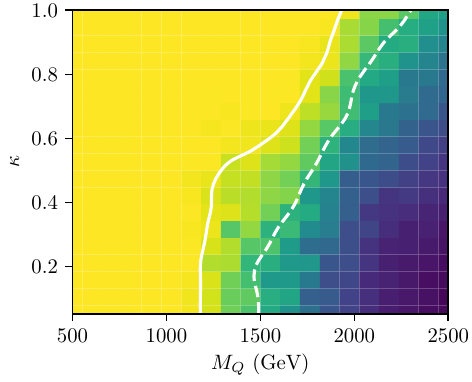}\label{fig:gen3:100}}
  \subfloat[]{
    \includegraphics[width=0.45\textwidth]{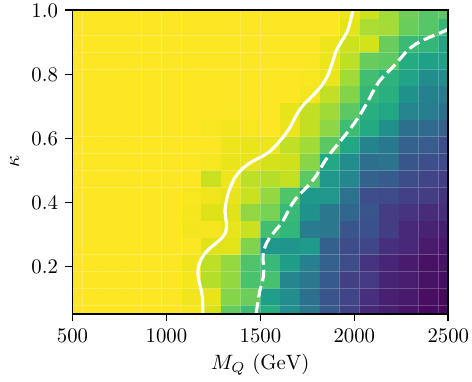}\label{fig:gen3:121}}
  \caption{ Sensitivity of LHC measurements to VLQ production in the $\kappa$ vs
    VLQ mass plane, where $\kappa$ is the coupling to third-generation SM quarks.
    All VLQ ($B, T, X, Y$) masses are set to be degenerate. The green to yellow
    regions are disfavoured at 68\%~CL to 95\%~CL respectively, with the dashed
    and solid white contours delineating the boundaries. The VLQ branching
    fractions to \WZH are \protect\subref{fig:gen3:001}~\WZHzzo
    \protect\subref{fig:gen3:010}~\WZHzoz \protect\subref{fig:gen3:100}~\WZHozz
    and \protect\subref{fig:gen3:121}~\WZHtoo.} %
  \label{fig:3gen}
\end{figure}

\clearpage

\section{Uncorrelated 3rd generation dominant-analyses maps} 
\label{app:dommaps3nocorr}
\begin{figure}[h]
  \centering
  \subfloat[]{
    \includegraphics[width=0.41\textwidth]{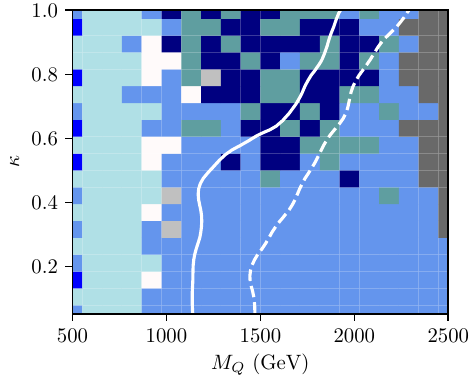}\label{fig:gen3:001:dom:nocorr}}
  \subfloat[]{
    \includegraphics[width=0.41\textwidth]{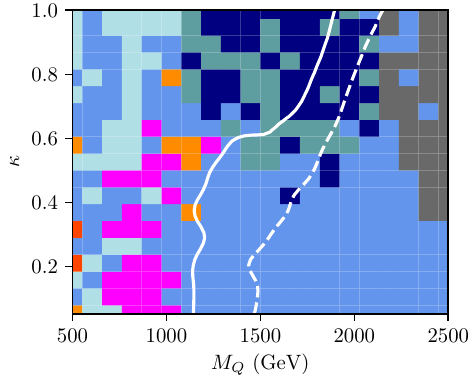}\label{fig:gen3:010:dom:nocorr}} \\
  \subfloat[]{
    \includegraphics[width=0.41\textwidth]{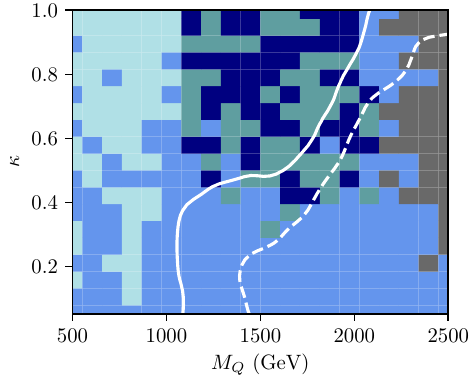}\label{fig:gen3:100:dom:nocorr}}
  \subfloat[]{
    \includegraphics[width=0.41\textwidth]{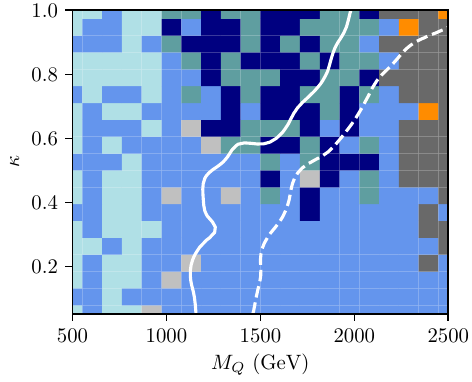}\label{fig:gen3:121:dom:nocorr}}
  \vspace*{2ex}
  \begin{tabular}{lll}
    \swatch{cornflowerblue}~ATLAS $WW$ & \swatch{navy}~ATLAS $\mu$+\MET{}+jet & \swatch{cadetblue}~ATLAS $e$+\MET{}+jet \\
    \swatch{powderblue} CMS $\ell$+\MET{}+jet & \swatch{darkorange}~ATLAS $\mu\mu$+jet & \swatch{magenta} ATLAS~4$\ell$ \\
    \swatch{silver}~ATLAS jets & \swatch{dimgrey}~CMS jets & \swatch{snow}~ATLAS $t\bar{t}$ hadronic
  \end{tabular}
  \vspace*{2ex}
  \caption{Dominant LHC analysis pools contributing to VLQ limit-setting in the $\kappa$ vs
    VLQ mass plane, where $\kappa$ is the coupling to third-generation SM quarks, \emph{without}
    bin-to-bin correlations included in the \CLs calculations.
    All VLQ ($B, T, X, Y$) masses are set to be degenerate. The disfavoured regions
    are located above and to the left of the dashed (68\%~CL)
    and solid (95\%~CL) white contours respectively. The VLQ branching
    fractions to \WZH are \protect\subref{fig:gen3:001:dom:nocorr}~\WZHzzo
    \protect\subref{fig:gen3:010:dom:nocorr}~\WZHzoz \protect\subref{fig:gen3:100:dom:nocorr}~\WZHozz
    and \protect\subref{fig:gen3:121:dom:nocorr}~\WZHtoo. %
  }
  \label{fig:3gen:dom:nocorr}
\end{figure}

\clearpage
\bibliography{vlq_contur.bib}

\begin{thebibliography}{10}
\providecommand{\url}[1]{\texttt{#1}}
\providecommand{\urlprefix}{URL }
\expandafter\ifx\csname urlstyle\endcsname\relax
  \providecommand{\doi}[1]{doi:\discretionary{}{}{}#1}\else
  \providecommand{\doi}{doi:\discretionary{}{}{}\begingroup
  \urlstyle{rm}\Url}\fi
\providecommand{\eprint}[2][]{\url{#2}}

\bibitem{Aguilar-Saavedra:2013qpa}
J.~Aguilar-Saavedra, R.~Benbrik, S.~Heinemeyer and M.~Pérez-Victoria,
\newblock \emph{{Handbook of vectorlike quarks: Mixing and single production}},
\newblock Phys. Rev. D \textbf{88}(9), 094010 (2013),
\newblock \doi{10.1103/PhysRevD.88.094010},
\newblock \eprint{1306.0572}.

\bibitem{Atre:2008iu}
A.~Atre, M.~Carena, T.~Han and J.~Santiago,
\newblock \emph{{Heavy Quarks Above the Top at the Tevatron}},
\newblock Phys. Rev. D \textbf{79}, 054018 (2009),
\newblock \doi{10.1103/PhysRevD.79.054018},
\newblock \eprint{0806.3966}.

\bibitem{Atre:2011ae}
A.~Atre, G.~Azuelos, M.~Carena, T.~Han, E.~Ozcan, J.~Santiago and G.~Unel,
\newblock \emph{{Model-Independent Searches for New Quarks at the LHC}},
\newblock JHEP \textbf{08}, 080 (2011),
\newblock \doi{10.1007/JHEP08(2011)080},
\newblock \eprint{1102.1987}.

\bibitem{Buchkremer:2013bha}
M.~Buchkremer, G.~Cacciapaglia, A.~Deandrea and L.~Panizzi,
\newblock \emph{{Model Independent Framework for Searches of Top Partners}},
\newblock Nucl. Phys. \textbf{B876}, 376 (2013),
\newblock \doi{10.1016/j.nuclphysb.2013.08.010},
\newblock \eprint{1305.4172}.

\bibitem{Aaboud:2018xpj}
{ATLAS Collaboration},
\newblock \emph{{Search for new phenomena in events with same-charge leptons
  and $b$-jets in $pp$ collisions at $\sqrt{s}= 13$ TeV with the ATLAS
  detector}},
\newblock JHEP \textbf{12}, 039 (2018),
\newblock \doi{10.1007/JHEP12(2018)039},
\newblock \eprint{1807.11883}.

\bibitem{Aaboud:2018saj}
{ATLAS Collaboration},
\newblock \emph{{Search for pair- and single-production of vector-like quarks
  in final states with at least one $Z$ boson decaying into a pair of electrons
  or muons in $pp$ collision data collected with the ATLAS detector at
  $\sqrt{s} = 13$ TeV}},
\newblock Phys. Rev. \textbf{D98}(11), 112010 (2018),
\newblock \doi{10.1103/PhysRevD.98.112010},
\newblock \eprint{1806.10555}.

\bibitem{Sirunyan:2018qau}
{CMS Collaboration},
\newblock \emph{{Search for vector-like quarks in events with two oppositely
  charged leptons and jets in proton-proton collisions at $\sqrt{s} =$ 13
  TeV}},
\newblock Submitted to: Eur. Phys. J.  (2018),
\newblock \eprint{1812.09768}.

\bibitem{Sirunyan:2018fjh}
{CMS Collaboration},
\newblock \emph{{Search for single production of vector-like quarks decaying to
  a b quark and a Higgs boson}},
\newblock JHEP \textbf{06}, 031 (2018),
\newblock \doi{10.1007/JHEP06(2018)031},
\newblock \eprint{1802.01486}.

\bibitem{Sirunyan:2018omb}
{CMS Collaboration},
\newblock \emph{{Search for vector-like T and B quark pairs in final states
  with leptons at $\sqrt{s} =$ 13 TeV}},
\newblock JHEP \textbf{08}, 177 (2018),
\newblock \doi{10.1007/JHEP08(2018)177},
\newblock \eprint{1805.04758}.

\bibitem{Butterworth:2016sqg}
J.~M. Butterworth, D.~Grellscheid, M.~Kr{\"a}mer, B.~Sarrazin and D.~Yallup,
\newblock \emph{{Constraining new physics with collider measurements of
  Standard Model signatures}},
\newblock JHEP \textbf{03}, 078 (2017),
\newblock \doi{10.1007/JHEP03(2017)078},
\newblock \eprint{1606.05296}.

\bibitem{Brooijmans:2020yij}
G.~Brooijmans \emph{et~al.},
\newblock \emph{{Les Houches 2019 Physics at TeV Colliders: New Physics Working
  Group Report}} (2020), \eprint{2002.12220}.

\bibitem{Buckley:2010ar}
A.~Buckley, J.~Butterworth, L.~Lonnblad, D.~Grellscheid, H.~Hoeth
  \emph{et~al.},
\newblock \emph{{Rivet user manual}},
\newblock Comput.Phys.Commun. \textbf{184}, 2803 (2013),
\newblock \doi{10.1016/j.cpc.2013.05.021},
\newblock \eprint{1003.0694}.

\bibitem{Bierlich:2019rhm}
C.~Bierlich \emph{et~al.},
\newblock \emph{{Robust Independent Validation of Experiment and Theory: Rivet
  version 3}},
\newblock SciPost Phys. \textbf{8}, 026 (2020),
\newblock \doi{10.21468/SciPostPhys.8.2.026},
\newblock \eprint{1912.05451}.

\bibitem{Read:2002hq}
A.~L. Read,
\newblock \emph{{Presentation of search results: The CL(s) technique}},
\newblock J. Phys. \textbf{G28}, 2693 (2002),
\newblock \doi{10.1088/0954-3899/28/10/313},
\newblock [,11(2002)].

\bibitem{Bahr:2008pv}
M.~Bahr \emph{et~al.},
\newblock \emph{{Herwig++ Physics and Manual}},
\newblock Eur. Phys. J. \textbf{C58}, 639 (2008),
\newblock \doi{10.1140/epjc/s10052-008-0798-9},
\newblock \eprint{0803.0883}.

\bibitem{Bellm:2019zci}
J.~Bellm \emph{et~al.},
\newblock \emph{{Herwig 7.2 Release Note}}  (2019),
\newblock \eprint{1912.06509}.

\bibitem{Cacciapaglia:2018qep}
G.~Cacciapaglia, A.~Carvalho, A.~Deandrea, T.~Flacke, B.~Fuks, D.~Majumder,
  L.~Panizzi and H.-S. Shao,
\newblock \emph{{Next-to-leading-order predictions for single vector-like quark
  production at the LHC}},
\newblock Phys. Lett. B \textbf{793}, 206 (2019),
\newblock \doi{10.1016/j.physletb.2019.04.056},
\newblock \eprint{1811.05055}.

\bibitem{Fuks:2016ftf}
B.~Fuks and H.-S. Shao,
\newblock \emph{{QCD next-to-leading-order predictions matched to parton
  showers for vector-like quark models}},
\newblock Eur. Phys. J. C \textbf{77}(2), 135 (2017),
\newblock \doi{10.1140/epjc/s10052-017-4686-z},
\newblock \eprint{1610.04622}.

\bibitem{delAguila:2000aa}
F.~del Aguila, M.~Perez-Victoria and J.~Santiago,
\newblock \emph{{Effective description of quark mixing}},
\newblock Phys. Lett. B \textbf{492}, 98 (2000),
\newblock \doi{10.1016/S0370-2693(00)01071-6},
\newblock \eprint{hep-ph/0007160}.

\bibitem{Aaboud:2018pii}
{ATLAS Collaboration},
\newblock \emph{{Combination of the searches for pair-produced vector-like
  partners of the third-generation quarks at $\sqrt{s} =$ 13 TeV with the ATLAS
  detector}},
\newblock Phys. Rev. Lett. \textbf{121}(21), 211801 (2018),
\newblock \doi{10.1103/PhysRevLett.121.211801,
  https://doi.org/10.17182/hepdata.83541},
\newblock \eprint{1808.02343}.

\bibitem{Cacciapaglia:2009cu}
G.~Cacciapaglia, S.~Choudhury, A.~Deandrea, N.~Gaur and M.~Klasen,
\newblock \emph{{Dileptonic signatures of T-odd quarks at the LHC}},
\newblock JHEP \textbf{03}, 059 (2010),
\newblock \doi{10.1007/JHEP03(2010)059},
\newblock \eprint{0911.4630}.

\bibitem{Maguire:2017ypu}
E.~Maguire, L.~Heinrich and G.~Watt,
\newblock \emph{{HEPData: a repository for high energy physics data}},
\newblock J. Phys. Conf. Ser. \textbf{898}(10), 102006 (2017),
\newblock \doi{10.1088/1742-6596/898/10/102006},
\newblock \eprint{1704.05473}.

\bibitem{ZllSearch}
{ATLAS Collaboration},
\newblock \emph{{Search for pair production of vector-like quarks in final
  states with at least one $Z$ boson decaying into a pair of electrons or muons
  in $pp$-collision data collected with the ATLAS detector at $\sqrt{s}$ = 13
  TeV}},
\newblock Physical Review D \textbf{98}(11) (2018),
\newblock \doi{10.1103/physrevd.98.112010}.

\bibitem{HadSearch}
{ATLAS Collaboration},
\newblock \emph{{Search for pair production of heavy vector-like quarks
  decaying into hadronic final states in $pp$ collisions at $\sqrt{s}$ = 13 TeV
  with the ATLAS detector}},
\newblock Physical Review D \textbf{98}(9) (2018),
\newblock \doi{10.1103/physrevd.98.092005}.

\bibitem{WtSearch}
{ATLAS Collaboration},
\newblock \emph{{Search for pair production of heavy vector-like quarks
  decaying into high-$p_T$ $W$ bosons and top quarks in the lepton-plus-jets
  final state in $pp$ collisions at $\sqrt{s}$ = 13 TeV with the ATLAS
  detector}},
\newblock Journal of High Energy Physics \textbf{2018}(8) (2018),
\newblock \doi{10.1007/jhep08(2018)048}.

\bibitem{TriLepSearch}
{ATLAS Collaboration},
\newblock \emph{{Search for new phenomena in events with same-charge leptons
  and $b$-jets in $pp$ collisions at $\sqrt{s}$ = 13 TeV with the ATLAS
  detector}},
\newblock Journal of High Energy Physics \textbf{2018}(12) (2018),
\newblock \doi{10.1007/jhep12(2018)039}.

\bibitem{Aad:2015auj}
{ATLAS Collaboration},
\newblock \emph{{Measurement of the transverse momentum and $\phi ^*_{\eta }$
  distributions of Drell–Yan lepton pairs in proton–proton collisions at
  $\sqrt{s}=8$ TeV with the ATLAS detector}},
\newblock Eur. Phys. J. \textbf{C76}(5), 291 (2016),
\newblock \doi{10.1140/epjc/s10052-016-4070-4},
\newblock \eprint{1512.02192}.

\bibitem{Aaboud:2017hox}
{ATLAS Collaboration},
\newblock \emph{{Measurement of the $k_\mathrm{t}$ splitting scales in $Z \to
  \ell\ell$ events in $pp$ collisions at $\sqrt{s} = 8$ TeV with the ATLAS
  detector}},
\newblock JHEP \textbf{08}, 026 (2017),
\newblock \doi{10.1007/JHEP08(2017)026},
\newblock \eprint{1704.01530}.

\bibitem{Aaboud:2017hbk}
{ATLAS Collaboration},
\newblock \emph{{Measurements of the production cross section of a $Z$ boson in
  association with jets in pp collisions at $\sqrt{s} = 13$ TeV with the ATLAS
  detector}},
\newblock Eur. Phys. J. \textbf{C77}(6), 361 (2017),
\newblock \doi{10.1140/epjc/s10052-017-4900-z},
\newblock \eprint{1702.05725}.

\bibitem{Aaboud:2019jcc}
{ATLAS Collaboration},
\newblock \emph{{Searches for scalar leptoquarks and differential cross-section
  measurements in dilepton-dijet events in proton-proton collisions at a
  centre-of-mass energy of $\sqrt{s}$ = 13 TeV with the ATLAS experiment}},
\newblock Eur. Phys. J. \textbf{C79}(9), 733 (2019),
\newblock \doi{10.1140/epjc/s10052-019-7181-x},
\newblock \eprint{1902.00377}.

\bibitem{Aaboud:2017fye}
{ATLAS Collaboration},
\newblock \emph{{Measurements of electroweak $Wjj$ production and constraints
  on anomalous gauge couplings with the ATLAS detector}},
\newblock Eur. Phys. J. \textbf{C77}(7), 474 (2017),
\newblock \doi{10.1140/epjc/s10052-017-5007-2},
\newblock \eprint{1703.04362}.

\bibitem{Aaboud:2018eqg}
{ATLAS Collaboration},
\newblock \emph{{Measurements of $t\bar{t}$ differential cross-sections of
  highly boosted top quarks decaying to all-hadronic final states in $pp$
  collisions at $\sqrt{s}=13\,$ TeV using the ATLAS detector}},
\newblock Phys. Rev. \textbf{D98}(1), 012003 (2018),
\newblock \doi{10.1103/PhysRevD.98.012003},
\newblock \eprint{1801.02052}.

\bibitem{Sirunyan:2018wem}
{CMS Collaboration},
\newblock \emph{{Measurement of differential cross sections for the production
  of top quark pairs and of additional jets in lepton+jets events from pp
  collisions at $\sqrt{s} =$ 13 TeV}},
\newblock Phys. Rev. \textbf{D97}(11), 112003 (2018),
\newblock \doi{10.1103/PhysRevD.97.112003},
\newblock \eprint{1803.08856}.

\bibitem{Khachatryan:2016mnb}
{CMS Collaboration},
\newblock \emph{{Measurement of differential cross sections for top quark pair
  production using the lepton+jets final state in proton-proton collisions at
  13 TeV}},
\newblock Phys. Rev. \textbf{D95}(9), 092001 (2017),
\newblock \doi{10.1103/PhysRevD.95.092001},
\newblock \eprint{1610.04191}.

\bibitem{Sirunyan:2018ptc}
{CMS Collaboration},
\newblock \emph{{Measurements of differential cross sections of top quark pair
  production as a function of kinematic event variables in proton-proton
  collisions at $ \sqrt{s}=13 $ TeV}},
\newblock JHEP \textbf{06}, 002 (2018),
\newblock \doi{10.1007/JHEP06(2018)002},
\newblock \eprint{1803.03991}.

\bibitem{HbbSearch}
{ATLAS Collaboration},
\newblock \emph{{Search for pair production of up-type vector-like quarks and
  for four-top-quark events in final states with multiple $b$-jets with the
  ATLAS detector}},
\newblock Journal of High Energy Physics \textbf{2018}(7) (2018),
\newblock \doi{10.1007/jhep07(2018)089}.

\bibitem{ZnunuSearch}
{ATLAS Collaboration},
\newblock \emph{{Search for pair production of vector-like top quarks in events
  with one lepton, jets, and missing transverse momentum in $\sqrt{s}$ = 13 TeV
  TeV $pp$ collisions with the ATLAS detector}},
\newblock Journal of High Energy Physics \textbf{2017}(8) (2017),
\newblock \doi{10.1007/jhep08(2017)052}.

\bibitem{WbSearch}
{ATLAS Collaboration},
\newblock \emph{{Search for pair production of heavy vector-like quarks
  decaying to high-$p_T$ $W$ bosons and $b$ quarks in the lepton-plus-jets
  final state in $pp$ collisions at $\sqrt{s}$ = 13 TeV with the ATLAS
  detector}},
\newblock Journal of High Energy Physics \textbf{2017}(10) (2017),
\newblock \doi{10.1007/jhep10(2017)141}.

\bibitem{Deandrea:1997wk}
A.~Deandrea,
\newblock \emph{{Atomic parity violation in cesium and implications for new
  physics}},
\newblock Phys. Lett. B \textbf{409}, 277 (1997),
\newblock \doi{10.1016/S0370-2693(97)00843-5},
\newblock \eprint{hep-ph/9705435}.

\bibitem{Wood:1997zq}
C.~Wood, S.~Bennett, D.~Cho, B.~Masterson, J.~Roberts, C.~Tanner and C.~E.
  Wieman,
\newblock \emph{{Measurement of parity nonconservation and an anapole moment in
  cesium}},
\newblock Science \textbf{275}, 1759 (1997),
\newblock \doi{10.1126/science.275.5307.1759}.

\bibitem{ALEPH:2005ab}
S.~Schael \emph{et~al.},
\newblock \emph{{Precision electroweak measurements on the $Z$ resonance}},
\newblock Phys. Rept. \textbf{427}, 257 (2006),
\newblock \doi{10.1016/j.physrep.2005.12.006},
\newblock \eprint{hep-ex/0509008}.

\bibitem{Sirunyan:2018xdh}
A.~M. Sirunyan \emph{et~al.},
\newblock \emph{{Measurements of the differential jet cross section as a
  function of the jet mass in dijet events from proton-proton collisions at $
  \sqrt{s}=13 $ TeV}},
\newblock JHEP \textbf{11}, 113 (2018),
\newblock \doi{10.1007/JHEP11(2018)113},
\newblock \eprint{1807.05974}.

\bibitem{Aaboud:2017wsi}
{ATLAS Collaboration},
\newblock \emph{{Measurement of inclusive jet and dijet cross-sections in
  proton-proton collisions at $\sqrt{s}=13$ TeV with the ATLAS detector}},
\newblock JHEP \textbf{05}, 195 (2018),
\newblock \doi{10.1007/JHEP05(2018)195},
\newblock \eprint{1711.02692}.

\bibitem{Aaboud:2019lxo}
{ATLAS Collaboration},
\newblock \emph{{Measurement of the four-lepton invariant mass spectrum in 13
  TeV proton-proton collisions with the ATLAS detector}},
\newblock Tech. rep., CERN (2019), \eprint{1902.05892}.

\bibitem{Aaboud:2017rwm}
{ATLAS Collaboration},
\newblock \emph{{$ZZ \to \ell^{+}\ell^{-}\ell^{\prime +}\ell^{\prime -}$
  cross-section measurements and search for anomalous triple gauge couplings in
  13 TeV $pp$ collisions with the ATLAS detector}},
\newblock Phys. Rev. \textbf{D97}(3), 032005 (2018),
\newblock \doi{10.1103/PhysRevD.97.032005},
\newblock \eprint{1709.07703}.

\bibitem{Aaboud:2018uzf}
{ATLAS Collaboration},
\newblock \emph{{Measurements of differential cross sections of top quark pair
  production in association with jets in ${pp}$ collisions at $\sqrt{s}=13$ TeV
  using the ATLAS detector}},
\newblock JHEP \textbf{10}, 159 (2018),
\newblock \doi{10.1007/JHEP10(2018)159},
\newblock \eprint{1802.06572}.

\bibitem{Aaboud:2017fha}
{ATLAS Collaboration},
\newblock \emph{{Measurements of top-quark pair differential cross-sections in
  the lepton+jets channel in $pp$ collisions at $\sqrt{s}=13$ TeV using the
  ATLAS detector}},
\newblock JHEP \textbf{11}, 191 (2017),
\newblock \doi{10.1007/JHEP11(2017)191},
\newblock \eprint{1708.00727}.

\bibitem{Aaboud:2018eki}
{ATLAS Collaboration},
\newblock \emph{{Measurements of inclusive and differential fiducial
  cross-sections of $ t\overline{t} $ production with additional heavy-flavour
  jets in proton-proton collisions at $ \sqrt{s} $ = 13 TeV with the ATLAS
  detector}},
\newblock JHEP \textbf{04}, 046 (2019),
\newblock \doi{10.1007/JHEP04(2019)046},
\newblock \eprint{1811.12113}.

\bibitem{Araque:2015cna}
J.~P. Araque, N.~F. Castro and J.~Santiago,
\newblock \emph{{Interpretation of Vector-like Quark Searches: Heavy Gluons in
  Composite Higgs Models}},
\newblock JHEP \textbf{11}, 120 (2015),
\newblock \doi{10.1007/JHEP11(2015)120},
\newblock \eprint{1507.05628}.

\bibitem{Abdallah:2020pec}
W.~Abdallah \emph{et~al.},
\newblock \emph{{Reinterpretation of LHC Results for New Physics: Status and
  Recommendations after Run 2}}  (2020),
\newblock \eprint{2003.07868}.

\end{thebibliography}

\nolinenumbers

\end{document}